\documentclass[useAMS,usenatbib]{mn2e}
\pdfoutput=1
\usepackage{graphicx, color}
\usepackage[]{hyperref}
\usepackage{amsmath}
\usepackage{pdflscape}
\usepackage{multirow,bigdelim}
\usepackage{calc}
\usepackage{threeparttable}

\def\aj{AJ}%
\def\apj{ApJ}%
\def\apjl{ApJ}%
\def\apjs{ApJS}%
\def\aap{A\&A}%
\def\aaps{A\&AS}%
\def\mnras{MNRAS}%
\def\pasa{PASA}%
\def\prc{Phys.~Rev.~C}%
\def\solphys{Sol.~Phys.}%
\def\zap{ZAp}%
\usepackage[]{caption}
\usepackage[font=small]{caption}
\usepackage[labelformat = empty,position=top]{subcaption}
\usepackage[export]{adjustbox}

\usepackage[T1]{fontenc}
\usepackage{ae,aecompl}
{}

\usepackage{mathptmx}
\usepackage{txfonts}

\title[MNRAS \LaTeX\ guide for authors]{$^{22}$Ne and $^{23}$Na ejecta from
intermediate-mass stars: \\ 
The impact of the new LUNA rate for $^{22}$Ne$(p,\gamma)^{23}$Na}

\author[A. Slemer, P. Marigo, D. Piatti et al.]
  {A. Slemer$^1$, 
  P. Marigo$^{1,2}$, D. Piatti$^{1,2}$, M. Aliotta$^{3}$, D. Bemmerer$^{4}$, A. Best$^{5}$, \newauthor A. Boeltzig$^{6}$, A. Bressan$^{7}$, C. Broggini $^{2}$, C.G.Bruno$^{3}$, A. Caciolli$^{1,2}$, F. Cavanna$^{8}$,\newauthor G.F. Ciani$^{6}$, P. Corvisiero$^{8}$, T. Davinson$^{3}$, R. Depalo$^{1,2}$, A. Di Leva$^{5}$, Z. Elekes$^{9}$, \newauthor F. Ferraro$^{8}$, A. Formicola$^{10}$, Zs. F\"{u}l\"{o}p$^{9}$, G. Gervino$^{11}$, A. Guglielmetti$^{12}$, C. Gustavino$^{13}$,\newauthor G. Gy\"{u}rky$^{9}$, G. Imbriani$^{5}$, M. Junker$^{10}$, R. Menegazzo$^{2}$, V. Mossa$^{14}$, F.R. Pantaleo$^{14}$, 
  \newauthor P. Prati$^{8}$, O. Straniero$^{15,10}$, T. Sz\"{u}cs$^{9}$, M.P. Tak\'{a}cs$^{4}$, D. Trezzi$^{11}$\\
  $^1$ Department of Physics and Astronomy G.\ Galilei, University of Padova
        Vicolo dell'Osservatorio 3, I-35122 Padova, Italy\\
  $^2$ INFN of Padova, Via Marzolo 8, I-35131 Padova, Italy\\
  $^3$ SUPA, School of Physics and Astronomy, University of Edinburgh, EH9 3FD Edinburgh, United Kingdom\\
  $^4$ Helmholtz-Zentrum Dresden-Rossendorf, Bautzner Landstr. 400, 01328 Dresden, Germany\\
  $^5$ Universit\`a di Napoli "Federico II" and INFN, Sezione di Napoli, 80126 Napoli, Italy\\
  $^6$ Gran Sasso Science Institute, Viale F.~Crispi 7, 67100 L'Aquila, Italy \\   
  $^7$ SISSA, via Bonomea 265, Trieste, Italy \\
  $^8$ Universit\`a degli Studi di Genova and INFN, Sezione di Genova, Via Dodecaneso 33, 16146 Genova, Italy\\
  $^9$ Institute for Nuclear Research (MTA ATOMKI), PO Box 51, HU-4001 Debrecen, Hungary\\
  $^{10}$ INFN, Laboratori Nazionali del Gran Sasso (LNGS), 67100 Assergi, Italy\\
  $^{11}$ Universit\`a degli Studi di Torino and INFN, Sezione di Torino, Via P. Giuria 1, 10125 Torino, Italy\\
  $^{12}$ Universit\`a degli Studi di Milano and INFN, Sezione di Milano, Via G. Celoria 16, 20133 Milano, Italy\\
  $^{13}$ INFN, Sezione di Roma La Sapienza, Piazzale A. Moro 2, 00185 Roma, Italy\\
  $^{14}$ Universit\`a degli Studi di Bari and INFN, Sezione di Bari, 70125 Bari, Italy\\
  $^{15}$ INAF, Osservatorio Astronomico di Teramo, 6410 Teramo, Italy} 
\date{Accepted 2016 November 21. Received 2016 November 21; in original form 2016 July 11}

\pubyear{2016}

\begin{document}
\label{firstpage}
\pagerange{\pageref{firstpage}--\pageref{lastpage}}
\maketitle

% Abstract of the paper
\begin{abstract}
We investigate the impact of the new LUNA rate for the nuclear
reaction $^{22}$Ne$(p,\gamma)^{23}$Na on the chemical ejecta of
intermediate-mass stars, with particular focus on the
thermally-pulsing asymptotic giant branch (TP-AGB) stars that
experience hot-bottom burning.  To this aim we use the \texttt{PARSEC}
and \texttt{COLIBRI} codes to compute the complete evolution, from the
pre-main sequence up to the termination of the TP-AGB phase, of a 
set of stellar models with initial masses in the range
$3.0\,M_{\odot} - 6.0\,M_{\odot}$, and metallicities
$Z_{\rm i}=0.0005$, $Z_{\rm i}=0.006$, and $Z_{\rm i} = 0.014$.  We find that the 
new LUNA measures have much reduced the nuclear uncertainties 
of the $^{22}$Ne and $^{23}$Na AGB ejecta, which drop from factors 
of $\simeq 10$ to only a factor of
few for the lowest metallicity models.  
Relying on the most recent estimations
for the destruction rate of $^{23}$Na, the uncertainties that still
affect the $^{22}$Ne and $^{23}$Na AGB ejecta are mainly dominated by
evolutionary aspects (efficiency of mass-loss, third dredge-up,
convection). Finally, we discuss how the LUNA
results impact on the hypothesis that invokes massive AGB
stars as the main agents of the observed O-Na anti-correlation in Galactic 
globular clusters.
We derive quantitative indications on the efficiencies of key physical 
processes (mass loss, third dredge-up, sodium destruction)
in order to simultaneously reproduce both the  Na-rich, O-poor extreme of
the anti-correlation, and the observational constraints on the CNO abundance.
Results for the corresponding chemical ejecta are made publicly available.
\end{abstract}


%%%%%%%%%%%%%%%%%%%%%%%%%%%%%%%%%%%%%%%%%%%%%%%%%%

%%%%%%%%%%%%%%%%% BODY OF PAPER %%%%%%%%%%%%%%%%%%

\section{Introduction}
Low- and intermediate-mass stars (with initial masses up to 6-8 M$_\odot$) play a key role in the chemical evolution of the Universe. During their lives they experience a rich nucleosynthesis and various mixing episodes, eventually ejecting 
significant amounts of newly synthesized elements into the interstellar medium.
Quantifying their chemical contribution is of key relevance to understand the  chemical enrichment of galaxies
and several theoretical studies were carried out to this purpose \citep{Cristallo_etal15, Doherty_etal14a, Doherty_etal14b, Cristallo_etal11, Siess_10, Ventura_etal13, KarakasLattanzio_14, VenturaMarigo_10, Marigo_01, ForestiniCharbonnel_97}.

Despite the valuable efforts, large uncertainties still affect 
the yields of various elements, due to the uncertainties of the physical
processes (i.e., mass loss, convection, mixing, nuclear reactions) 
that characterize the advanced evolutionary stages, in particular the thermally-pulsing asymptotic giant branch  (TP-AGB).

In this study we will focus on the nucleosynthesis of $^{22}$Ne and $^{23}$Na and their
ejecta produced by stars massive enough to experience the process of hot-bottom burning (hereinafter also HBB) during the AGB phase (M$_{\rm i}$ > 3-4 M$_{\odot}$).
When, during the quiescent AGB evolution, the temperature at the base of convective envelope exceeds $\mathrm{\simeq 0.07\, GK}$, the CNO, NeNa and MgAl cycles are efficiently activated \citep{ForestiniCharbonnel_97}, with the effect of significantly altering the abundances of the catalysts involved in the proton-capture reactions.
The NeNa cycle is responsible for affecting the abundances of isotopes between $\mathrm{^{20}Ne}$ and $\mathrm{^{24}Mg}$.
The current uncertainties of the  $^{22}$Ne and $^{23}$Na ejecta are dramatically high, up
to factors of $\simeq 10$,  given the large uncertainties that affect a few reaction rates involved in the NeNa cycle 
\citep[e.g.,][]{Karakas_10, Izzard_etal07, VenturaDantona_05a}.
The poor knowledge of resonances in
$\mathrm{^{23}Na(p,\alpha)^{20}Ne}$ and
$\mathrm{^{23}Na(p,\gamma)^{24}Mg}$ is critical \citep{Hale_etal04}.
The rate of the NeNa cycle is determined by the slowest reaction of
the chain, the $\mathrm{^{20}Ne(p,\gamma)^{21}Na}$
\citep{Rolfs&Rodney88}, and most uncertainties are caused by the
$\mathrm{^{22}Ne(p,\gamma)^{23}Na}$ reaction.  In fact, the systematic
analysis carried out by \citet{Izzard_etal07} has shown that the
ejecta of $^{23}$Na is dominated by the uncertainties in the
$\mathrm{^{22}Na(p,\gamma)^{23}Na}$ rate, with the destruction rates
of $\mathrm{^{23}Na(p,\gamma)^{24}Mg}$ and
$\mathrm{^{23}Na(p,\alpha)^{20}Ne}$ playing a lesser role.

The contribution of resonances to the
$\mathrm{^{22}Na(p,\gamma)^{23}Na}$ rate is still uncertain because of
the large number of levels of $\mathrm{^{23}Na}$, the complexity of
direct measurements and the interpretation of indirect data. This is
particularly true for resonances at energies corresponding to the
typical temperatures of hot-bottom burning in AGB stars,
i.e. $\mathrm{0.07\, GK \la T \la 0.11\, GK}$
\citep[e.g.,][]{Marigo_etal13, Boothroyd_etal95}.

In this paper we  analyze the impact on $^{22}$Ne and $^{23}$Na ejecta 
of the new rate for $\mathrm{^{22}Ne(p,\gamma)^{23}Na}$ that has been recently
revised following accurate  measurements at LUNA \citep{Cavanna_etal15}.
We computed a large set of evolutionary models for stars
that experience HBB and the third dredge-up during the AGB phase.
The results are compared to those obtained with other versions 
of the rate in the literature, and also by
varying other parameters that are critical
for the evolution of AGB stars.
The final aim is to re-evaluate the uncertainties that affect the $^{22}$Ne and $^{23}$Na ejecta, as well as to explore the implications we may draw on the hypothesis that
metal-poor AGB stars are promising candidates to explain the O-Na
anti-correlation exhibited by Galactic globular clusters' stars 
\citep[e.g.,][]{Dantona_etal16, Dercole_etal12, VenturaDantona_09}.

The structure of the paper is organized as follows.
In Section~\ref{sec_measures} we recall the main results and improvements
obtained with recent LUNA data for the S-factor of the 
$^{22}$Ne(p,$\gamma)^{23}$Na reaction.
In Section~\ref{sec_evol} we outline the most relevant characteristics and 
input physics of the stellar evolutionary models. 
In Section~\ref{sec_abund} we discuss
the evolution of the surface abundance of neon, sodium and magnesium isotopes 
in stars that experience HBB and the third dredge-up during the TP-AGB phase.
A quantitative comparison of the $^{22}$Ne and $^{23}$Na ejecta as a function of 
the initial stellar mass and metallicity is provided in Section~\ref{sec_ejecta}.
In the context of the origin of the O-Na anti-correlation
 in Galactic globular clusters (GGCs), Section~\ref{sec_ggcs} analyzes the 
 impact of the new LUNA rate on the AGB star hypothesis.
 Section~\ref{sec_sum} closes the paper providing a summary and  a few 
 final remarks.

\section{The new LUNA rate for $^{22}$N\lowercase{e}(\lowercase{p},$\gamma)^{23}$N\lowercase{a}}
\label{sec_measures}
\begin{figure}
\resizebox{\hsize}{!}{\includegraphics{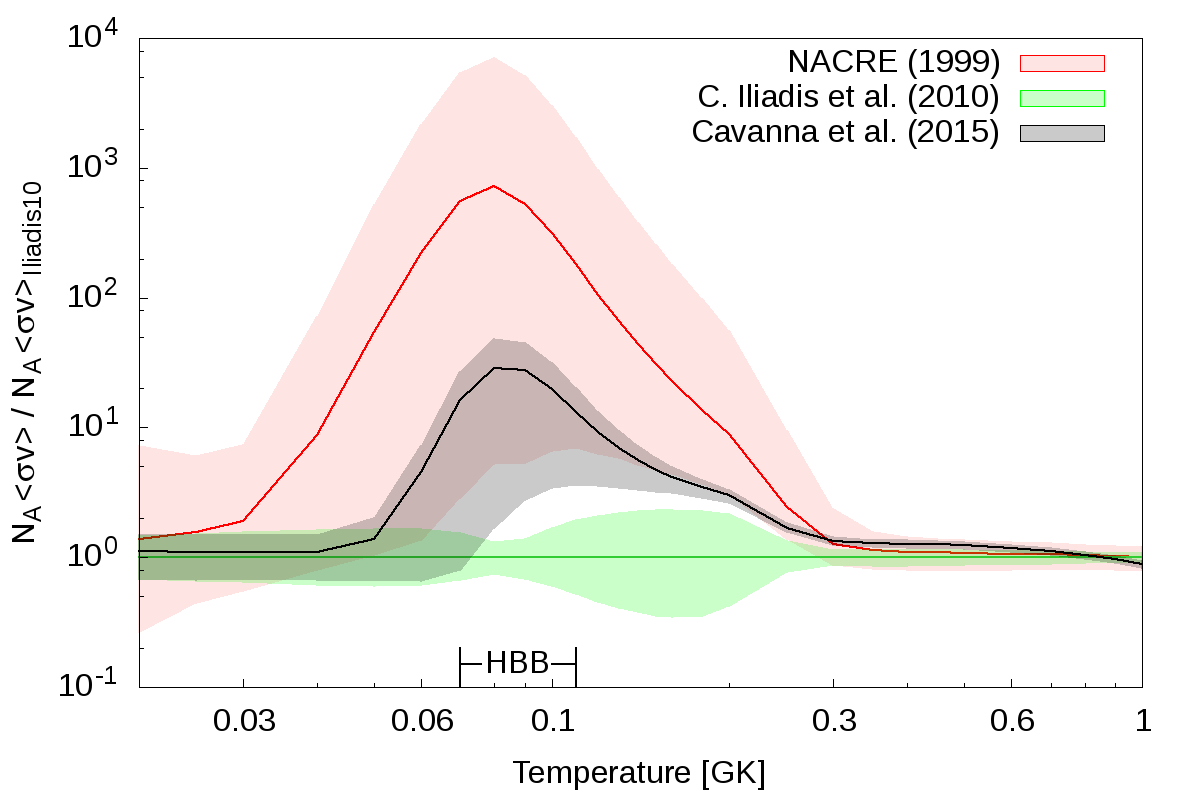}}
\caption{Total reaction rate normalized to IL10, as a function of temperature, calculated by NACRE (red line), IL10 (green line) and \citet{Cavanna_etal15} (black line). The colored regions show the corresponding uncertainties. The range of temperatures relevant for the occurrence of HBB  is also indicated.}
\label{fig_Cavanna_etal15}
\end{figure}

In stellar models the $^{22}$Ne$(p,\gamma)^{23}$Na reaction has
usually been described according to one of the two popular rate
compilations quoted in Table~\ref{tab:models}. They collect direct and
indirect data on $\mathrm{^{22}Ne(p,\gamma)^{23}Na}$ resonance
strengths, namely: \citet[][hereinafter NACRE]{Angulo_etal99};
\citet[][hereinafter IL10]{Iliadis_etal10a, Iliadis_etal10b}. The
latter was recently updated by the STARLIB group including new
indirect data \citep{Sallaska_etal13}. It differs from the previous
version by less than $3\%$ in the range of temperatures explored in
this paper and we will still refer to IL10. 
Because of the
uncertainties that affect some resonance strengths and the different
treatment of other debated resonances 
\citep[][and references therein]{Cavanna_etal15}, 
the discrepancy between the NACRE
and the IL10 total reaction rate is up 
to a factor of $\simeq$ 1000 at $T \sim 0.08$ GK, well inside the range that 
is relevant for HBB (see figure~\ref{fig_Cavanna_etal15}).

This situation was recently improved by direct measurements performed at LUNA in the underground facility of the Gran Sasso National Laboratory, where the low-background environment \citep{Broggini_etal10, Costantini_etal09} and the available setup \citep{Cavanna_etal14} offer the possibility to investigate nuclear reactions down to very low energies \citep{Cavanna_etal15}.

In \citet{Cavanna_etal15} three new resonances were observed for the first time, at $156.2, 189.5$ and $\mathrm{259.7\,keV}$ laboratory energy. In addition, more precise $\mathrm{^{23}Na}$ excitation energies corresponding to the new resonances were found, except for the $\mathrm{189.5\,keV}$ resonance. For other three resonances, at $71, 105$, and $\mathrm{215\,keV}$, new upper limits to the strengths were obtained. 

In order to estimate the new total reaction rate a Monte Carlo method was used \citep[see for more details][]{Cavanna_etal15}. The new data were combined with previous direct measurement results for higher energy resonances \citep{Depalo_etal15} and with literature resonant and non-resonant contributions \citep{Iliadis_etal10a, Iliadis_etal10b}.

The new central value of the reaction rate lies between those of NACRE
and IL10, see Fig. \ref{fig_Cavanna_etal15}. The more precise
excitation energies found for the $\mathrm{156.2\,keV}$ and
$\mathrm{259.7\,keV}$ resonances are responsible for the increase of
the reaction rate by a factor of $3-5$ with respect to IL10 at
temperatures $\mathrm{0.12\, GK \la T \la 0.20\, GK}$. For
$\mathrm{0.08\, GK \la T \la 0.25\, GK}$ the $\mathrm{1\sigma}$ lower
limit of the new reaction rate is above the upper limit calculated by
IL10. Another effect of the direct observation of three new resonances
in the range of temperatures $\mathrm{1.7\, GK \la T \la 2.5\, GK}$ is
the reduction of the error bars of the total reaction rate, in
comparison to NACRE and IL10. Nevertheless the new reaction rate has
still larger uncertainties than IL10 for $\mathrm{0.05\, GK \la T \la
0.1\, GK}$. This is because of the different treatment of the
$\mathrm{71\, and\, 105\,keV}$ resonances, for which further
investigation is necessary.  As a matter of fact, in the range of
temperatures of HBB in TP-AGB stars, see
Fig. \ref{fig_Cavanna_etal15}, the new reaction rate is higher than
IL10 by about a factor of 20, which will significantly impact on model
predictions.

\section{Stellar evolutionary models}
\label{sec_evol}
Stellar evolutionary models for intermediate-mass stars were calculated with the \texttt{PARSEC} and \texttt{COLIBRI} codes 
\citep[][]{Bressan_etal12, Marigo_etal13}. The \texttt{PARSEC} tracks cover the evolution from the pre-main sequence up to the initial stages of the thermally-pulsing asymptotic giant branch (TP-AGB) phase.
Starting from the first thermal pulse computed with \texttt{PARSEC}, the subsequent evolution up to the almost complete ejection of the envelope is followed with the \texttt{COLIBRI} code.
The reader should refer to the aforementioned papers for all details about the two codes.

We shortly recall the prescriptions for the adopted input physics which are mostly relevant for this work, common to both \texttt{PARSEC}  and \texttt{COLIBRI}.
Stellar convection is described by means of the classical mixing length theory \citep{mlt_58}.
The mixing length is assumed to scale linearly with the pressure scale height, $H_{\rm p}$, according to 
setting the proportionality factor $\alpha_{\rm MLT}=1.74$, following our recent calibration of the solar model \citep{Bressan_etal12}.

Overshoot is applied to the borders of convective cores as well as at the base of the convective envelope,  and is described through the parameter $\Lambda$ which sets its extension in units of $H_{\rm p}$. In the range of intermediate stellar masses under consideration our default choice is $\Lambda_{\rm c} = 0.5$ for convective core overshoot (across the classical Schwarzschild border)
and $\Lambda_{\rm e} = 0.7$ for envelope overshoot. 

The network of nuclear reaction rates includes the proton-proton chains, the CNO tri-cycle, the
NeNa and MgAl cycles, and the most important $\alpha$-capture reactions, together with few $\alpha$-n reactions.  
In the burning regions, at each time step, the network is integrated to derive
the abundances of 26 chemical species: $^1$H, D,
$^3$He, $^4$He,$^7$Li, $^8$Be, $^4$He, $^{12}$C, $^{13}$C, $^{14}$N,
$^{15}$N, $^{16}$N, $^{17}$N, $^{18}$O, $^{19}$F, $^{20}$Ne,
$^{21}$Ne, $^{22}$Ne, $^{23}$Na, $^{24}$Mg, $^{25}$Mg, $^{26}$Mg,
$^{26}$Al$^m$, $^{26}$Al$^g$, $^{27}$Al, $^{28}$Si.
Our initial reference set of nuclear reaction rates is taken 
from the JINA reaclib database
\citep{Cyburt_etal10}, from which we also take the $Q$-value of each
reaction. In total we consider $42$ reaction rates \citep[for the complete list and references see Table~1 of][]{Marigo_etal13}.
No neutron-capture reactions are included.

We computed the evolution of intermediate-mass stars with initial
masses in the range between 3.0$M_{\odot}$ and $M_{\rm up}$ (in
incremental steps of $0.2\,M_{\odot}$), the upper limit being the
maximum mass for a star to develop a carbon-oxygen degenerate core at
the end of the core helium burning phase. With the adopted input
physics and prescriptions (e.g. the extension of convective core
overshoot) we find that $M_{\rm up} \simeq 5-6\, M_{\odot}$ for the
metallicity range under consideration.  As for the chemical
composition we consider three choices of the initial metallicity
$Z_{\rm i}$ and helium content $Y_{\rm i}$, namely: $(Z_{\rm i},
Y_{\rm i}) = (0.0005,0.249); (0.006, 0.0259); (0.014, 0.273)$. For
each $Z_{\rm i}$ the corresponding $Y_{\rm i}$ is derived assuming a
linear relation with a helium-to-metals enrichment ratio $\Delta
Y/\Delta Z=1.78$, a primordial helium abundance $Y_{\rm p }=0.2485$,
a Sun's metallicity at its birth $Z_{\odot}=0.01774$, and a
present-day Sun's metallicity $Z_{\odot}=0.01524$ \citep[more details
  can be found in][]{Bressan_etal12}.  The initial distribution of
metals is assumed to follow a scaled-solar pattern
\citep{Caffau_etal11} for $Z_{\rm i}=$0.006, 0.014, while we adopt an
$\alpha$-enhanced mixture with [$\alpha$/Fe]=0.4 for $Z_{\rm i}=0.0005$. This
latter is suitable to describe the chemical pattern of low-metallicity
Halo stars and first-generation stars in Galactic globular clusters.  
Considering that the initial metallicity $Z_{\rm i}=0.0005$
includes the abundances of all elements heavier than helium (hence
also the $\alpha$-elements), the iron content of our $\alpha$-enhanced
mixture corresponds to [Fe/H]$\sim -1.56$ (see also Section~\ref{sec_ggcs}).
The assumed chemical composition of the
evolutionary models is summarized in Table~\ref{tab:models}.

Major effects on the NeNa nucleosynthesis show up during the TP-AGB phase of stars with hot-bottom burning. Therefore it is worth recalling the main features of the \texttt{COLIBRI} code, and our reference 
set of prescriptions according to \citet[][hereinafter also $M13$]{Marigo_etal13}.
Other model assumptions, summarized in Table~\ref{tab_test},  will be 
tested and discussed later in the paper
(Sects.~\ref{ssec_totunc} and \ref{sec_ggcs}).

The evolution of the models presented in this work is followed 
at constant mass until the onset of the TP-AGB phase.
To compute the mass-loss rate along the TP-AGB we first adopt the semi-empirical relation  by \citet{SchroderCuntz_05}, modified according to \citet{Rosenfield_etal14}, and then, as the star enters the dust-driven wind regime, we adopt  an exponential form $\dot M \propto \exp({M^{a} R^{b}})$,  
as a function of stellar mass and radius \citep[see for more details][]
{Bedijn88, Girardi_etal10, Rosenfield_etal14}. The latter was calibrated on a sample of Galactic long-period variables with measured mass-loss rates, pulsation periods, masses, effective temperatures, and radii. 
We emphasize  that the combination of the two mass-loss laws was calibrated through observations of resolved AGB stars in a large sample nearby galaxies of low  metallicities and various star-formation histories,  
observed with the HST/ACS Nearby Galaxy
Survey Treasury \citep{Rosenfield_etal16, Rosenfield_etal14, Dalcanton_etal09}, leading to a satisfactory reproduction of the measured star counts and luminosity functions. 

In \texttt{COLIBRI} we account for the changes in the surface chemical composition caused by the occurrence of the third dredge-up and hot-bottom burning.
As for the third dredge-up we adopt a hybrid approach that involves detailed
physics as well as free parameters.
We perform numerical integrations of the envelope structure at the stage of the post-flash luminosity peak to determine if and when the third dredge-up is expected to take place according to a temperature criterion \citep{MarigoGirardi_07}.
The chemical composition of the pulse-driven convection zone is predicted by solving a nuclear network that includes the main $\alpha$-capture reactions.
The efficiency of the third dredge-up as a function of stellar mass and metallicity is computed with an analytic formalism based on full stellar models \citep{Karakas_etal02}. It includes adjustable parameters which are suitably modified in order to reproduce basic observables of AGB stars, such as carbon star luminosity functions, M-C transition luminosities, surface C/O ratios  
\citep[e.g.,][]{Marigo15, Rosenfield_etal14, Marigo_etal13, Girardi_etal10, Marigo_etal08, MarigoGirardi_07, Marigo_etal03}.

The process of hot-bottom burning experienced by massive AGB stars (with initial masses $M_{\rm i} \ge 3-4\, M_{\odot}$, depending on metallicity and model details) is consistently taken into account in terms of energetics and nucleosynthesis.The nucleosynthesis of all species is coupled in time and in 
space with a diffusive description  of convection.
\begin{figure*}
\centering
\begin{minipage}{0.48\textwidth}
  \resizebox{\hsize}{!}{\includegraphics{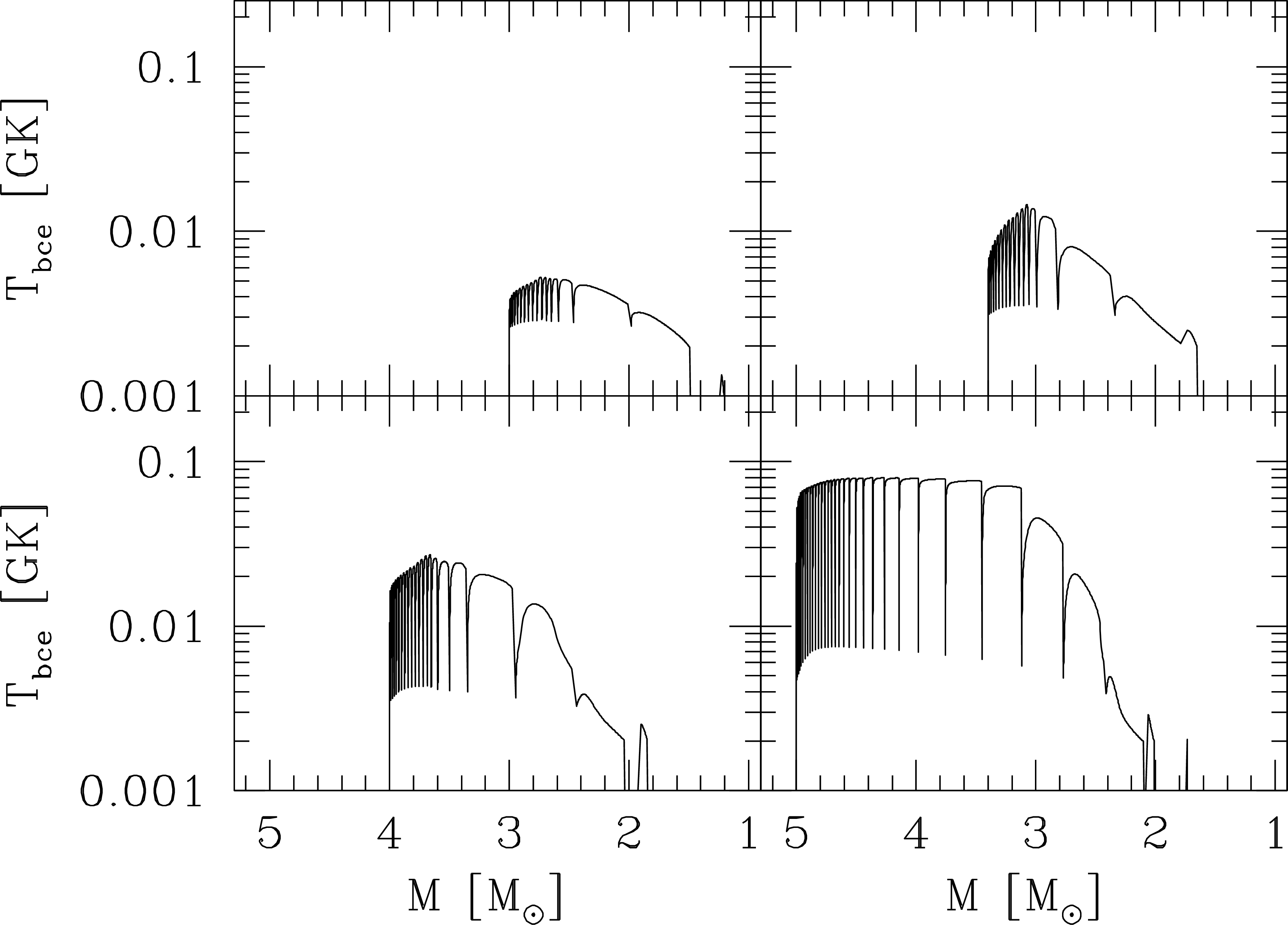}}
\end{minipage}
\hfill
\begin{minipage}{0.48\textwidth}
\resizebox{\hsize}{!}{\includegraphics{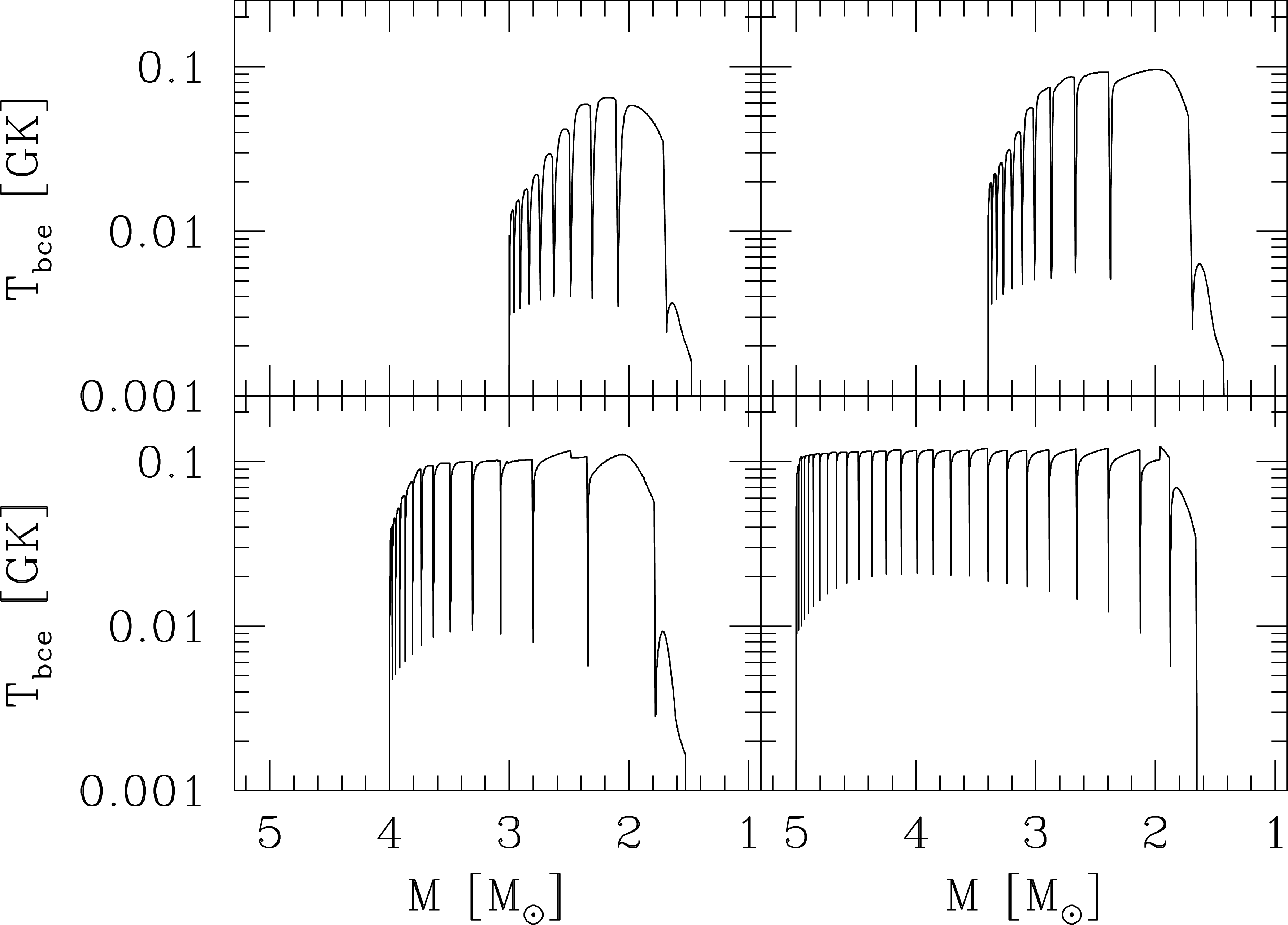}}
\end{minipage}
\caption{Evolution of the temperature at the base of the convective envelope as a function of the current stellar mass, during the TP-AGB phase of a few selected models with initial masses of 3.0, 3.4, 4.0, and 5.0 $M_{\odot}$, and metallicities $Z_{\rm i}=0.014$ (left plot) and $Z_{\rm i}=0.0005$ (right plot).}
\label{fig_tbot}
\end{figure*}

%The velocity and the mean-free path 
%of the convective eddies are computed in the framework of the  standard mixing %length theory \citep{mlt_58}. 
%The mixing-length parameter is $\alpha_{\rm MLT}=%1.74$, the same as in \texttt{PARSEC}.

A key characteristic of the \texttt{COLIBRI} code is that the 
equation of state for $\simeq 800$ atomic and molecular species,
and the Rosseland mean of 
the gas opacities across the atmosphere and the deep envelope 
are computed on-the-fly, ensuring a full consistency with the changing abundances of all involved chemical elements \citep{MarigoAringer_09}.

As for the nuclear reaction $^{22}$Ne$(p,\gamma)^{23}$Na we mainly investigated three different experimental rates, namely: LUNA, IL10, and NACRE (see Table~\ref{tab:models}). Each selected option is adopted throughout the evolutionary calculations, from the main sequence to the end of the TP-AGB phase. For comparison, we also tested the theoretical rate from 
\citet{Cyburt_etal10}, which was calculated with the version 5.0w of the  \texttt{NON-SMOKER$^{\rm WEB}$} code \citep{RauscherThielemann_00}.
We note that in the temperature range of interest for HBB, $T \approx 0.07-0.12$ GK, the theoretical CYB10 rate is larger than IL10 by factors of $\sim 1\,000$.

\begin{table}
\caption{Prescriptions adopted in the stellar evolutionary models (\texttt{PARSEC} and \texttt{COLIBRI} codes), namely: initial metallicity $Z_{\rm i}$, initial helium abundance $Y_{\rm i}$ (both in mass fraction), distribution of metals, range of initial masses $M_{\rm i}$. The upper mass limit corresponds to $M_{\rm up}$, that is the maximum mass for a star to develop an electron-degenerate C-O at the end of the He-burning phase. 
Three experimental versions, together with a theoretical version for
the rate of $^{22}$Ne$(p,\gamma)^{23}$Na, are reported.  The
ratio $\mathrm{\frac{<\sigma\upsilon>}{<\sigma\upsilon>_{IL10}}}$
gives the value of a given rate at a temperature of 0.1 GK, normalized
to the IL10 version.}
\centering
\begin{tabular}{ccccc}
\multicolumn{5}{c}{Stellar parameters}\\
\hline 
$Z_{\rm i}$ & $Y_{\rm i}$ & initial partition & \multicolumn{2}{c}{$M_{\rm i}$ $[M_{\odot}]$ range}  \\
 & & of metals  & \multicolumn{2}{c}{(in steps of $0.2\,M_{\odot}$)} \\
\hline
\multicolumn{1}{l}{0.0005} & 0.249 & $[\alpha/{\rm Fe}]$=0.4 &  \multicolumn{2}{c}{3.0-5.0}  \\
\multicolumn{1}{l}{0.006} & 0.259 & scaled-solar &  \multicolumn{2}{c}{3.0-5.4  } \\
\multicolumn{1}{l}{0.014} & 0.273 & scaled-solar &  \multicolumn{2}{c}{3.0-5.6  }\\
\hline 
\multicolumn{5}{c}{Rate for $^{22}$Ne$(p,\gamma)^{23}$Na}\\
\hline 
\multicolumn{2}{l}{Reference} & type & acronym & 
$\mathrm{\frac{<\sigma\upsilon>}{<\sigma\upsilon>_{IL10}}}$\\
\hline
\multicolumn{2}{l}{\citet{Iliadis_etal10a}}  & experimental & IL10 
& 1.00e00\\
\multicolumn{2}{l}{\citet{Cavanna_etal15}}  & experimental & LUNA 
& 1.80e01\\
\multicolumn{2}{l}{\citet{Angulo_etal99}} & experimental & NACRE 
& 3.13e02\\
\multicolumn{2}{l}{\citet{Cyburt_etal10}} &  theoretical &  CYB10 
& 4.35e03\\
\hline
\end{tabular}
\label{tab:models}
\end{table}

\begin{table*}
\captionsetup{justification=centering}  
\caption{Prescriptions for convection, mass loss and third dredge-up assumed 
in our TP-AGB models.
The $M13$ set corresponds to our reference choice, initially 
adopted for all stellar models considered in this work. The $A$-$F$ 
combinations are tested in stellar models with the lowest metallicity, 
i.e. $Z_{\rm i} =0.0005$, [$\alpha$/Fe]=0.4, for which HBB is most efficient 
(see Section~\ref{sec_ggcs}).}
\label{tab_test}
\begin{tabular}{cccclc p{1cm}}
\hline 
model & $\alpha_{\rm ML}$ & $\dot{M}$ & $\lambda_{\rm max}$ & notes \\  
class & & & \\
\hline
$M13$  & 1.74 & $M13$ & $M13$ & reference set $^{a}$ &\rdelim\}{5}{2mm}[\parbox{2cm}{\flushleft{very efficient third dredge-up} $\mathrm{\lambda_{\rm max} up\,\, to \simeq 1}$}]\\
$A$ & 1.74 & VW93 & $M13$ & popular mass-loss law \\
$B$ & 1.74 & B95   & $M13$ & efficient mass loss \\
    &      &       &       & with $\eta=0.02$ \\
$C$ & 2.00 & $M13$ & $M13$ & efficient HBB \\
$D$ & 1.74 & $M13$ & $\lambda=0$ & no third dredge-up &\\
\hline
$E$ & 2.00 & $M13$ & 0.5 & efficient HBB \\
    &   &        &     & moderate third dredge-up\\ 
    &   &        &     & $^{23}$Na$(p,\alpha)^{20}$Na reduced by 5\\ 
\hline
$F$ & 1.74 & B95 & $\lambda=0$ & efficient mass loss \\
    &      &       &   & with $\eta=0.03$ \\
    &   &        &     & no third dredge-up\\ 
    &   &        &     & $^{23}$Na$(p,\alpha)^{20}$Na reduced by 3\\ 
\hline
\multicolumn{6}{l}{$^a$ Input prescriptions as in \citet{Marigo_etal13}} \\
\hline
\end{tabular}
\end{table*}

\section{Changes in the surface $^{22}$N\lowercase{e} and $^{23}$N\lowercase{a} abundances}
\label{sec_abund}
\subsection{Prior to the TP-AGB: the second dredge-up}
\label{ssec_2dup}
We will briefly discuss here the predicted changes in the surface Ne-Na abundances that may take place before the development of thermal pulses in intermediate masses, with $3 M_{\odot}\la  M_{\rm i } \la\, 6 M_{\odot}$.
We focus on the first and second dredge-up processes in the context of classical models, i.e. neglecting the possible contribution of extra-mixing events.

The first dredge-up takes place at the base of the red giant branch (RGB) when material that was processed through partial hydrogen burning via the CNO cycle and p-p chains is brought up to the surface.
Models predict an increase of surface nitrogen, and a reduction of the isotopic ratio $^{12}$C/$^{13}$C.
As for the Ne-Na isotopes minor changes are expected, and their abundances remain essentially those of the initial chemical composition.

\begin{figure*}
\centering
\begin{minipage}{0.47\textwidth}
  \resizebox{\hsize}{!}{\includegraphics{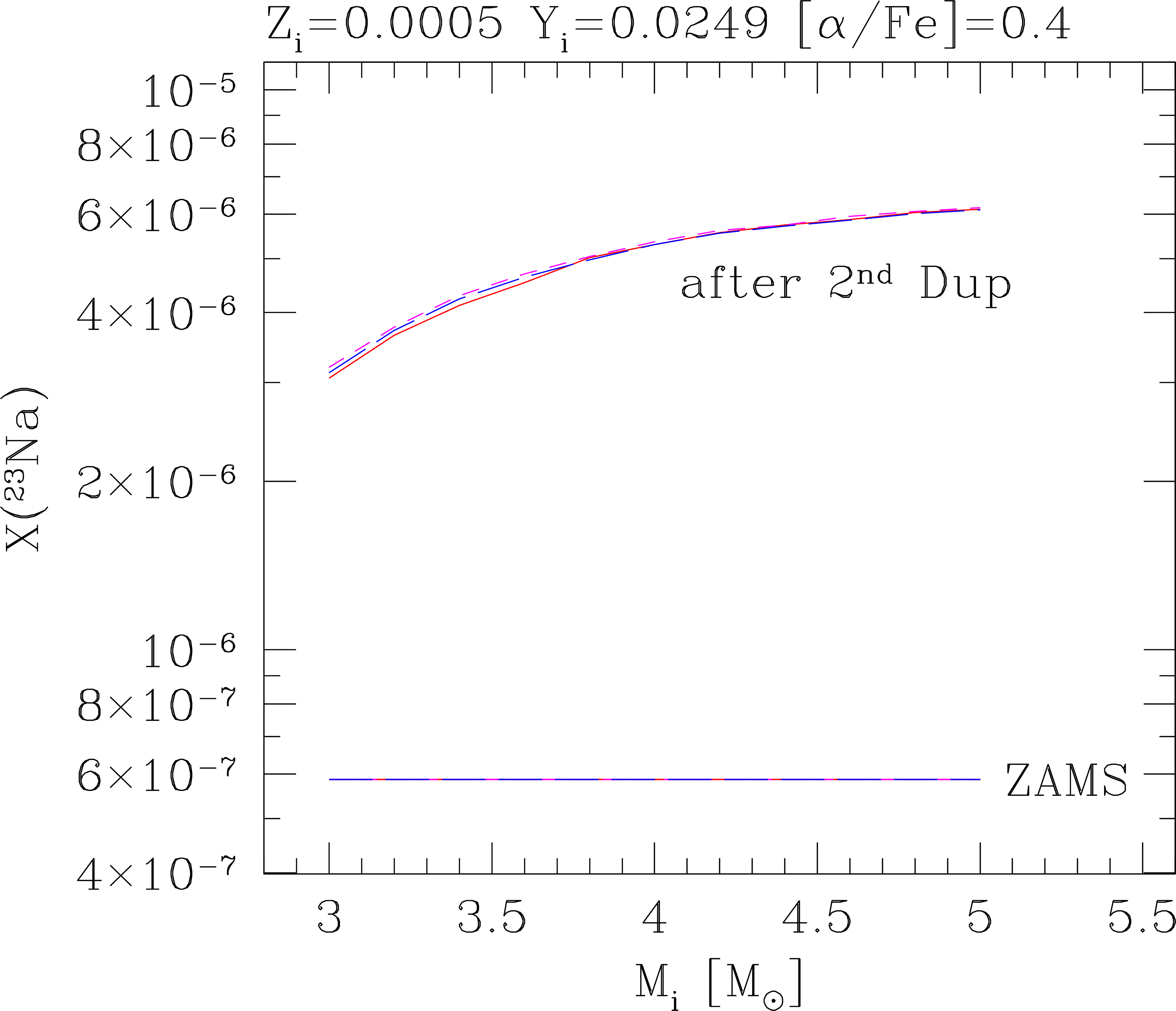}}
\end{minipage}
\hfill
\begin{minipage}{0.48\textwidth}
\resizebox{\hsize}{!}{\includegraphics{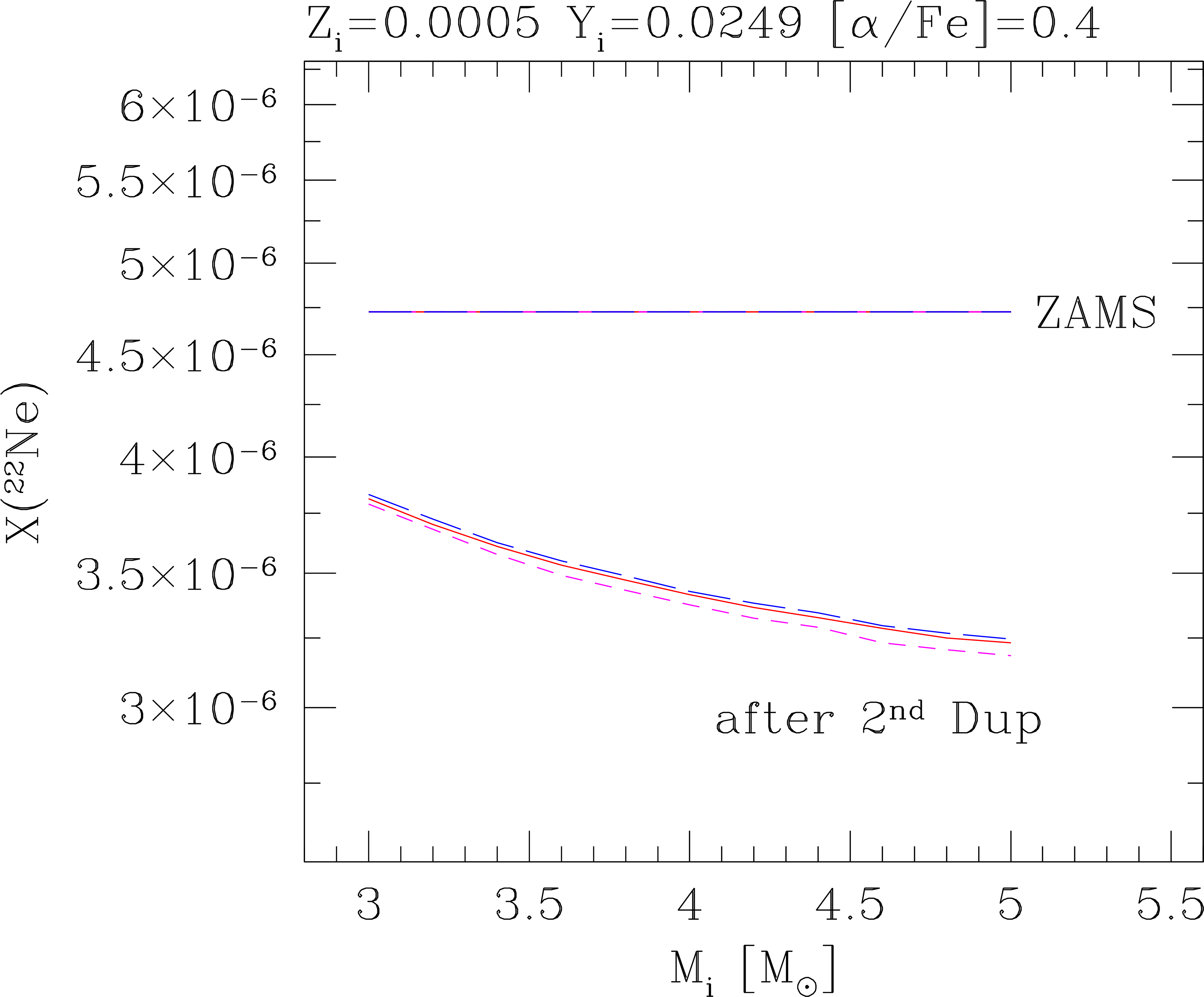}}
\end{minipage}
\caption{Surface abundances of $^{22}$Ne (right) and $^{23}$Na (left) as a function of the initial stellar mass at the
zero-age main sequence (ZAMS) and after the second dredge-up on the E-AGB.
Predictions are shown for three choices of the $^{22}$Ne$(p,\gamma)^{23}$Na rate, namely: NACRE (magenta short-dashed line),  IL10 (blue long-dashed line), LUNA (red solid line).}
\label{fig_2dup}
\end{figure*}

The situation is different when the second dredge-up occurs during the early-AGB (hereinafter also E-AGB) of stars with initial masses $M_{\rm i} > 3-4\, M_{\odot}$ (depending on metallicity and other model details). In these stars the base of the convective envelope deepens into the layers previously processed by the temporarily extinguished H-burning shell. The surface is enriched with material containing the products of complete H-burning.
Significant variations in the surface concentrations of the Ne-Na isotopes are expected, as illustrated in Fig.~\ref{fig_2dup} for a set of models
 with initial metal-poor composition.
We see that $^{23}$Na increases by a factor of 5-10 (larger for higher stellar masses), while $^{22}$Ne is correspondingly reduced.
These trends agree with the predictions of other stellar models
  in the literature \citep{KarakasLattanzio_14, VenturaDantona_06, Smiljanic_etal09, Mowlavi_99a, ForestiniCharbonnel_97}.

The effects of different $^{22}$Ne$(p,\gamma)^{23}$Na rates on the final Ne-Na abundances after the second dredge-up is minor.
%as the bottom of the envelope
%extends into layers where H was completely burnt and  
%the nuclei involved in the NeNa cycle reached full nuclear equilibrium, and the 
Comparing the results obtained with the rates quoted in Table~\ref{tab:models}, we find that the relative differences with respect to the NACRE rate span a range $\simeq 1-4 \%$ for $^{23}$Na and to $\simeq 0.5-1 \%$ for $^{22}$Ne.
This means that the Ne-Na surface concentrations after the second dredge-up are mainly controlled by the depth of the envelope penetration (e.g. through the mixing length, and/or the overshoot parameter). 
Conversely, the nuclear rates have a dramatic impact during the TP-AGB phase, when intermediate-mass stars are affected by the third dredge-up and hot-bottom burning.
This aspect is discussed next, in Section~\ref{sec_nena}.

\subsection{During the TP-AGB: HBB nucleosynthesis and the third dredge-up} 
\label{sec_nena}

Evolutionary calculations of the TP-AGB phase indicate that the activation of the NeNa cycle at the base of the convective envelope requires relatively high temperatures, $T > 0.05$ GK, which can be attained in luminous and massive AGB and super-AGB stars,  preferably at low metallicity
\citep[e.g.,][]{Doherty_etal14b, Marigo_etal13, KarakasLattanzio_07, ForestiniCharbonnel_97}. 
Figure~\ref{fig_tbot} compares the predicted temperatures at the
  base of the convective envelope, $T_{\rm bce}$, in TP-AGB models of
  various initial masses and two choices of the metallicity. Higher
  temperatures are reached by stars of larger mass and lower
  metallicity. The model with $\mathrm{M_{\rm i}=5.0\, M_{\odot}}$ and $\mathrm{Z_{i}=0.0005}$ attains 
  the highest temperatures, up to $\mathrm{T_{bce} \sim 0.12}$ K.
  In all models the final drop in temperature is caused
   by the reduction of the envelope mass by stellar winds, which
   eventually extinguishes HBB.

Provided that the NeNa cycle operates for sufficiently long time, the main result is the synthesis of $^{23}$Na at the expenses of the Ne isotopes.
In general, 
the competition between production (through the reaction $^{22}$Ne$(p,\gamma)^{23}$Na) and destruction (through the reactions $^{23}$Na$(p,\alpha)^{20}$Ne and $^{23}$Na$(p,\gamma)^{24}$Mg) depends on the temperature of the burning zone and the duration of the process.

The picture above becomes more complex if, in addition to HBB,  the star experiences also the third dredge-up. During the power-down phase of a thermal pulse the base of the convective envelope may reach the region that was previously affected by the pulse-driven convective zone (hereafter PDCZ), which causes a rapid change in the surface chemical composition.
The standard  chemical composition of the PDCZ mainly consists of $^{12}$C ($\simeq 20\%-25\%$), $^{16}$O
($\simeq 1\%-2\%$), $^{22}$Ne ($\simeq 1\%-2\%$),
with $^{4}$He practically comprising all the rest 
\citep{BoothroydSackmann_88, Mowlavi_99a},
almost regardless of metallicity and core mass.

\begin{figure*}
\centering
\begin{minipage}{0.48\textwidth}
\resizebox{\hsize}{!}{\includegraphics{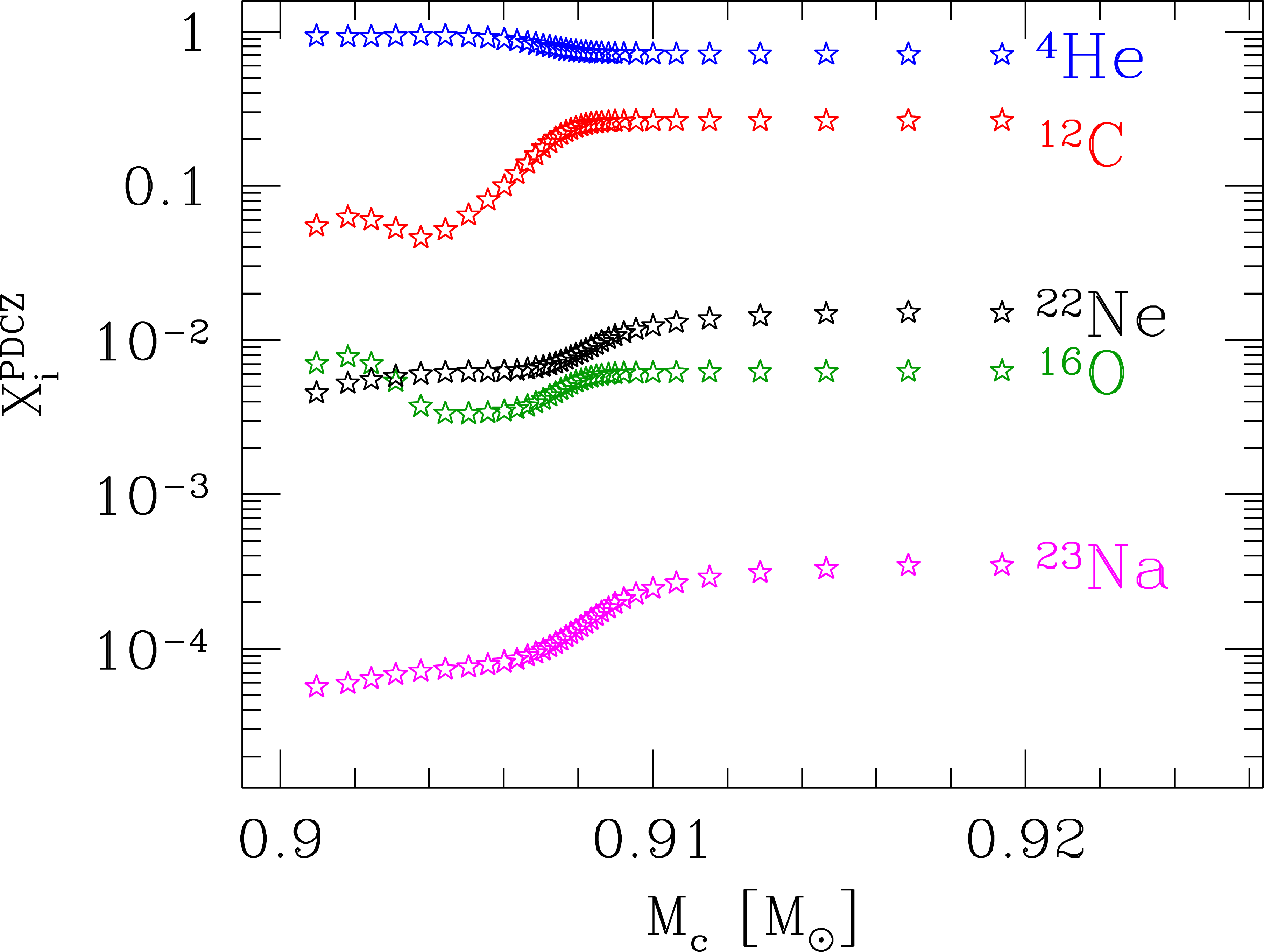}}
\end{minipage}
\hfill
\begin{minipage}{0.48\textwidth}
\resizebox{\hsize}{!}{\includegraphics{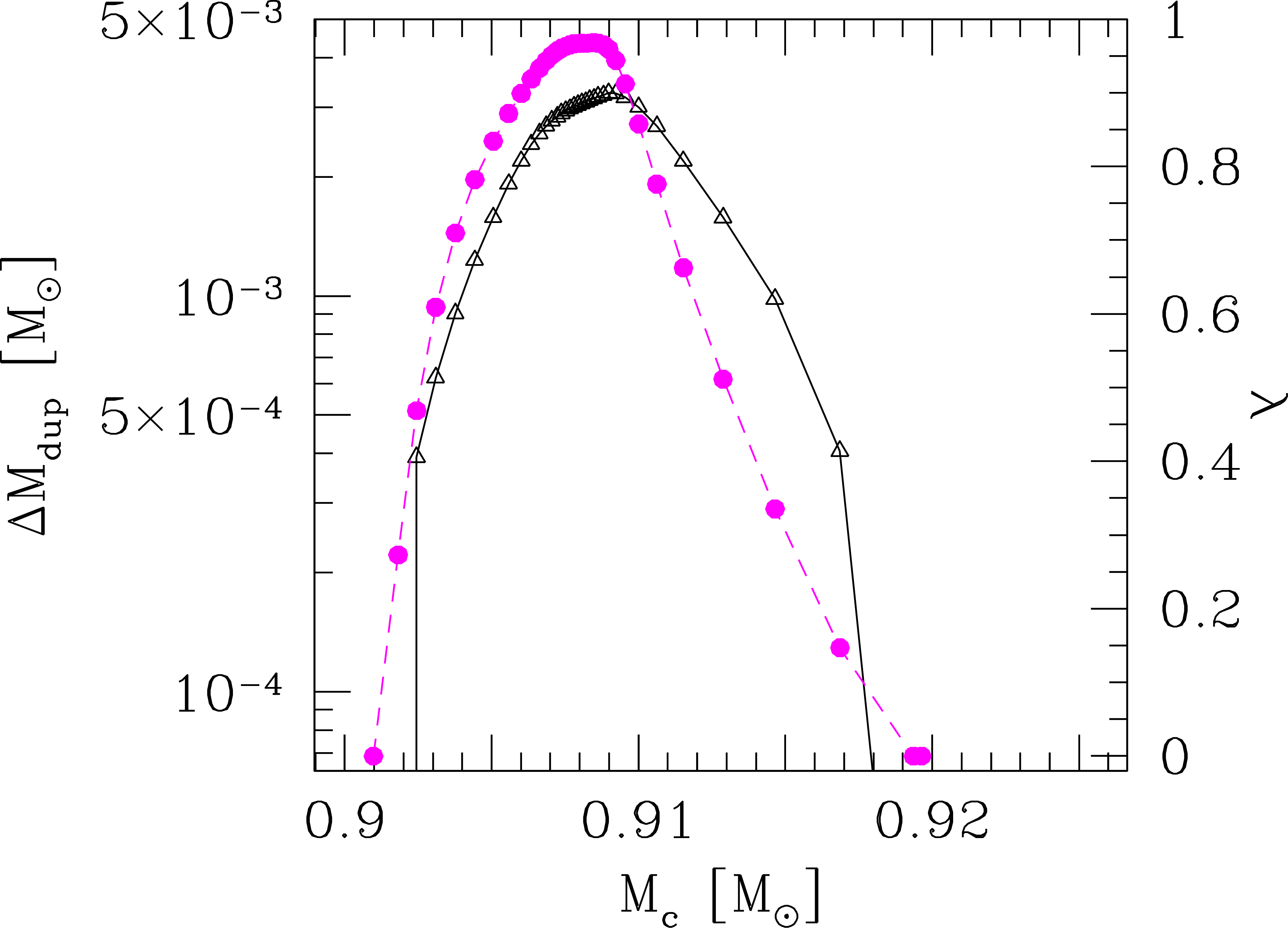}}
\end{minipage}
\caption{Characteristics of the third dredge-up as a function of the core mass 
during the TP-AGB evolution of a star with initial mass 
$M_{\rm i}=5\, M_{\odot}$ and metallicity Z$_{\rm i}=0.006$.
Input prescriptions correspond to our reference set ($M13$; see Table~\ref{tab_test}), while other assumptions for the third dredge-up are discussed later in the paper (see Sections~\ref{ssec_totunc}, \ref{sec_ggcs} and Table~\ref{tab_test}). 
Left panel: Abundances (in mass fraction) left in the PDCZ after the development of each thermal pulse as a function of the core mass.
Right panel: Amount of dredged-up material at each thermal pulse (black triangles connected by solid line), and efficiency parameter $\lambda$ (filled magenta circles connected by dashed line). Similar trends 
hold for the other $Z_{\rm i}$ considered in this work.}
\label{fig_3dup}
\end{figure*}

Figure~\ref{fig_3dup} (left panel) shows the predicted abundances in the PDCZ developing at each thermal pulse in TP-AGB stars with initial mass $M_{\rm i}=5\, M_{\odot}$ and initial metallicity $Z_{\rm i}=0.006$, computed with the \texttt{COLIBRI} code.
We note that $^{4}$He, $^{12}$C, and $^{16}$O achieve the typical concentrations that characterize the classical PDCZ composition.  
The amount of mass dredged-up at each thermal pulse and the corresponding efficiency $\lambda$\footnote{According to a standard notation the efficiency of the third dredge-up is expressed with $\lambda = \Delta M_{\rm dup}/\Delta M_{\rm c}$, which is the fraction of the core mass increment over an inter-pulse period that is dredged-up to the surface at the next thermal pulse.} 
are also illustrated in Fig.~\ref{fig_3dup} (right panel). Similar results apply to  the other metallicities here considered. In all models with $M_{\rm i}> 4\, M_{\odot}$ the
the third dredge-up is predicted to become quite deep as thermal pulses develop, reaching a maximum around $\lambda \simeq 1$. These trends are obtained following the predictions of full stellar AGB calculations of \citet{Karakas_etal02}, which are characterized by very efficient third dredge-up. Different prescriptions,
i.e. lower values of $\lambda$, are adopted in additional sets of AGB models, which are discussed in Sections~\ref{ssec_evunc} and \ref{sec_ggcs}.
The rapid decrease of $\lambda$ 
takes place over the last stages, when the envelope mass is dramatically reduced by stellar winds.

In the context of this study it is interesting to analyze the abundances of $^{22}$Ne and $^{23}$Na in the PDCZ, and the effect of the envelope chemical composition on them.
The $^{22}$Ne isotope is relatively abundant in the PDCZ, increasing up to nearly $1\%$ in mass fraction in the $Z_{\rm i} = 0.0005$ models, while it reaches up to $\simeq 2\%$ in the $Z_{\rm i} = 0.014$ models, where it exceeds the 
$^{16}$O abundance.
In the PDCZ $^{22}$Ne is the product of the chain of $\alpha$-capture reactions that starts from the $^{14}{\rm N}$, left over by the H-burning shell at the end of the inter-pulse period, i.e.  $^{14}{\rm N}(^4{\rm He},\gamma)^{18}{\rm F}(\beta^+ \nu)^{18}{\rm O}(^4{\rm He},\gamma)^{22}{\rm Ne}$. 
Therefore, at each thermal pulse the abundance $^{22}$Ne in the PDCZ depends on the current CNO content in the envelope, and positively correlates 
with the efficiency of the third dredge-up. In fact,  
the injection of primary $^{12}$C into the envelope by the third dredge-up increases the CNO abundance available to the H-burning shell, which will be mainly 
converted into $^{14}{\rm N}$ during the quiescent inter-pulse periods.

Conversely, the abundance of $^{23}$Na in the PDCZ is largely unaffected by He-burning nucleosynthesis during the thermal pulse \citep{ForestiniCharbonnel_97}, while it is essentially determined by the shell H-burning during the previous inter-pulse period. 
In fact, when a thermal pulse develops, the associated PDCZ can reach the
inter-shell region where some unburnt $^{23}$Na survived against proton captures.
Then, this secondary $^{23}$Na is mixed out in the PDCZ and eventually injected into the envelope during the third dredge-up \citep[see][for a thorough analysis]{Mowlavi_99a}. 
More recently, \citet{Cristallo_etal09} discussed the formation of a $^{23}$Na-pocket in the transition region between the core and the envelope, which may provide an additional source of sodium. However, those results apply to low-mass stars and should not affect the ejecta of sodium 
from more massive AGB stars considered here.

In view of the above, it is clear that the third dredge-up and HBB nucleosynthesis are closely coupled and affect the  surface abundances of $^{22}$Ne and $^{23}$Na, \citep[see, e.g., ][for similar results discussed in the past literature]{VenturaDantona_06, KarakasLattanzio_03, Mowlavi_99b, ForestiniCharbonnel_97}.
Each time a third dredge-up event takes place, some  amounts of $^{22}$Ne and $^{23}$Na are injected into the convective envelope where they will be subsequently involved in the NeNa cycle when HBB is re-activated during the quiescent inter-pulse periods.

\begin{figure*}
\centering
\begin{minipage}{0.48\textwidth}
\resizebox{\hsize}{!}{\includegraphics{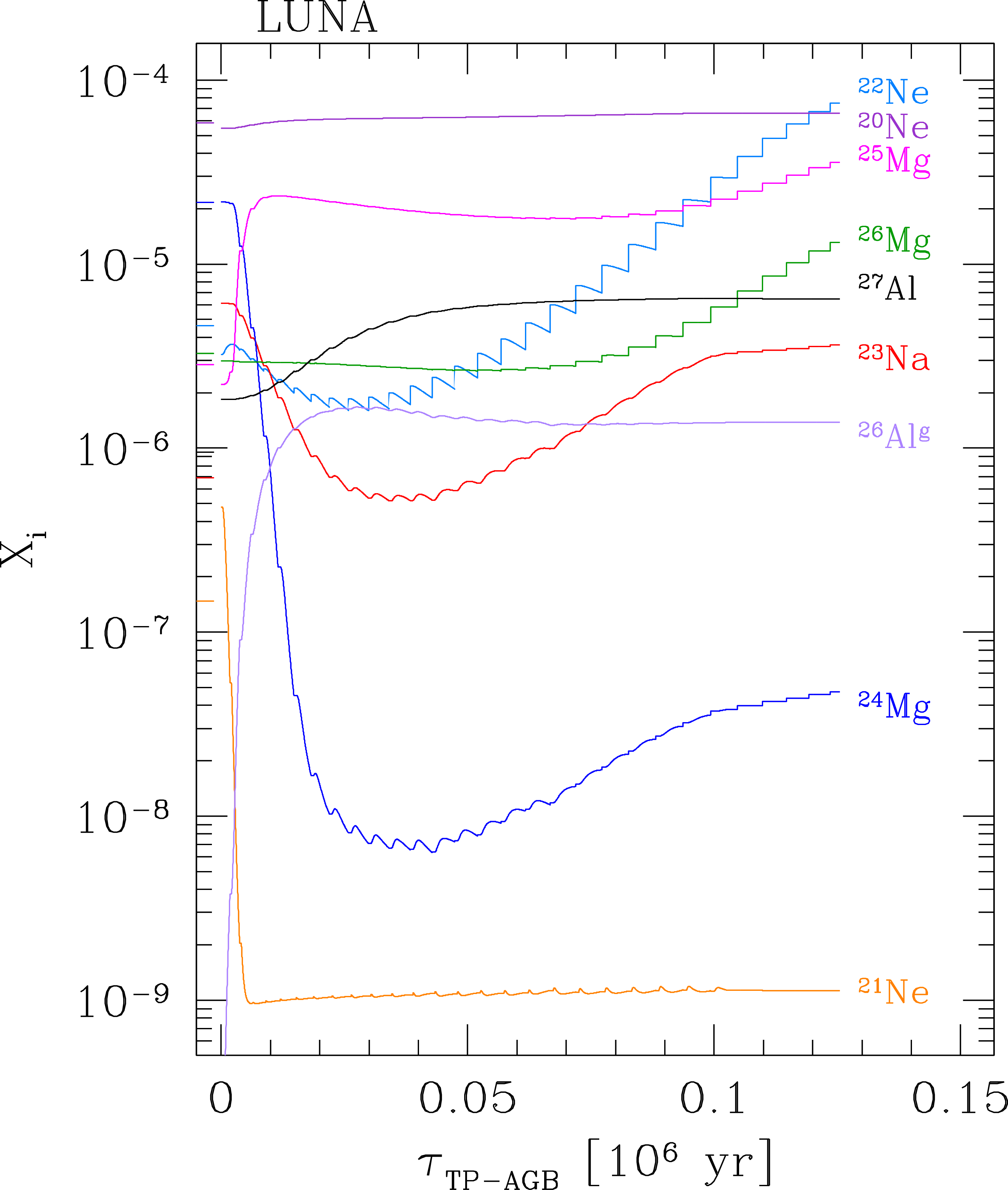}}
\end{minipage}
\hfill
\begin{minipage}{0.48\textwidth}
\resizebox{\hsize}{!}{\includegraphics{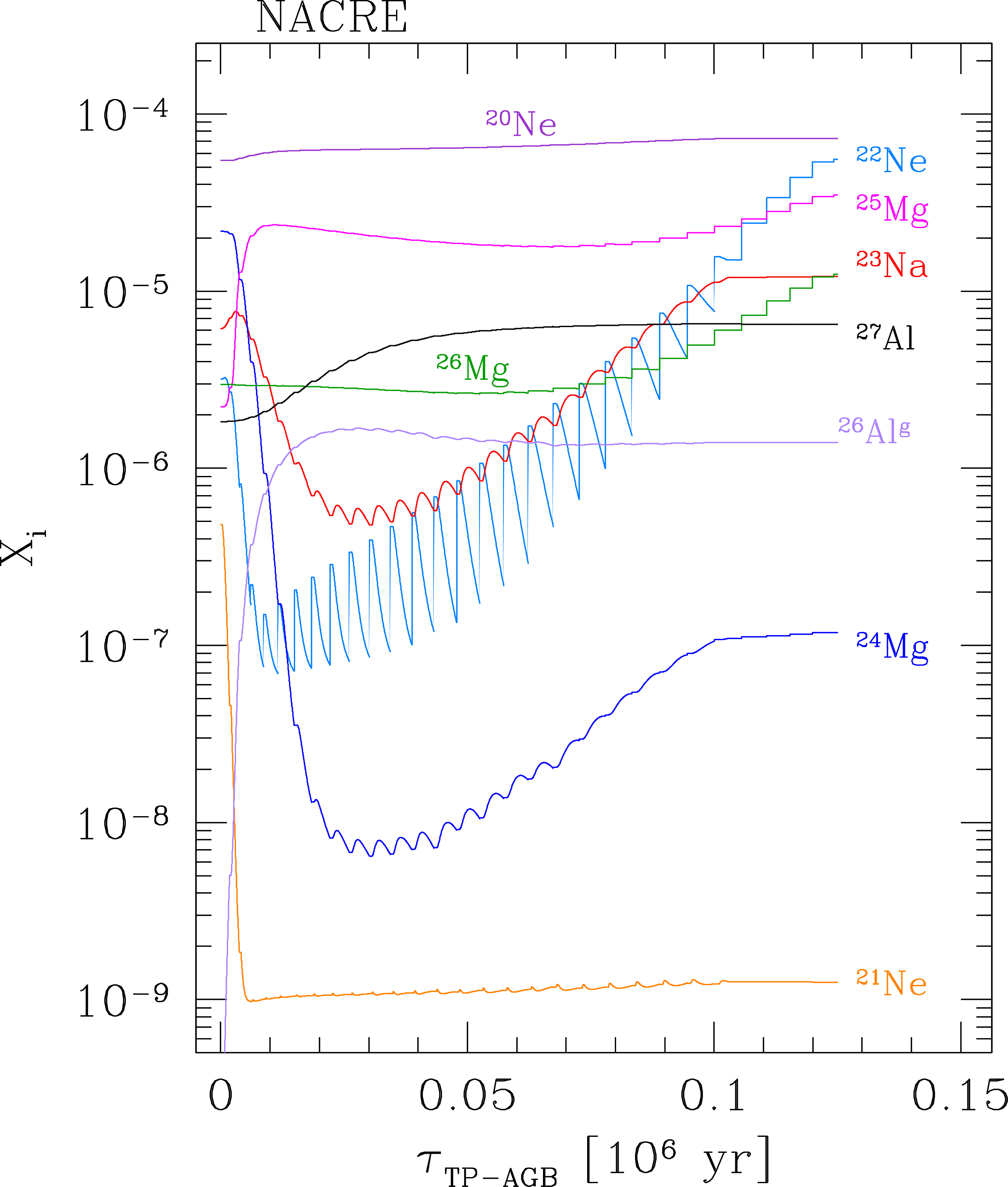}}
\end{minipage}
\begin{minipage}{0.48\textwidth}
\resizebox{\hsize}{!}{\includegraphics{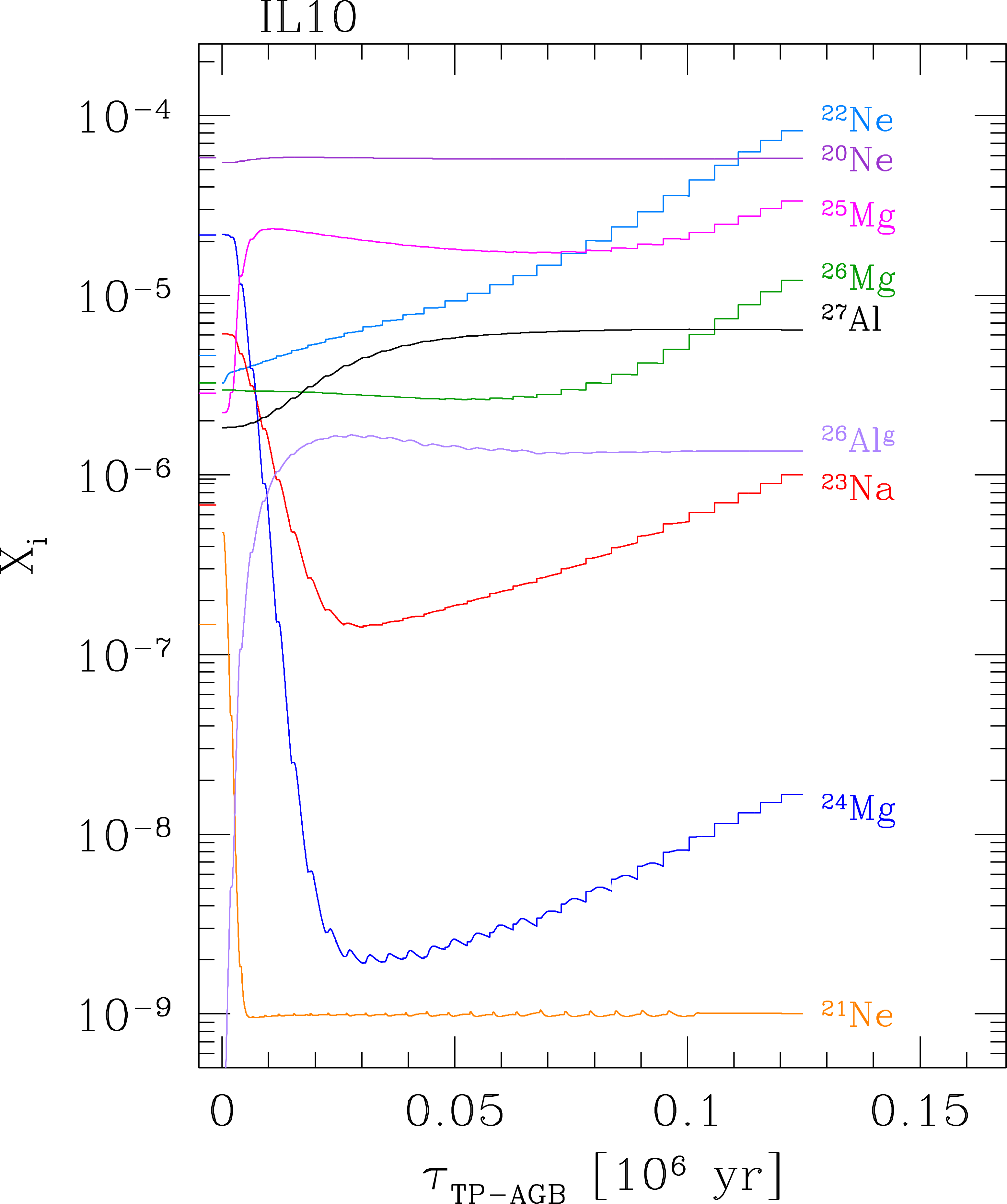}}
\end{minipage}
\hfill
\begin{minipage}{0.48\textwidth}
\resizebox{\hsize}{!}{\includegraphics{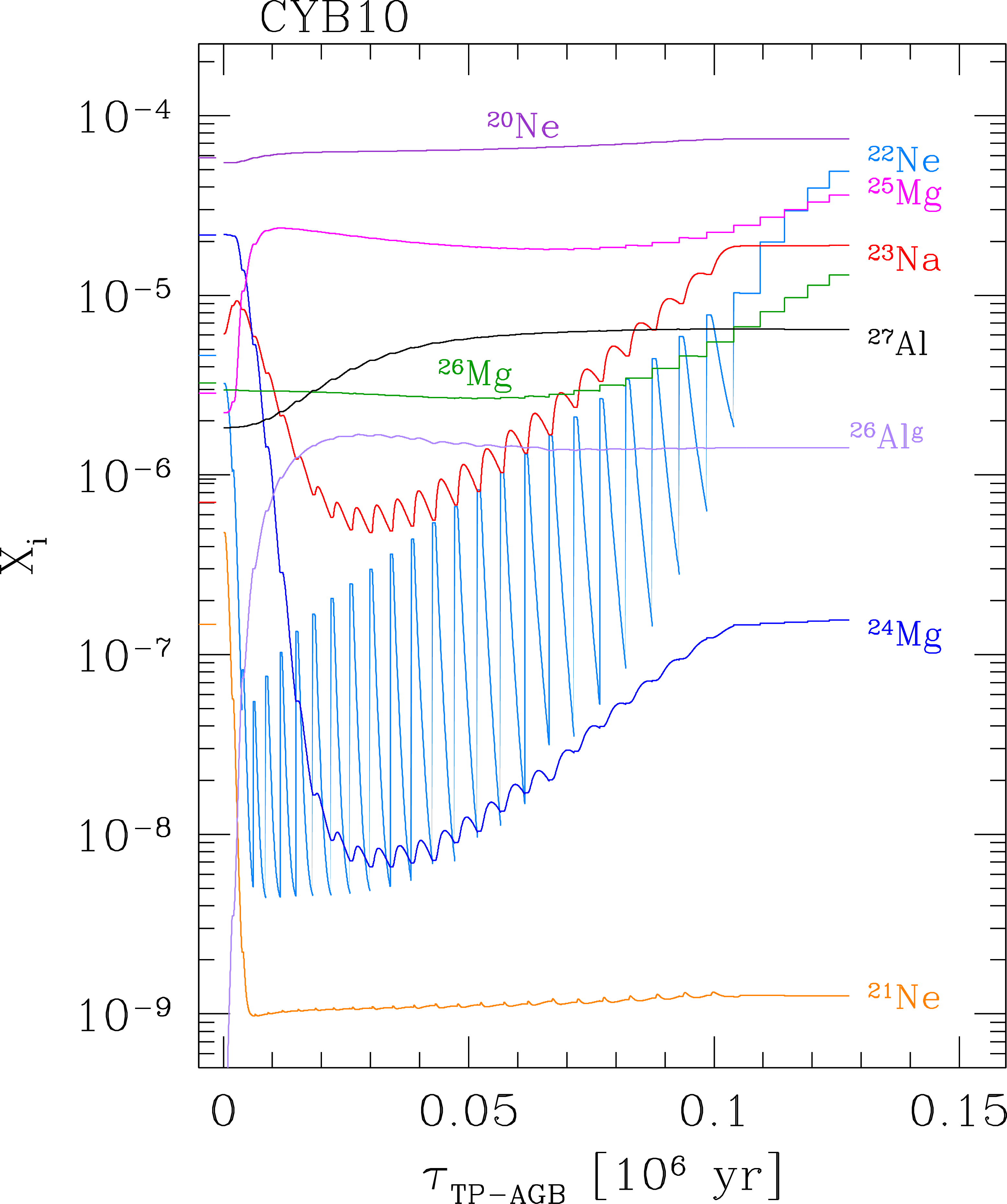}}
\end{minipage}
\caption{Evolution of envelope abundances  of Ne, Na, and Mg isotopes (in mass fraction) during the whole TP-AGB phase of a star with initial mass
$M_{\rm i}=5\, M_{\odot}$, metallicity  Z$_{\rm i}=0.0005$, and $\alpha$-
enhancement $[\alpha/{\rm Fe}]=0.4$. Time is counted since the first TP. The model experiences  both HBB and third dredge-up events.
All models share the same input physics but for the rate of $^{22}$Ne$(p,\gamma)^{23}$Na, as indicated in the labels (see also Table~\ref{tab:models}). Major differences show up in the evolution  of $^{22}$Ne, $^{23}$Na, and $^{24}$Mg.}
\label{fig_hbbnuc}
\end{figure*}

This is exemplified in Fig.~\ref{fig_hbbnuc}, which shows the evolution of the surface abundances in low-metallicity stars that undergo both 
HBB during the quiescent inter-pulse periods and
recurrent third dredge-up episodes at thermal pulses.
The spikes of $^{22}$Ne
correspond to the quasi-periodic enrichment caused by the third dredge-up, while the subsequent decrease (particularly evident in the bottom-left panel) shows the destruction due to $^{22}$Ne$(p,\gamma)^{23}$Na when HBB is reignited.

Comparing the four panels of Fig.~\ref{fig_hbbnuc}, each corresponding to a different choice of the rate for $^{22}$Ne$(p,\gamma)^{23}$Na, it is also evident that the abundance trends of $^{22}$Ne, $^{23}$Na, and $^{24}$Mg are critically affected by 
this reaction. Note, for instance, how much the amplitude of the saw-teeth trend
for $^{22}$Ne is reduced when passing from CYB10 to LUNA.
This simply reflects the fact that with the new LUNA rate proton captures on $^{22}$Ne nuclei are much less frequent than predicted by CYB10 when HBB is active.

Besides the evolution of the surface abundances, it is particularly relevant to quantify the amount of the processed material AGB stars expel
via stellar winds.
Therefore, in the next section we will analyze the ejecta of $^{22}$Ne and $^{23}$Na 
and their uncertainties, with particular focus on the impact of the new LUNA rate.

\section{AGB ejecta of $^{22}$N\lowercase{e} and $^{23}$N\lowercase{a}}
\label{sec_ejecta}
\begin{figure*}
\centering
\begin{minipage}{0.48\textwidth}
\resizebox{\hsize}{!}{\includegraphics{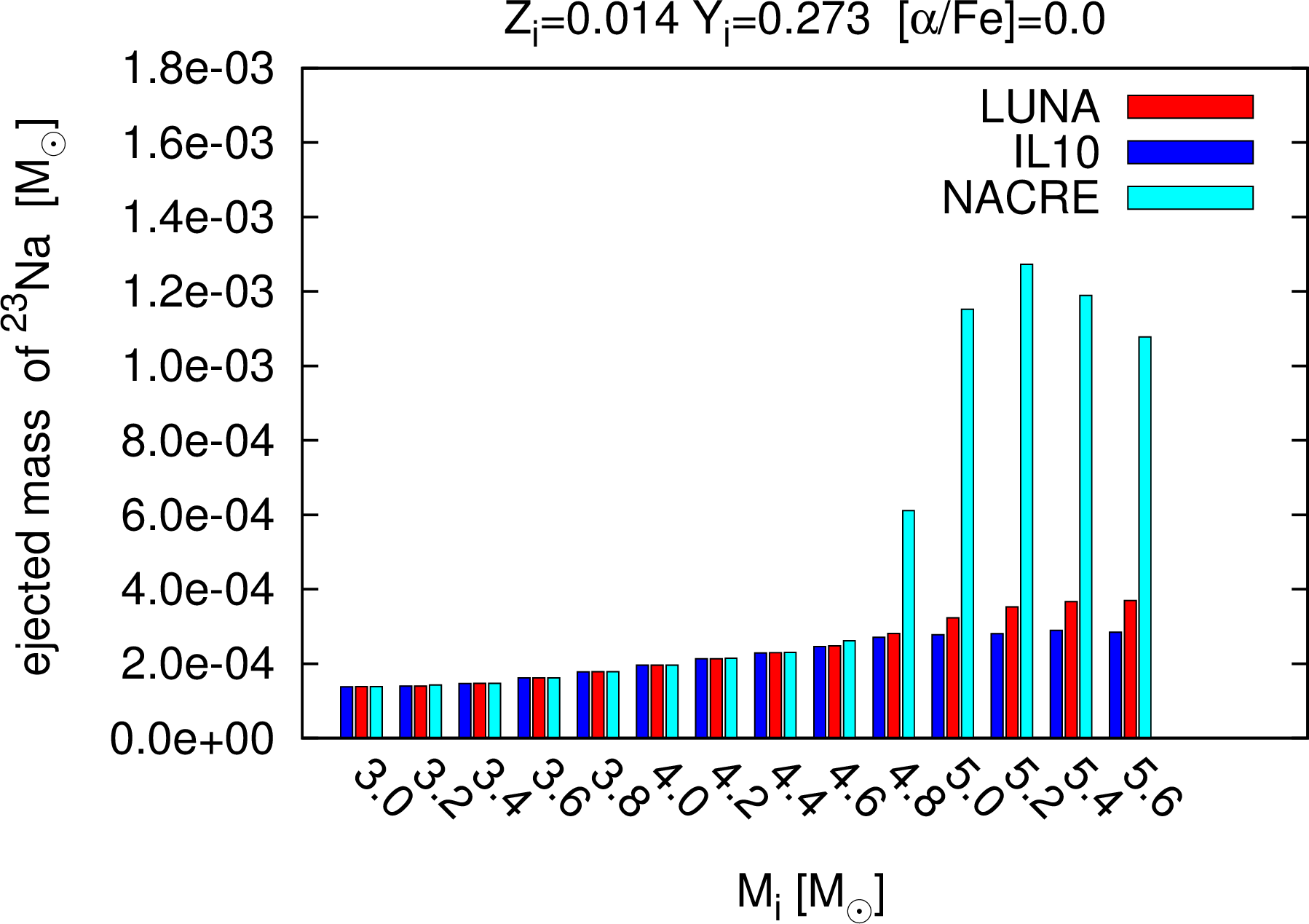}}
\end{minipage}
\hfill
\begin{minipage}{0.48\textwidth}
\resizebox{\hsize}{!}{\includegraphics{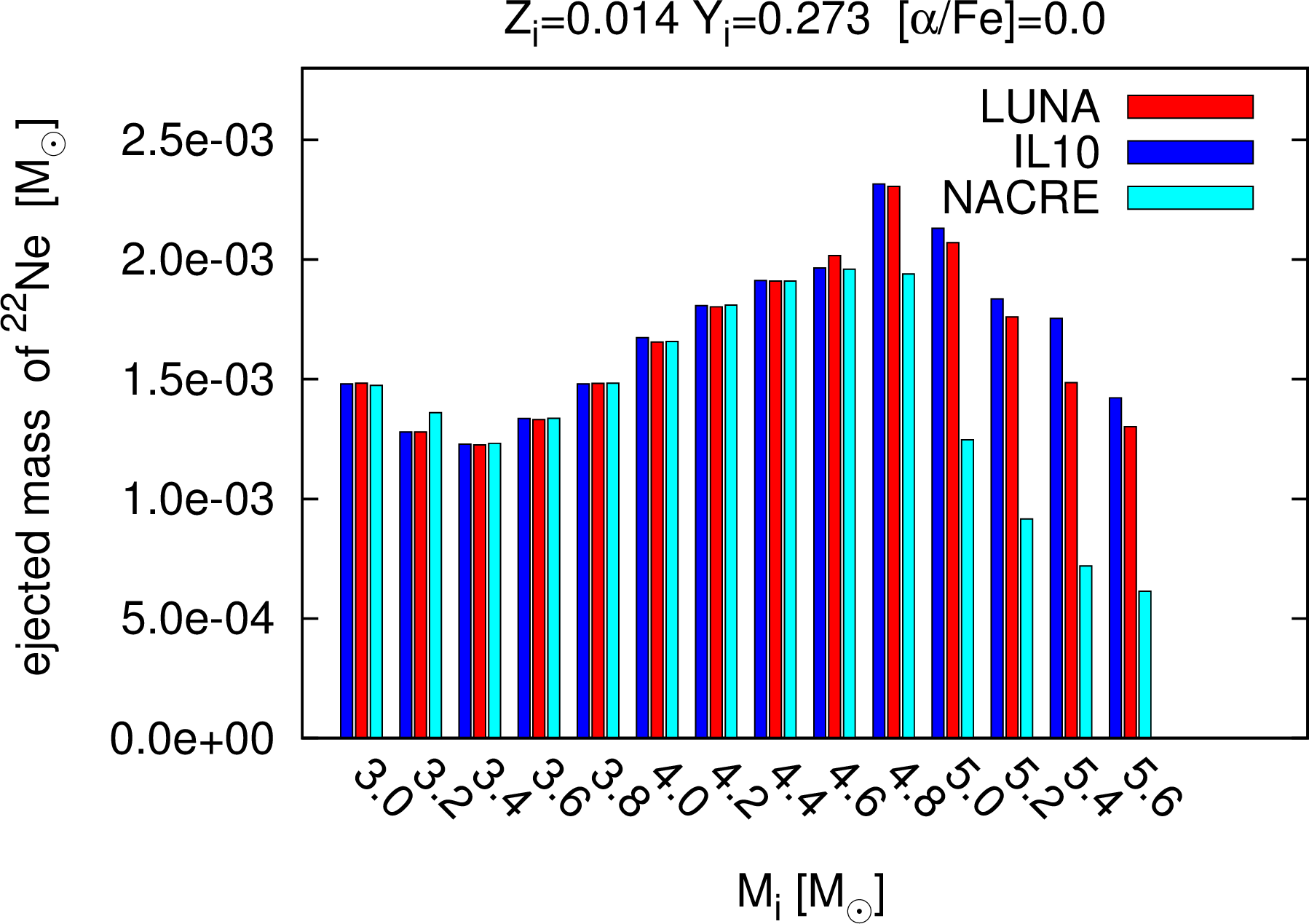}}
\end{minipage}
\hfill
\begin{minipage}{0.48\textwidth}
\resizebox{\hsize}{!}{\includegraphics{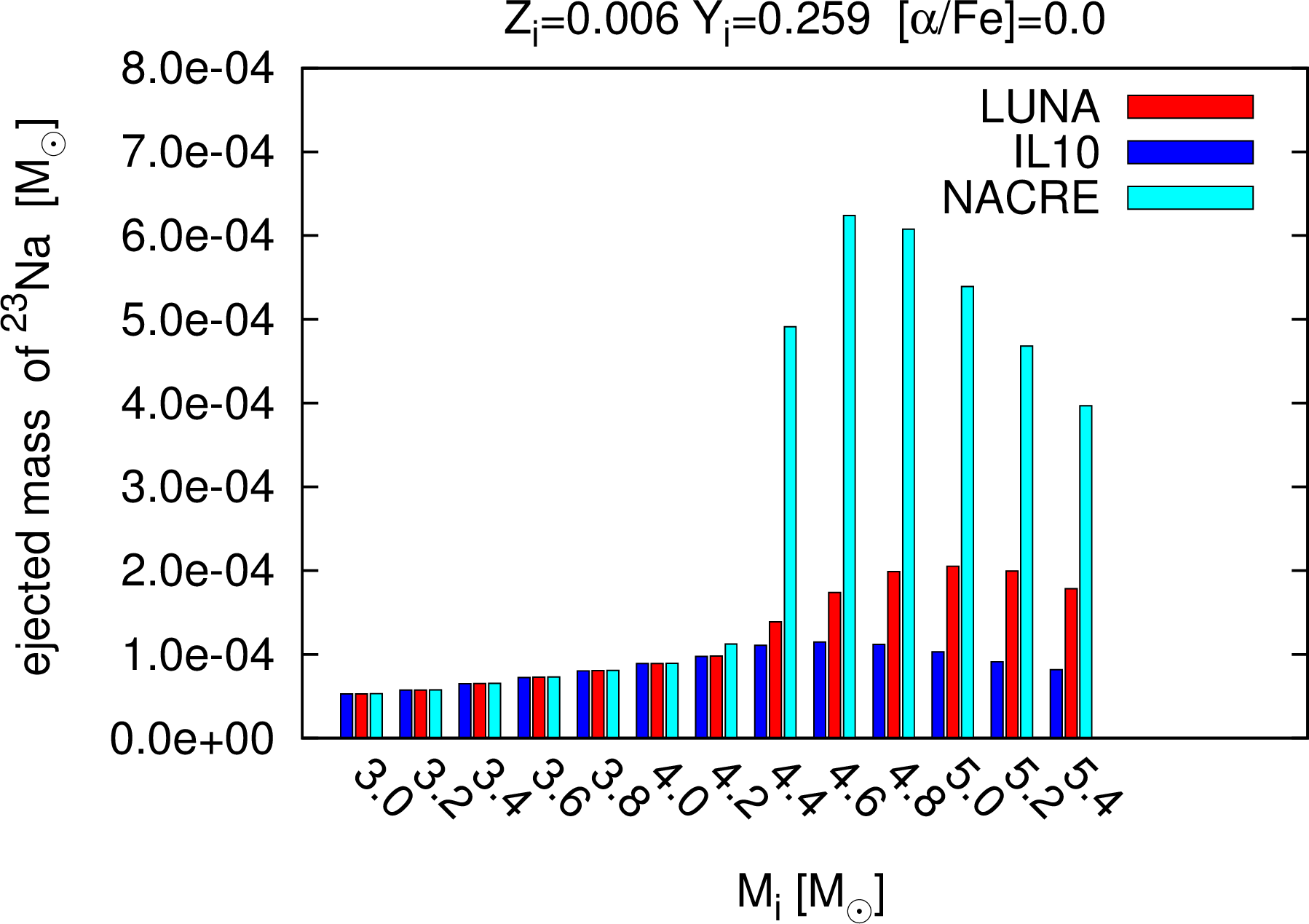}}
\end{minipage}
\hfill
\begin{minipage}{0.48\textwidth}
\resizebox{\hsize}{!}{\includegraphics{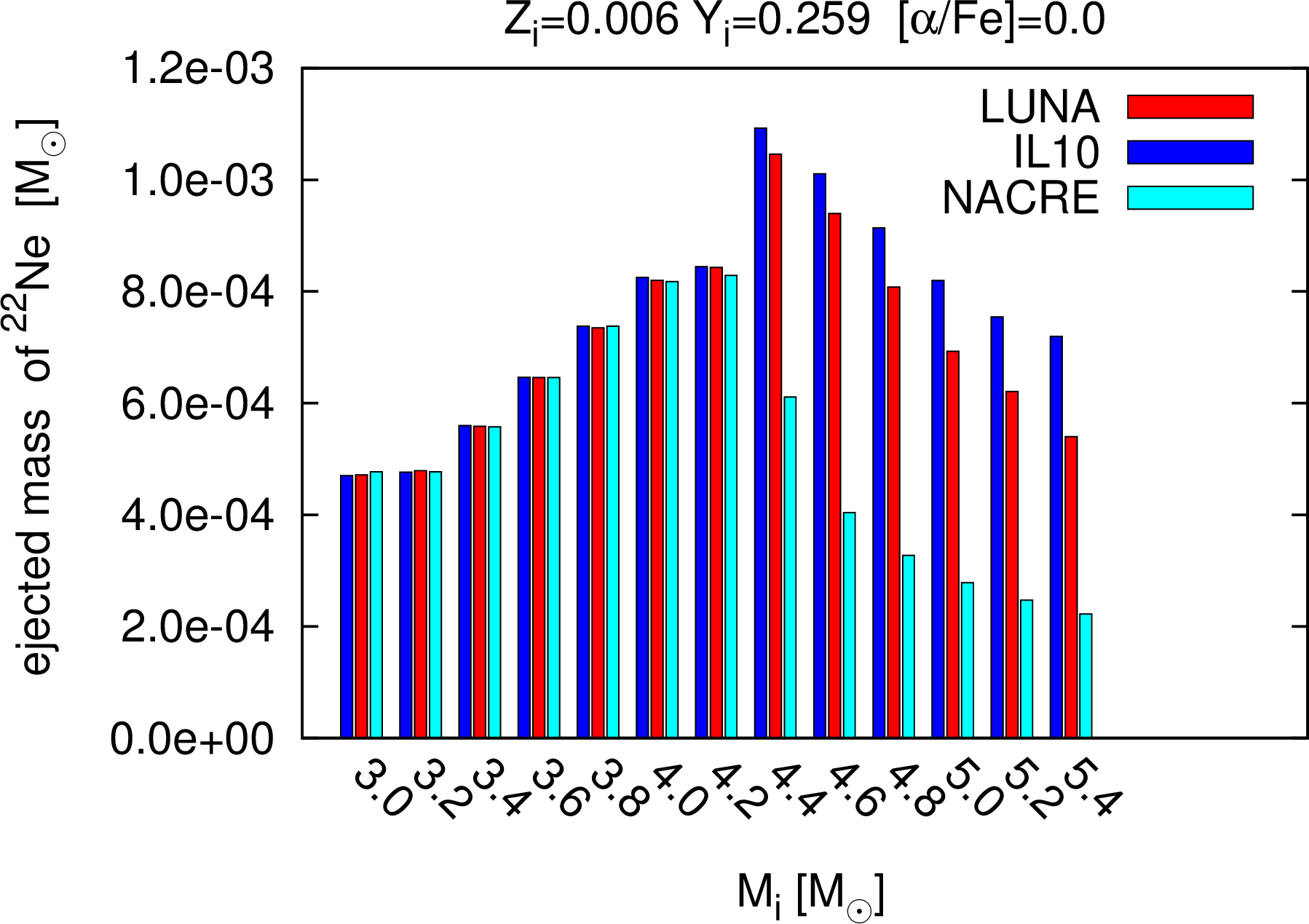}}
\end{minipage}
\hfill
\begin{minipage}{0.48\textwidth}
\resizebox{\hsize}{!}{\includegraphics{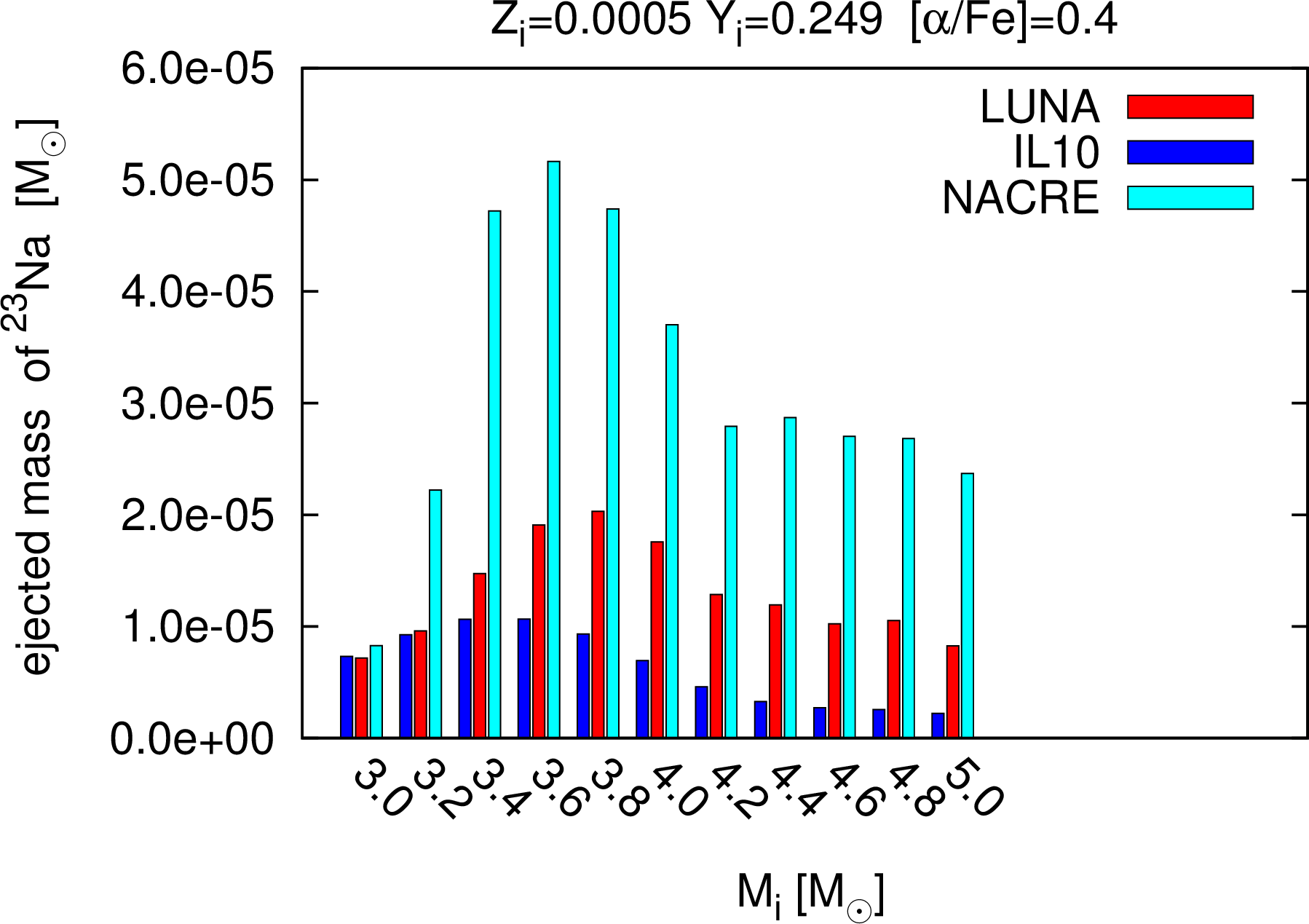}}
\end{minipage}
\hfill
\begin{minipage}{0.48\textwidth}
\resizebox{\hsize}{!}{\includegraphics{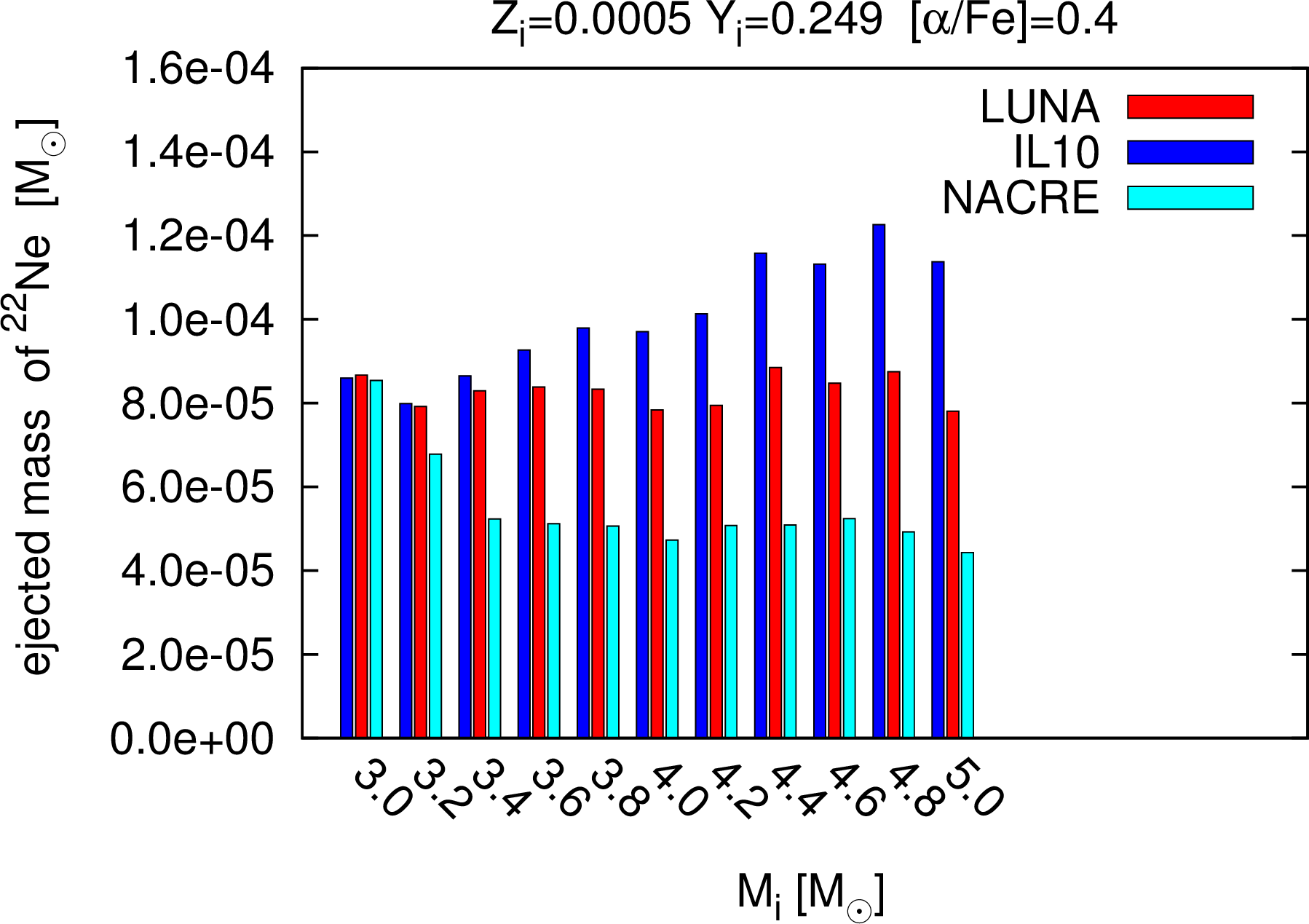}}
\end{minipage}
\caption{$^{22}$Ne and $^{23}$Na ejecta expelled into the interstellar medium by stellar winds during the whole TP-AGB phase by intermediate-mass stars  with HBB as a function of the initial mass and for three choices of the original metallicity, namely: Z$_i=0.014$, Z$_i=0.006$, and  Z$_i=0.0005$. The plots compare the results obtained with four choices for the $^{22}$Ne$(p,\gamma)^{23}$Na rate (as indicated in the upper labels).}
\label{fig_yields}
\end{figure*}

Figure~\ref{fig_yields} illustrates the ejecta of $^{22}$Ne and $^{23}$Na produced by all stellar models in our reference grid \citet[][$M13$, see also Table~\ref{tab_test}]{Marigo_etal13}, for three choices of the initial composition and 
three choices of the $^{22}$Ne$(p,\gamma)^{23}$Na rate.
We do not present the results for $^{24}$Mg since, contrarily 
  to the evolution of the abundance, the time-integrated ejecta are
  found to be little affected by the adopted rate.  This is due to two
  reasons. In stars of relatively low mass or high metallicity the
  temperature at the base of the convective envelope may not reach the
  values necessary to activate the Mg-Al cycle.  In
  more massive and metal-poor stars, that attain the suitable
  temperature conditions, the main contribution to the time-integrated
  $^{24}$Mg ejecta comes from the very initial stages when 
 the abundance of this isotope starts to be quickly reduced 
 by proton captures (see the initial steep decrease of 
$^{24}$Mg  in all panels of Fig.~\ref{fig_hbbnuc}). The initial drop of the $^{24}$Mg
abundance is practically independent of the assumed rate for the
$^{22}$Ne$(p,\gamma)^{23}$Na reaction. Then, when the abundance
evolution of $^{24}$Mg becomes affected by the $^{23}$Na production
rate (as the $^{24}$Mg curve reaches a minimum and starts to increase), the
$^{24}$Mg concentration has already decreased by orders of magnitude, 
  and the contribution to the ejecta remains small. For instance, 
the differences in the final  $^{24}$Mg ejecta  are within $\sim 2-5\%$
for the models in Fig.~\ref{fig_hbbnuc}.

We see that the LUNA results are intermediate between those predicted
with NACRE and IL10. At a given initial stellar mass, the LUNA ejecta
for $^{23}$Na are lower than NACRE, but somewhat larger than IL10. The
opposite is true for $^{22}$Ne. The differences become prominent
towards higher initial stellar masses and lower metallicities,
conditions that favor the development of HBB.

In this respect the bar diagrams also show the minimum mass for the
activation of HBB, in particular the NeNa cycle, in AGB stars as a
function of the metallicity.  We adopt an empirical definition,
  looking for the stellar mass above which the chemical yields of
  $^{22}$Ne and $^{23}$Na, calculated with different rates for 
  the $^{22}$Ne$(p,\gamma)^{23}$Na reaction, 
  start to differ in the bar diagram of
  Fig.~\ref{fig_yields}. At lower masses the yields are essentially the same  
because the nuclear rate remain too low during the TP-AGB phase.
We see that this mass limit is $\sim 4.8\,
M_{\odot}$ at $Z_{\rm i}=0.014$, $ \sim 4.2\, M_{\odot}$ at $Z_{\rm
  i}=0.006$, and $\sim 3.0\, M_{\odot}$ at $Z_{\rm i}=0.0005$.

We also see that the trend of the $^{22}$Ne and $^{23}$Na  ejecta with the stellar mass is not monotonic. At increasing stellar mass, the ejecta initially increase, reach a maximum, and then  decrease again.
The maximum $^{22}$Ne and $^{23}$Na ejecta do not occur at the same initial mass, but a lower mass for $^{22}$Ne,  both decreasing with the metallicity.

These behaviors are the combined result of the strength of HBB, 
the efficiency of the third dredge-up, the TP-AGB lifetime, 
and their dependencies on  stellar mass and metallicity.

\subsection{Nuclear versus stellar model uncertainties}
\label{ssec_totunc}
We discuss here the impact of the uncertainties associated to the
nuclear rate cross sections, as well as those produced by evolutionary
aspects that characterize the AGB evolution.
As to super-AGB stars, the reader may refer to the studies of 
\citet{Doherty_etal14a, Doherty_etal14b}.

\subsubsection{Nuclear uncertainties}
\label{sssec_nucunc}
\begin{figure}
\centering
\resizebox{0.84\hsize}{!}{\includegraphics{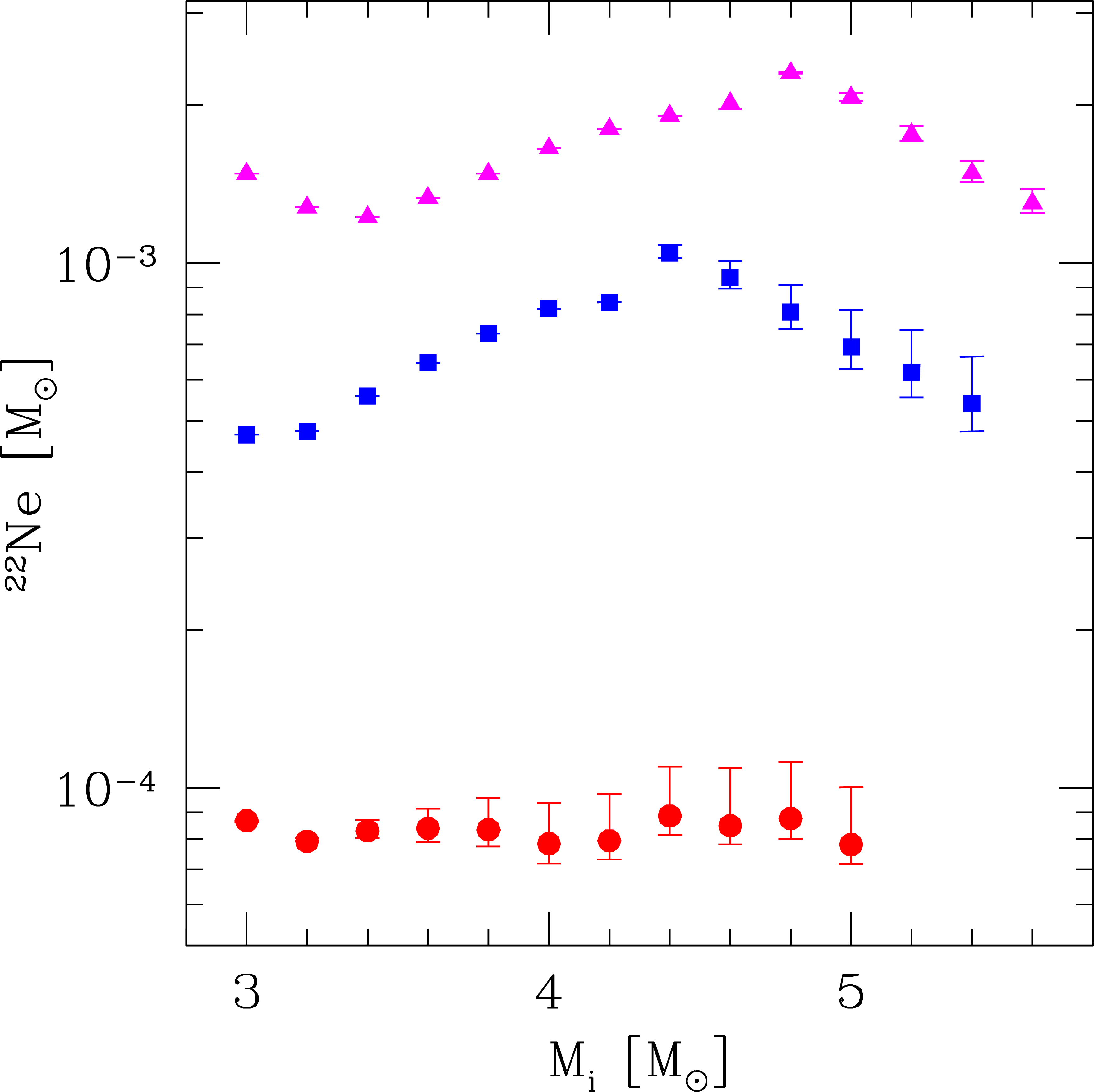}}
\resizebox{0.84\hsize}{!}{\includegraphics{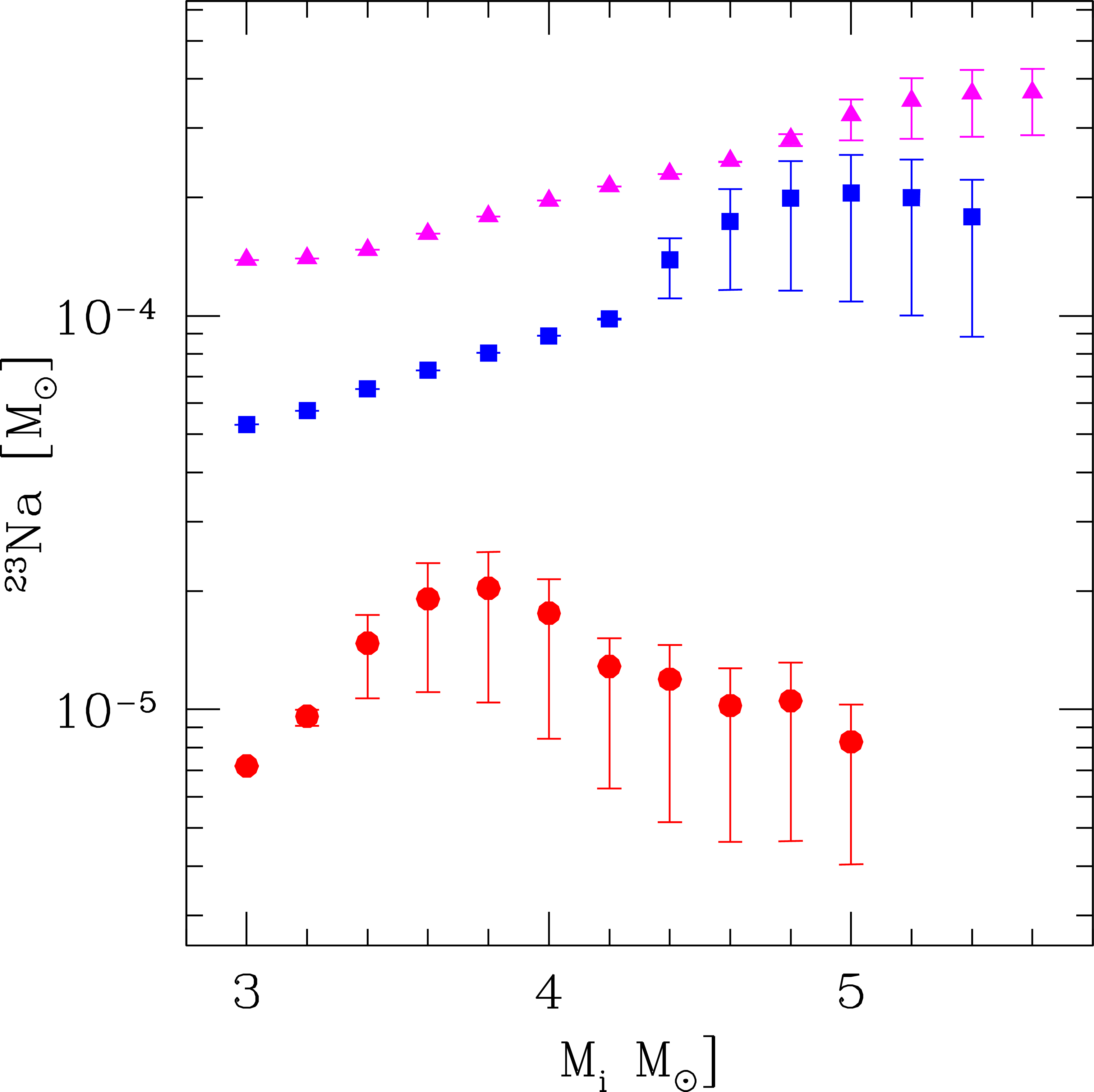}}
\resizebox{0.84\hsize}{!}{\includegraphics{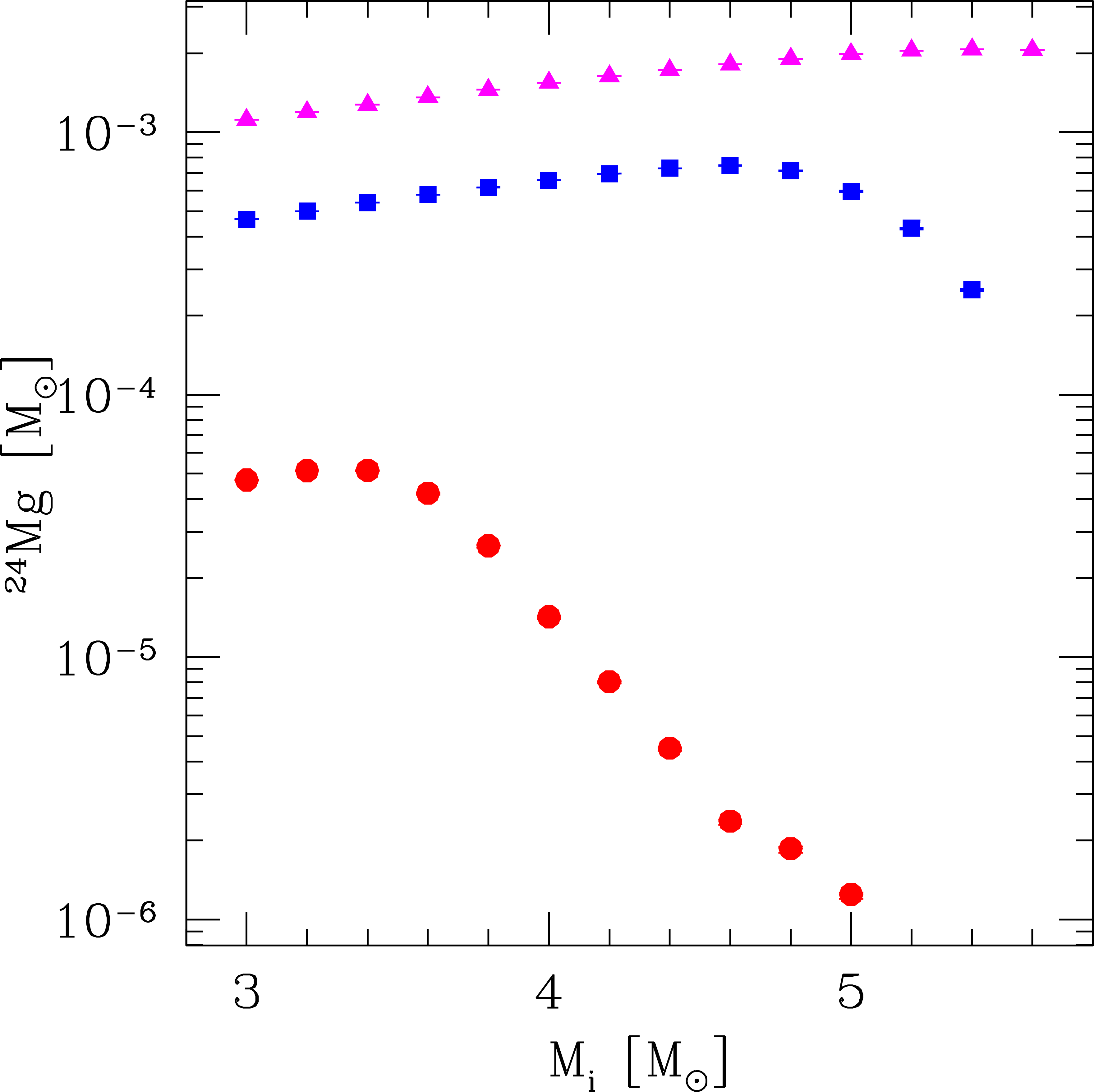}}
\caption{Ejecta and corresponding uncertainties of  $^{22}$Ne, $^{23}$Na, and 
$^{24}$Mg due to the uncertainties in the LUNA rate for the $^{22}$Ne$(p,\gamma)^{23}$Na nuclear reaction, as a function of the initial stellar mass  and metallicity (magenta triangles for $Z_{\rm i}=0.014$, blue squares for $Z_{\rm i}=0.006$, and read circles for  $Z_{\rm i}=0.0005$). Symbols show the results obtained with the recommended rate, while the error bars correspond to the use of the lower and upper limits for the rate (see Fig.~\ref{fig_Cavanna_etal15}).}
\label{fig_lunabars}
\end{figure}

\begin{figure}
\resizebox{0.95\hsize}{!}{\includegraphics{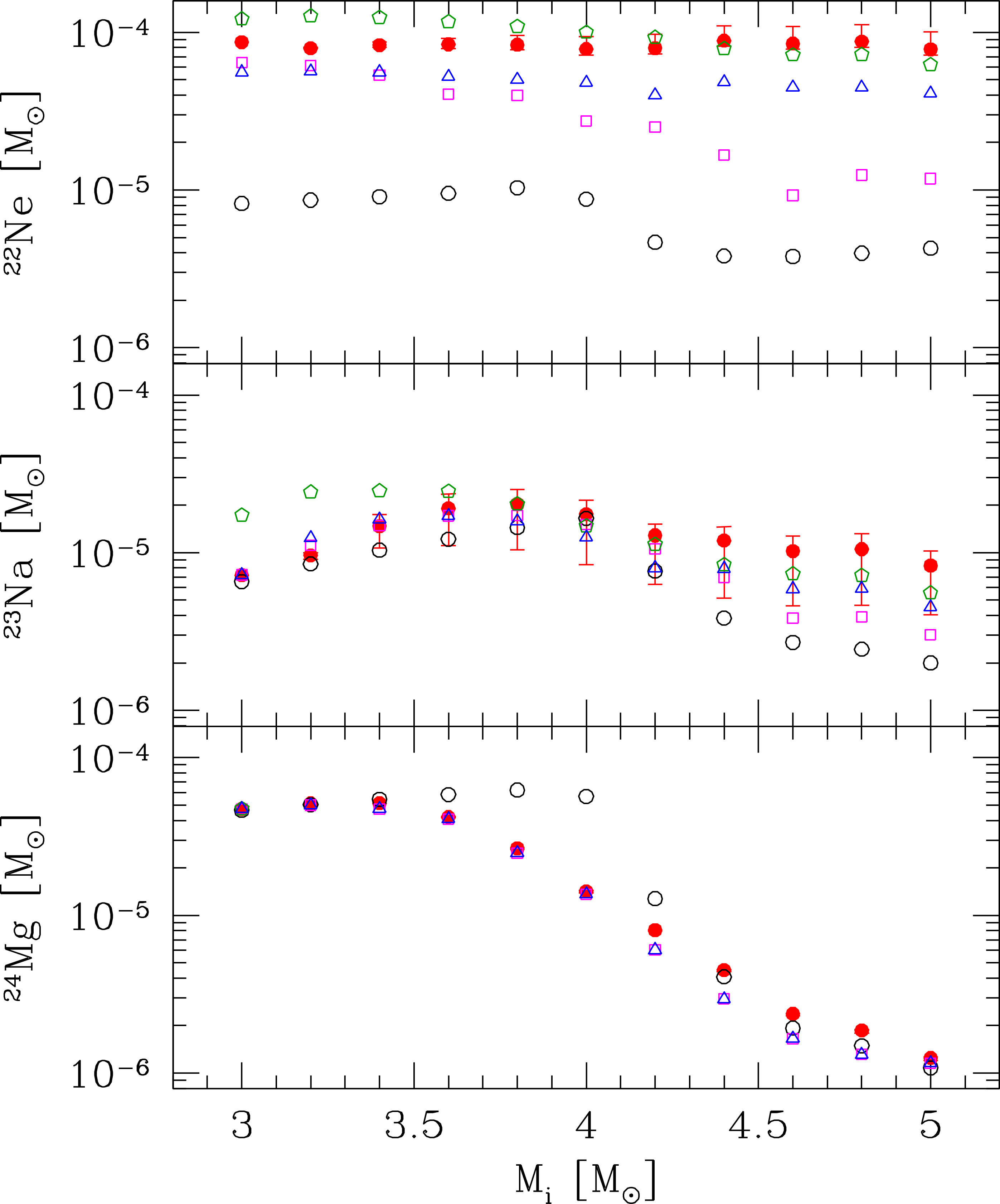}}
\caption{Uncertainties in the $^{22}$Ne, $^{23}$Na, and $^{24}$Mg
  ejecta contributed by stars with initial masses in the range $M_{\rm
    i}=3.0-5.0\, M_{\odot}$ and metallicity $Z_{\rm i}=0.0005$. The
  red error bars represent the uncertainties in the LUNA rate and are
  the same as in Fig.~\ref{fig_lunabars}. The empty symbols correspond
  to the ejecta obtained with the recommended LUNA rate while varying
  other model prescriptions, namely: \citet{VassiliadisWood_93}
  mass-loss law (green pentagons), \citet{Bloecker_95} mass-loss law
  (magenta squares), mixing-length parameter $\alpha_{\rm ML}=2.0$
  (blue triangles), no third dredge-up (black circles).}
\label{fig_evolbars}
\end{figure}

Figure \ref{fig_lunabars} displays the uncertainties in the $^{22}$Ne, 
$^{23}$Na and $^{24}$Mg ejecta ascribed only to the current uncertainties in the LUNA rate of the $^{22}$Ne$(p,\gamma)^{23}$Na reaction, for our reference set of stellar model
prescriptions.
The error bars for $^{22}$Ne and $^{23}$Na increase in models with larger initial mass and lower metallicity. This is not surprising since these conditions favor the development of HBB due to the higher temperatures attained at the base of the convective envelope.

Let us denote with $f_{\rm L}$  and $f_{\rm U}$  the ratios of the 
between the ejecta obtained with the lower and upper limits of the LUNA rate
and the those obtained with the recommended LUNA rate.
In the AGB models with $Z_{\rm i} =0.0005$ and initial masses in the range $3.0-5.0\, M_{\odot}$
the error bars for the $^{22}$Ne  and $^{23}$Na ejecta correspond 
to factor pairs
$(f_{\rm L},f_{\rm U})$ 
of $\simeq (0.92-0.97,1.01-1.28)$ and $(0.43-0.95,1.01-1.25 )$, respectively.
These values are significantly lower than the error bars estimated by \citet{Izzard_etal07}, who reported much wider ranges $\sim (0.14-0.17,1.00-1.01)$
and $\sim (0.53-0.62,33-106)$ for the $^{22}$Ne and $^{23}$Na ejecta
produced by the lowest metallicity set of their synthetic TP-AGB
models\footnote{The quoted results of \citet{Izzard_etal07} refer to
stellar models with $Z_{\rm i}=0.0001$ and $M_{\rm i}=4,5,6\,
M_{\odot}$.} when varying the $^{22}$Ne$(p,\gamma)^{23}$Na rate only.

The LUNA improvement is indeed striking for the upper limit of
$^{23}$Na ejecta, as the relative uncertainty has decreased from $\sim
100$ to $\sim 1.25$ in the worst case.  No significant 
effect is predicted for the ejecta of $^{24}$Mg.

To have a global evaluation of the nuclear uncertainties affecting the ejecta of $^{22}$Ne  and $^{23}$Na we should consider other relevant reactions involved in the NeNa cycle, in particular $\mathrm{^{20}Na(p,\gamma)^{21}Ne}$,  
$\mathrm{^{23}Na(p,\alpha)^{20}Ne}$ and $\mathrm{^{23}Na(p,\gamma)^{24}Mg}$.
To this aim  we refer to the results of detailed investigations carried out
by  \citet{Izzard_etal07} and more recently by \citet{Cesaratto_etal13}.

In the work of \citet{Izzard_etal07} all reaction  rates involved in the NeNa cycle were varied simultaneously
in all possible combinations of lower and upper limits, available at that time.
As to the  $^{23}$Na+p rates, the reference rates were taken from \citet{Rowland_etal04}, and multiplicative factors of $/1.3, \times 1.3$ and   $/40, \times 10$ were adopted to define the lower and upper limits for the rates of $\mathrm{^{23}Na(p,\alpha)^{20}Ne}$ and $\mathrm{^{23}Na(p,\gamma)^{24}Mg}$, respectively.

A  conclusion of the study by \citet{Izzard_etal07} was that the ejecta of $^{22}$Ne  and $^{23}$Na are mainly affected by the uncertainties of the $^{22}$Ne$(p,\gamma)^{23}$Na rate (see tables 6 and 7 of \citet{Izzard_etal07}). Only for  $^{23}$Na the lower-range uncertainties in the ejecta 
were found to be somewhat influenced  by the  uncertainties in the destruction rates $^{23}$Na+p (see their table 7).

More recently, \citet{Cesaratto_etal13} calculated new rates  for  $\mathrm{^{23}Na(p,\alpha)^{20}Ne}$ and $\mathrm{^{23}Na(p,\gamma)^{24}Mg}$ based on nuclear experiments which allowed, for the first time, to derive an upper limit estimate for the strength of a 138-keV resonance,
until then neglected in previous studies.
A consequence of this is that the recommended rate for $\mathrm{^{23}Na(p,\gamma)^{24}Mg}$ has been reduced significantly 
(by over one order of magnitude at $T \simeq 0.07$ GK), compared to the IL10 version. At the same time, the contribution of the 138-keV resonance is found to be negligible for the $\mathrm{^{23}Na(p,\alpha)^{20}Ne}$ reaction and the revised rate of \citet{Cesaratto_etal13}  is in excellent agreement with that of IL10. 

As a result,  the $^{23}$Na destruction due to proton captures appears to be totally dominated by  the $\mathrm{^{23}Na(p,\alpha)^{20}Ne}$ reaction over the temperature range relevant for HBB.
The $\mathrm{(p,\alpha)/(p,\gamma)}$  reaction rate ratio is 
$\ga 100$ all over the temperature interval characteristic of HBB, so that a minor leakage into the Mg-Al cycle is expected \citep[see figure~16 of][]{Cesaratto_etal13}.

Therefore, despite the large reduction of the $\mathrm{^{23}Na(p,\gamma)^{24}Mg}$ rate, the impact on the abundance of $^{23}$Na is quite small. In their test  nucleosynthesis calculations, applied to an AGB model with HBB, \citet{Cesaratto_etal13} derived an increase in the final $^{23}$Na abundance 
by only $\simeq 13\%$ compared the predictions obtained  with the IL10 rate.

Concerning the present estimates for the lower and  upper limit uncertainties of the $^{23}$Na+p reactions over the range temperature range 0.07-0.1 GK, the typical dividing/multiplicative factors with respect to the recommended rate do not exceed $\simeq 1.20-1.25$ in the case of the IL10 rate for $\mathrm{^{23}Na(p,\alpha)^{20}Ne}$, and  are within the range $\simeq 1.4-3.0$ in the case of the rate for $\mathrm{^{23}Na(p,\gamma)^{24}Mg}$ revised by \citet{Cesaratto_etal13}.
These values correspond to relatively small uncertainties and should be taken into consideration when discussing the role of AGB stars with HBB in the context of the observed O-Na anti-correlations of GGC stars (see Section~\ref{sec_ggcs}).

\subsubsection{Evolutionary uncertainties}
\label{ssec_evunc}
It is instructive to compare now the current nuclear uncertainties  with those that are driven by stellar evolution uncertainties.
It is well known that the most problematic aspects to treat on theoretical grounds are those related to mass loss, third dredge-up and HBB, due to our still defective knowledge of the complex physics involved. Basically, we lack an accurate determination of the efficiency of these processes, and how they vary with the mass and the composition of the star.

Mass loss is commonly parameterized in AGB stellar models and several possible
options are available. Depending on the adopted mass-loss rate prescription
quite significant differences arise in the evolutionary models, mainly in terms of lifetimes, number of thermal pulses, chemical enrichment, final core mass,
and HBB over-luminosity \citep[see, e.g.][]{Rosenfield_etal16, Kalirai_etal14, VenturaDantona_05a}.
HBB efficiency is also critically affected by the adopted theoretical framework
to treat convection and its related parameters
\citep[e.g.,][]{VenturaDantona_05b}.
The depth of the third dredge-up is still much debated among AGB modelers
\citep[e.g.,][for a review]{MarigoGirardi_07, Marigo15}, as it critically 
depends also on technical and numerical details \citep{Mowlavi_99b, FrostLattanzio_96}.
For massive AGB stars with $M_{\rm i} \ga 4\, M_{\odot}$, 
the situation is particularly heterogeneous,
as the predictions for the efficiency $\lambda$  vary from high
\citep[$\approx 1$ or larger, e.g.][]{Herwig_04, Karakas_etal02, 
VassiliadisWood_93}, to moderate 
\citep[e.g.,][]{Cristallo_etal15, VenturaDantona_08}.
In this mass range direct constraints from observations
are still lacking, making the overall picture 
rather unclear.

In view of the above, we estimated the impact of stellar evolution
assumptions computing additional TP-AGB models with ($Z_{\rm i}
=0.0005$, [$\alpha$/Fe]=0.4), each time changing an input
prescription.  The adopted prescriptions are summarized in
Table~\ref{tab_test}.  With respect to the reference model, calculated
following $M13$, the changes
were applied to the mixing-length parameter $\alpha_{\rm ML}$, the
mass-loss rate $\dot M$, and the third dredge-up efficiency $\lambda$.
The reference $M13$ model is characterized by a very efficient third
dredge-up (with a maximum $\lambda$ close to unity; see right panels
of Fig.~\ref{fig_3dup}), a relatively efficient HBB which leads to the
activation of the CNO, NeNa, MgAl cycles (see Fig.~\ref{fig_hbbnuc}),
and a mass-loss prescription that was calibrated on a sample of
Galactic Miras.

The sequence of the four models $A-B-C-D$ was chosen to test the
effect on the ejecta of $^{22}$Ne, $^{23}$Na, and $^{24}$Mg when
varying the strength of the aforementioned processes. It is worth
noting that there is a strong coupling among them so that a change in
one process may have a sizable impact also on the others.  The
main results are presented in Fig.~\ref{fig_evolbars} for the whole mass 
range considered and the lowest metallicity $Z=0.0005$, for which HBB is 
expected to be most efficient.  

\paragraph*{Efficiency of mass loss:}
Models $A$ and $B$ differ from  model $M13$ in terms of the mass-loss law. 
While model $A$ adopts the popular mass-loss formula proposed by \citet[][hereinafter also VW93]{VassiliadisWood_93}, model $B$ uses the \citet{Bloecker_95} prescription with the efficiency parameter $\eta=0.02$,
which gives much higher rates.
We find that the VW93 model predicts chemical ejecta that are comparable with those of the $M13$ reference models. In fact the two mass-loss prescriptions, though based on different approaches and different calibration samples, share a similar functional dependence that predicts an exponential increase of $\dot M$ during the initial stages of the TP-AGB evolution \citep[see the discussion in][]{Marigo_etal13}.

 Large differences show up, instead, between the $M13$, $A$
  models, and the models $B$. As to this latter group, the higher mass-loss
  rates lead to a reduction of the TP-AGB lifetimes,
  particularly significant for the most massive and luminous AGB stars.  For
  instance, the $B$ model with $M_{\rm i}=5.0\,M_{\odot}$ suffers a
  lower number of third dredge-up episodes (14 instead of 30) and HBB
  remains active for a shorter time. As a consequence, compared
  to the reference $M13$ models, the $B$ models
  predict ejecta of $^{22}$Ne and $^{23}$Na that are lower by factors 
  in the range $1.3-9.2$  and $1.1-2.7$, respectively.
  The reduction of the $^{24}$Mg yield is smaller, by factors in the
  range $\simeq 1.02-1.5$.

\paragraph*{Efficiency of HBB:}
 Models $C$ test the effect of increasing the strength of
  HBB. This is obtained setting the mixing length parameter to a
  higher value ($\alpha_{\rm ML}=2.00$) compared to the reference
  value ($\alpha_{\rm ML}=1.74$). As a consequence, hotter
  temperatures are attained in the deepest layers of the convective
  envelope so that nuclear reactions in NeNa cycle occur faster.
  Also, the maximum quiescent luminosity attained is larger (e.g.,
  $\log (L)_{\rm max} \simeq 4.81$ instead of $\simeq 4.76$ for the
  reference $M13$ model with $M_{\rm i}=5\, M_{\odot}$). Despite the
  stronger HBB, the integrated yields of $^{22}$Ne, $^{23}$Na, and
  $^{24}$Mg for $C$ models are found to be lower than the $M13$
  predictions (by factors in the range $\simeq 1.1-1.9$). This is
  explained considering that the higher luminosities reached by $C$
  models favor a more intense mass loss, which anticipates the
  termination of the AGB phase (e.g., 24 thermal pulses in $C$ model
  compared to 30 in $M13$ model with $M_{\rm i}=5\, M_{\odot}$).

\paragraph*{Efficiency of the third dredge-up:}
  As models $M13$, $A$, $B$, $C$ are all characterized by a very
  efficient third dredge-up, we explored in the $D$ models the case in
  which no dredge-up ($\lambda=0$) is expected to take place during
  the entire TP-AGB evolution. In this way we may sample the overall
  uncertainty in the chemical yields bracketed by two opposite
  conditions.  The main effect of taking $\lambda=0$ is that no newly
  synthesized $^{22}$Ne is injected into the convective envelope at
  thermal pulses.  As a consequence, the production of $^{23}$Na
  through the $^{22}$Ne$(p,\gamma)^{23}$Na reaction during the
  inter-pulse phase is greatly reduced as it involves only the
  cycling of the NeNa isotopes that are originally present in the
  envelope when HBB is activated. This is evident in
  Fig.~\ref{fig_evolbars} where  the $^{22}$Ne and $^{23}$Na yields
  predicted in models $D$ are found to be lower than those produced by 
  the reference models $M13$ by a factor in the ranges $\simeq 8-22$
  and $\simeq 1.1-4.1$, respectively.
  The variation in the $^{24}$Mg yields is not monotonic with the stellar 
  mass. The absence of the third dredge-up favors larger $^{24}$Mg yields
  at initial masses of $3.5-4.2\,M_{\odot}$, while smaller yields are predicted
  at larger masses, $M_{\rm i} \ga 4.5\, M_{\odot}$. This complex trend is the 
  time-integrated result of mass loss and HBB efficiency during the 
  TP-AGB evolution in stars of different initial masses.

In summary, from this exercise it is evident that the improvements in
the nuclear S-factor for the $^{22}$Ne$(p,\gamma)^{23}$Na reaction
achieved with LUNA have significantly reduced the uncertainties in the
chemical ejecta of $^{22}$Ne and $^{23}$Na produced by
intermediate-mass AGB stars with HBB.  On the other hand, we conclude
that remaining, not negligible, uncertainties are ascribed mainly to
evolutionary aspects that still urge a substantial theoretical effort.

To give some representative numbers we refer to the
($M_{\rm i} = 5.0\, M_{\odot}, Z_{\rm i}=0.0005$) model. The largest
uncertainty factors for the $^{22}$Ne yields due to the nuclear
S-factor of $^{22}$Ne$(p,\gamma)^{23}$Na have decreased 
from $\approx 5-7 $ to $\approx 10-30\%$. As
to the $^{23}$Na yields, we go from $\approx 100$ to $\approx 2$.  At the
same time, the evolutionary uncertainties still make a large
contribution, rising the factors up to $\approx 18$ for $^{22}$Ne
and to $\approx 4$ for $^{23}$Na.
As to the $^{24}$Mg yields, the impact of $^{22}$Ne$(p,\gamma)^{23}$Na
is found to be smaller than in previous estimates
\citep[e.g.,][]{Izzard_etal07}, and its nuclear uncertainties should be 
dominated by other  nuclear reactions in the NeNa cycle 
($^{23}$Na$(p,\gamma)^{24}$Mg, $^{24}$Mg$(p,\gamma)^{25}$Al), not analyzed here.

\section{The oxygen-sodium anti-correlation in GGCs} 
\label{sec_ggcs}
In recent years a number of studies have analyzed the hypothesis
of metal-poor intermediate-mass AGB and super-AGB stars experiencing
HBB as plausible candidates to explain the observed anti-correlations
between light elements (C-N,O-Na, Al-Mg) that characterize the
chemical patterns exhibited by the stars of Galactic globular clusters
\citep[e.g., ][and references therein]{Dantona_etal16, 
Renzini_etal15, Conroy_12, Dercole_etal10,
VenturaDantona_09, Renzini_08, Prantzos_etal07, Karakas_etal06, 
VenturaDantona_05, Fenner_etal04, Herwig_04, DenissenkovHerwig_03}.  
Though a uniform
consensus on the AGB scenario has not been reached \citep[other
stellar candidates are discussed, for instance,
by][]{DenissenkovHartwick_14, Krause_etal13, Demink_etal09, Prantzos_etal07,
Decressin_etal07}, it is interesting to look at the patterns of the
AGB chemical yields on the observed O-Na anti-correlation diagram.
Relevant properties of the AGB ejecta are provided in Table~\ref{tab_ejecta}.

\begin{figure}
\centering
\resizebox{\hsize}{!}{\includegraphics{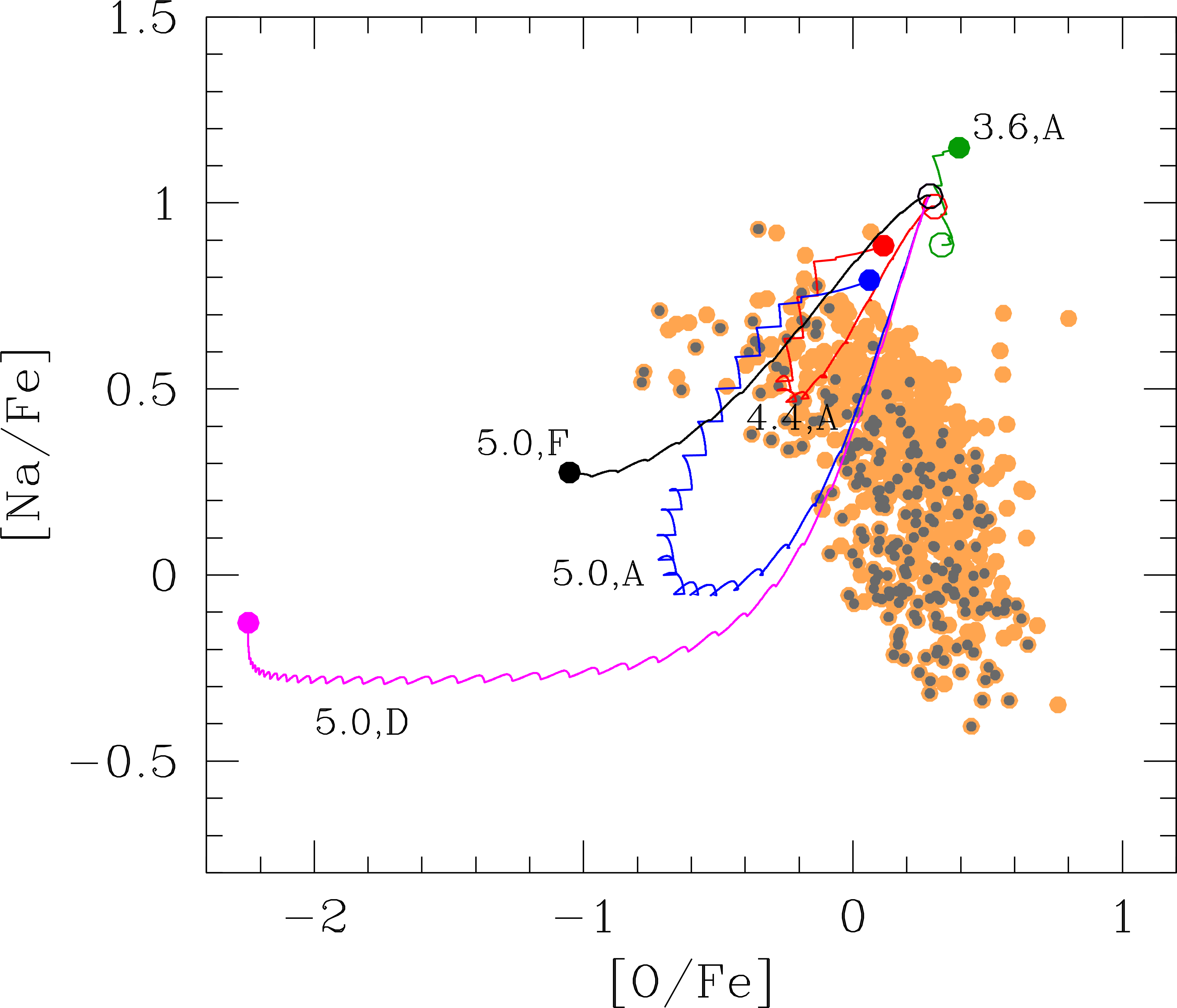}}
\caption{O-Na anti-correlation in stars of GGCs. Spectroscopic data
(orange dots) for 17 clusters are taken from the catalog of 
\citet{Carretta_etal09}.
The data for clusters with iron content $-1.51\la $[Fe/H]$ \la -1.58$ are 
marked with grey dots.
Standard spectroscopic notation is
adopted, i.e. $[Y_i/{\rm Fe}]=\log(n_i/n_{\rm
Fe})-\log(n_{i,\odot}/n_{{\rm Fe},\odot})$ (with $n_i$ being the number
density of the element $i$).
The curves display the evolution of 
abundance ratios during the whole TP-AGB phase for a few selected 
models with initial metallicity $Z_{\rm i}=0.0005$. The corresponding
stellar masses (in $M_\odot$) are indicated on the plot. 
All models correspond to the reference $M13$ prescriptions, except for 
those labeled with $F$ and $D$ (see Table~\ref{tab_test} for details).
In each curve the empty circle marks the abundances after the second dredge-up,
while the filled circle indicates the final ratios at
the termination of the TP-AGB phase.}
\label{fig_naoev}
\end{figure}

In Fig.~\ref{fig_naoev} we show the evolution drawn by a few selected low-metallicity models (with $Z_{\rm i} = 0.0005$, and [$\alpha/$Fe]$=0.4$),  
during their whole TP-AGB evolution, until the complete ejection of the envelope.
This is the result of the combined effect of both HBB and the third dredge-up
(if present), and mass loss.

Among the seventeen clusters included in the catalog 
of \citet{Carretta_etal09}, which span a large range 
in metallicity, four  
(NCG 1904, NGC 3201, NGC 6254, NGC 6752) have  iron abundances 
([Fe/H]$\simeq$ -1.579, -1.512, -1.575, -1.555, respectively) 
that are quite close (within the errors) 
to that of our set of low-metallicity models
([Fe/H]$\simeq$ -1.56)\footnote{Our reference solar mixture
\citep{Caffau_etal11}, and that from \citet{Kurucz_94} used in the spectroscopic  work of
 \citet{Carretta_etal09} are characterized by similar metal abundances,
corresponding to a total Sun's metallicity
$Z_{\odot} \simeq$ 0.0152 and 0.0158, respectively.}.
The abundance data for these clusters (grey dots), draw a 
well-defined O-Na anti-correlation, with a few stars extending into the 
upper region characterized by the highest Na enrichment, which is the main
focus of the analysis that follows.

We note that the $M_{\rm i}=3.6\, M_{\odot}$ model exhibits a modest abundance evolution, characterized by a little depletion of O, and some enrichment in $^{23}$Na due to a relatively mild HBB.
Moving to larger stellar mass (i.e. $M_{\rm i}=4.4, 5.0 \, M_{\odot}$) HBB becomes stronger and the models draw an extended loop,
along which $^{23}$Na is initially destroyed together with O, and later it is
efficiently produced thanks to the periodic injection of fresh
$^{22}$Ne by the third dredge-up at thermal pulses, followed by the
operation of the $^{22}$Ne$(p,\gamma)^{23}$Na reaction during the
inter-pulse periods (see also Fig.~\ref{fig_hbbnuc}). As HBB becomes weaker and eventually extinguishes (due to the reduction
of the envelope mass by stellar winds),
some additional O enrichment may occur if a few final third
dredge-up events take place before the termination of the TP-AGB phase.
Conversely, if no third dredge-up occurs
($\lambda=0$ as in models $F$ and $D$; Table~\ref{tab_test}) the source of $^{22}$Ne synthesized during thermal pulse is not at work so that the 
abundance loop does not show up and sodium is essentially destroyed by HBB with respect to its abundance after the second dredge-up. The significance of the
different trends is discussed further in Section~\ref{ssec_onarep}. 

\begin{figure*}
\centering
\begin{minipage}{0.48\textwidth}
\resizebox{0.8\hsize}{!}{\includegraphics{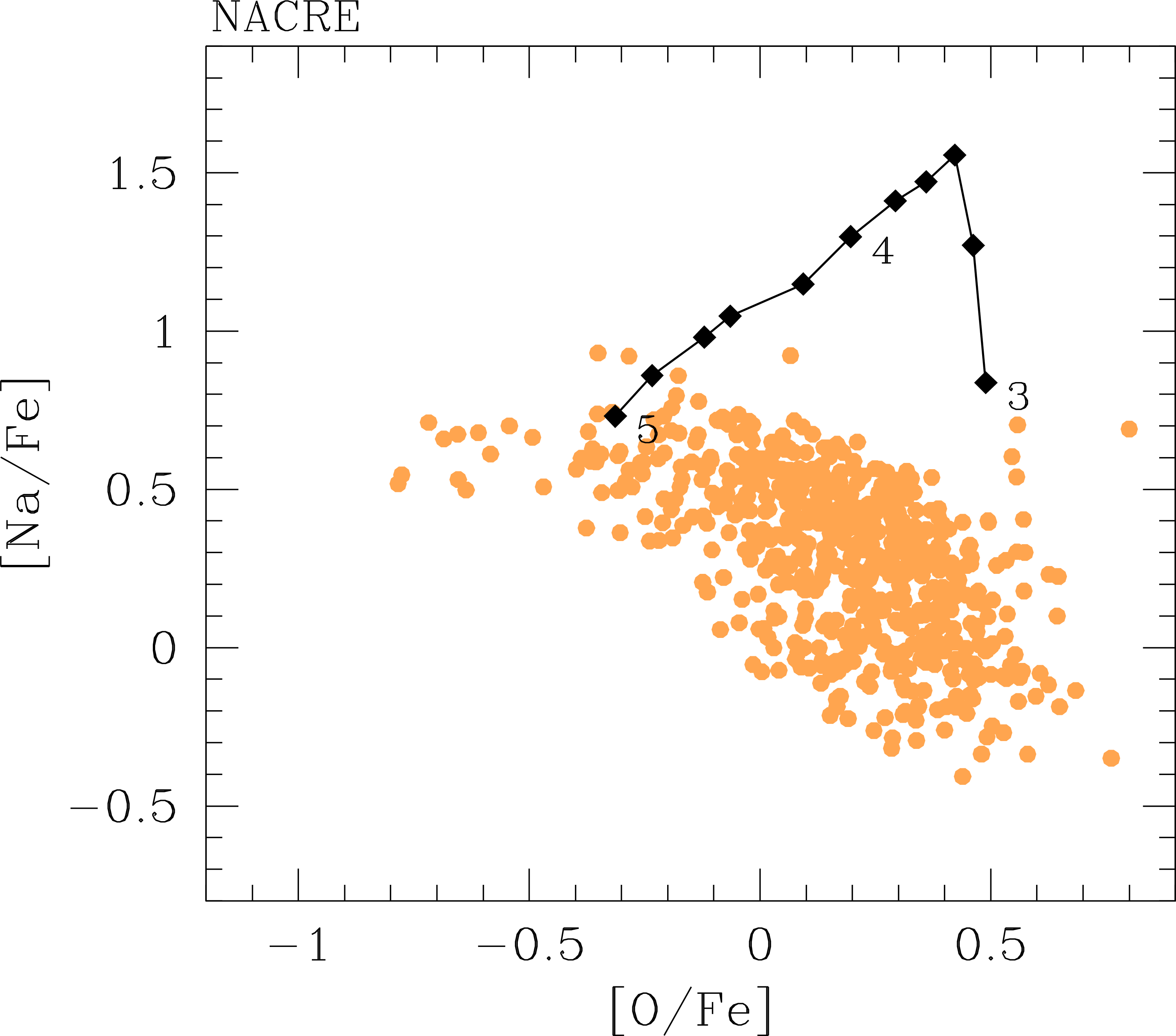}}
\end{minipage}
\hfill
\begin{minipage}{0.48\textwidth}
\resizebox{0.8\hsize}{!}{\includegraphics{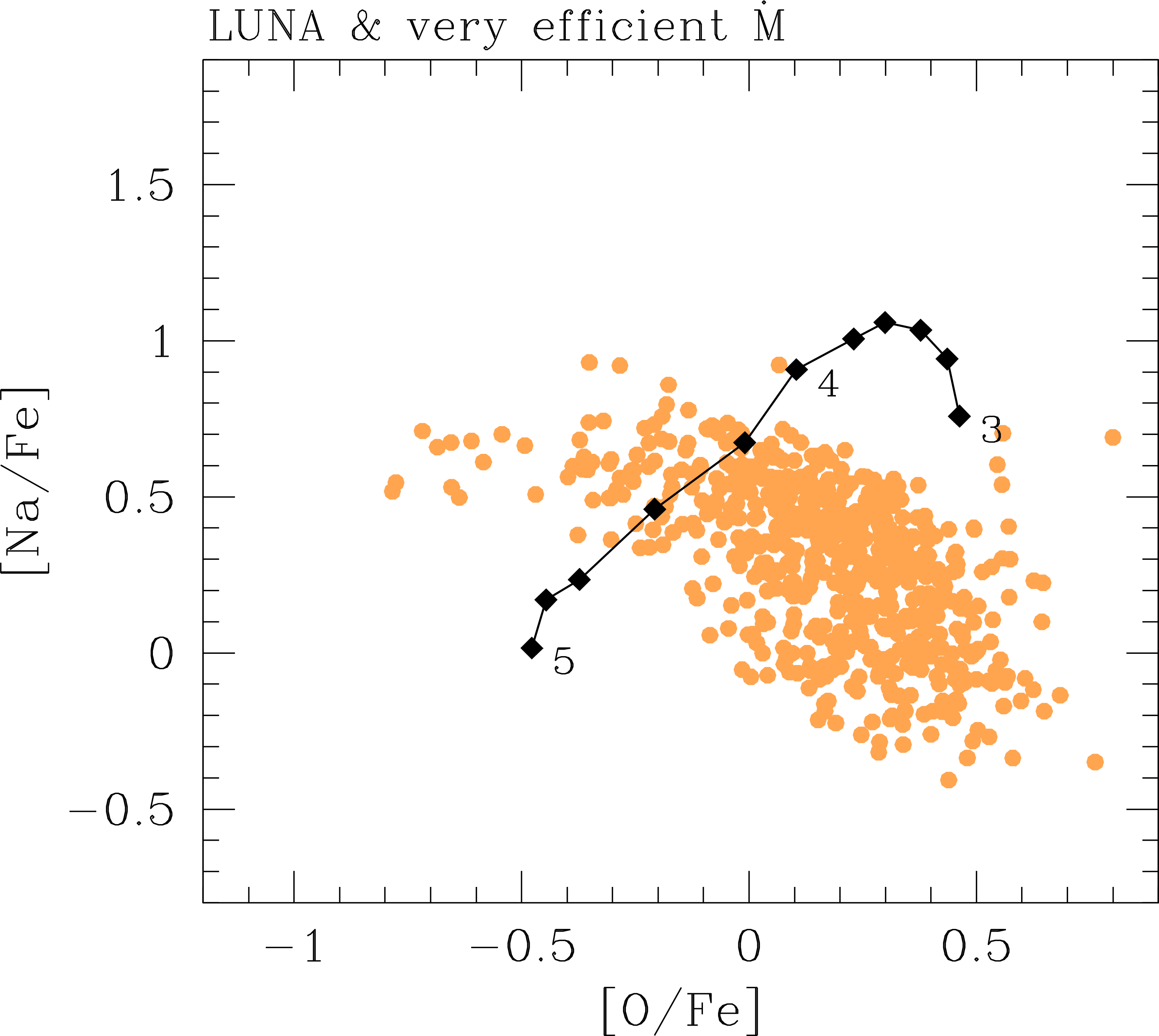}}
\put(-150,140){\Large{B}}
\end{minipage}
\hfill
\begin{minipage}{0.48\textwidth}
\resizebox{0.8\hsize}{!}{\includegraphics{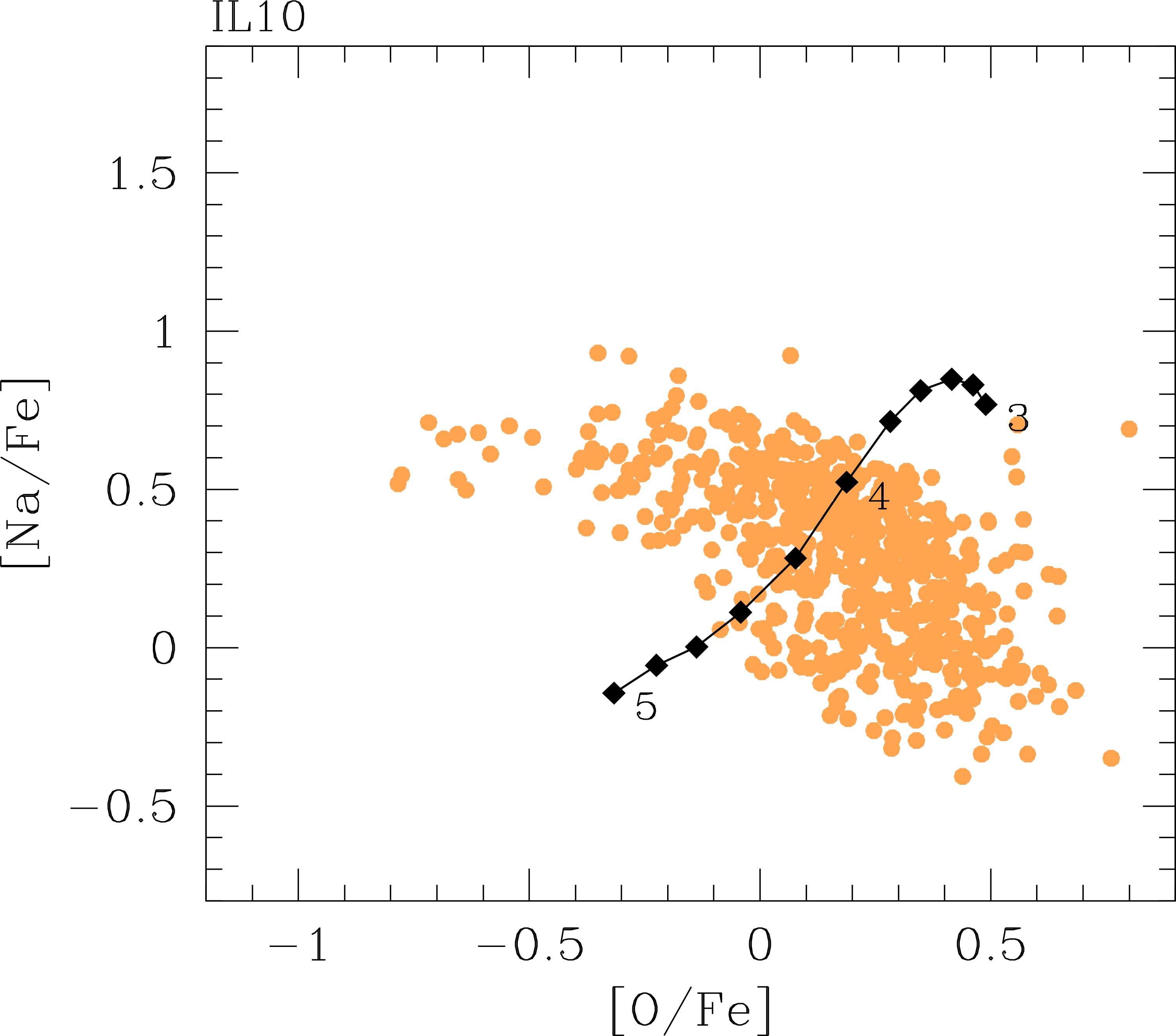}}
\end{minipage}
\hfill
\begin{minipage}{0.48\textwidth}
\resizebox{0.8\hsize}{!}{\includegraphics{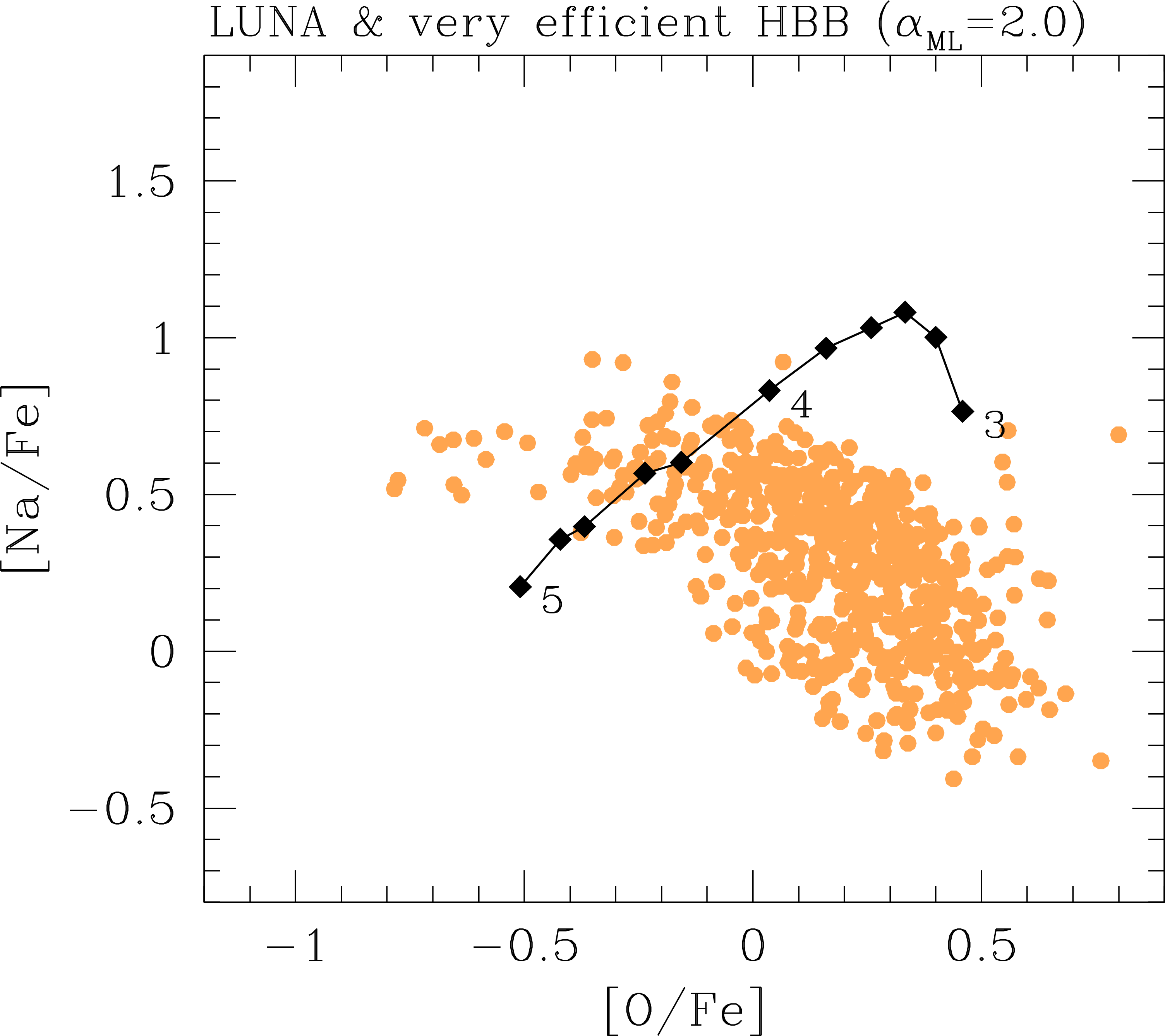}}
\put(-150,140){\Large{C}}
\end{minipage}
\hfill
\begin{minipage}{0.48\textwidth}
  \resizebox{0.8\hsize}{!}{\includegraphics{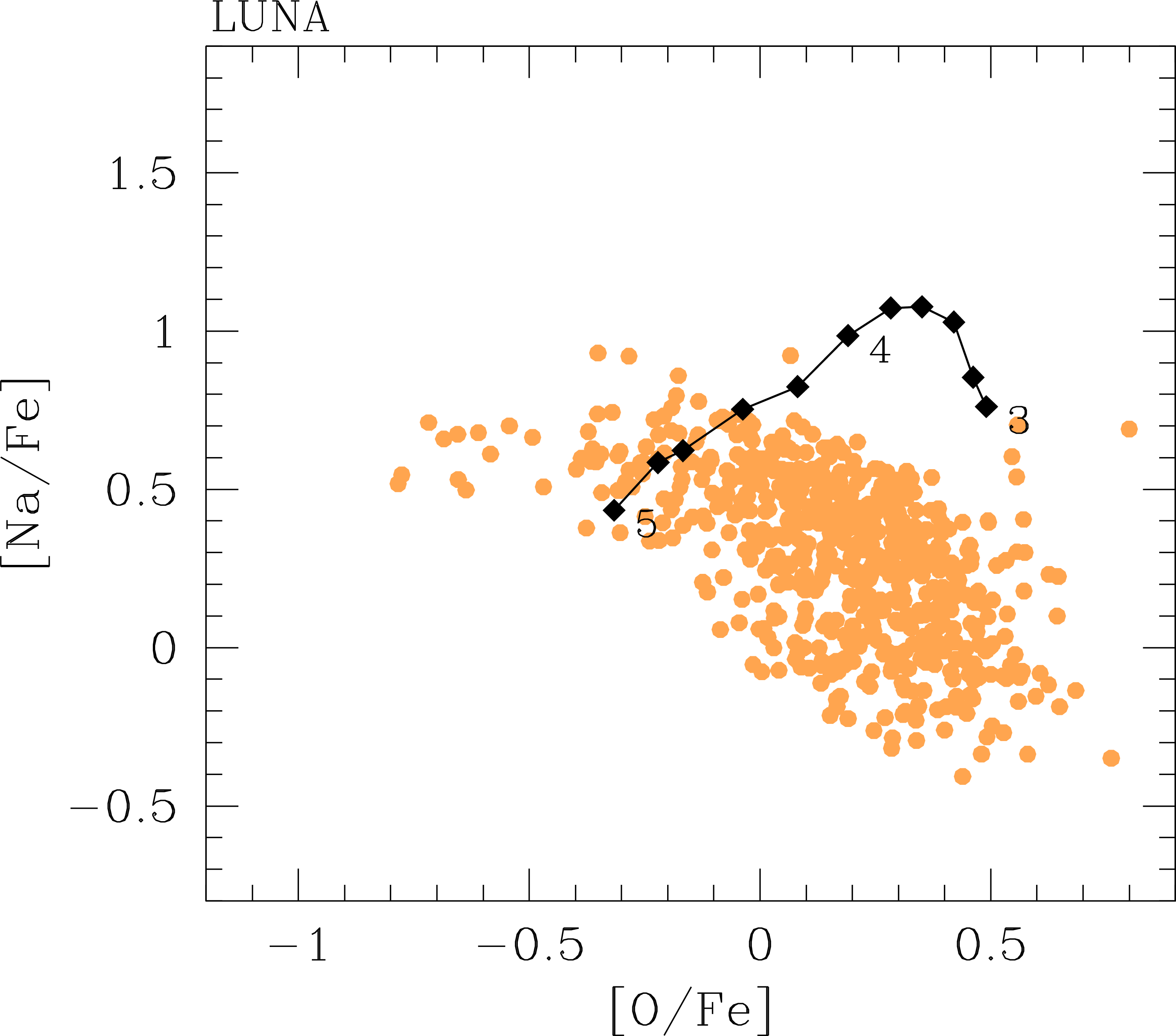}}
  \put(-150,140){\Large{A}}
\end{minipage}
\hfill
\begin{minipage}{0.48\textwidth}
\resizebox{0.8\hsize}{!}{\includegraphics{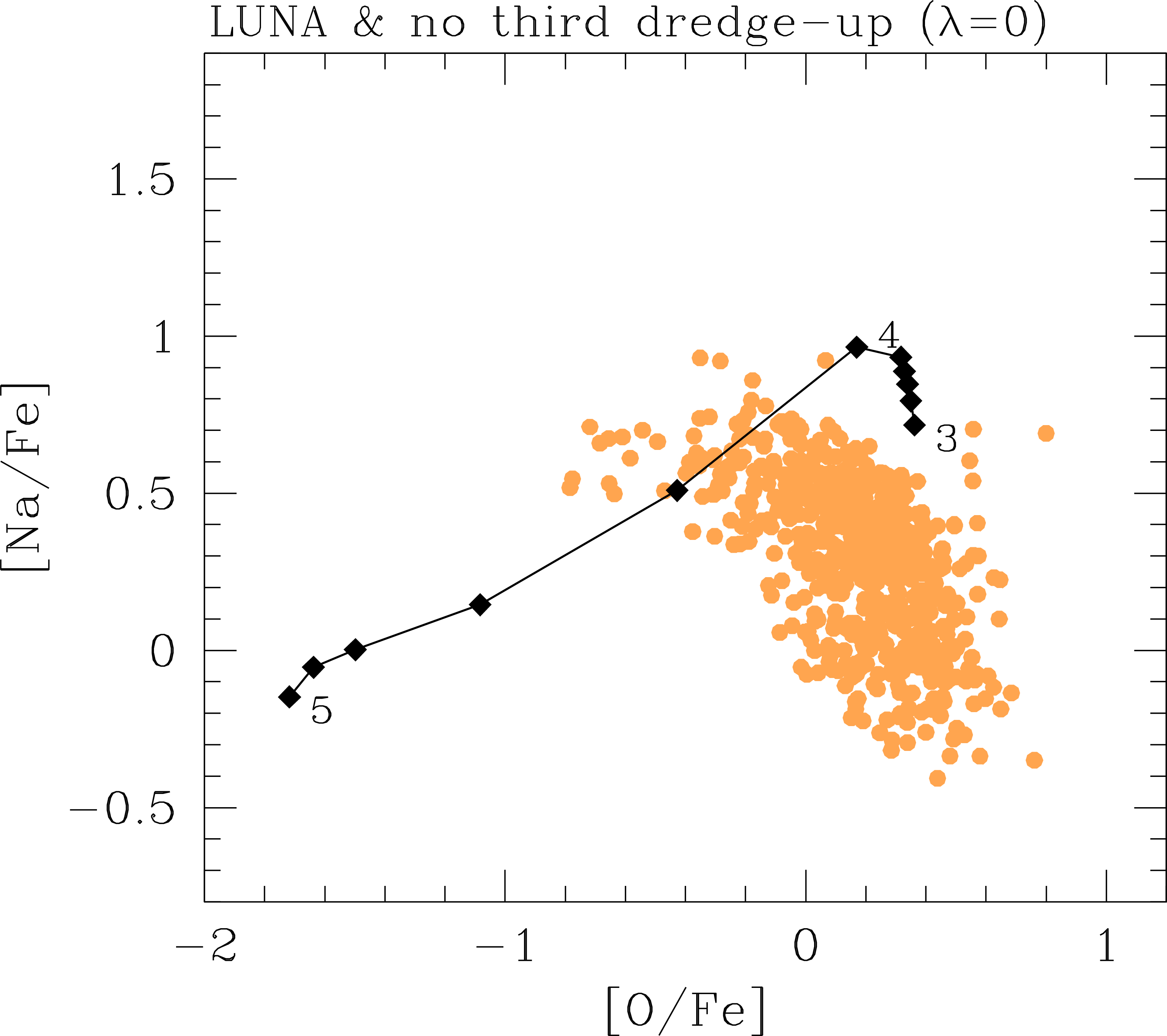}}
\put(-150,140){\Large{D}}
\end{minipage}
\caption{O-Na anti-correlation in stars of GGCs. Observed data are the same as in Fig.~\ref{fig_naoev}.
In each panel the sequence of filled squares (from right to left)
corresponds to the elemental ratios [Na/Fe] and [O/Fe] in the TP-AGB
ejecta of stars with initial composition Z$_{\rm i} =0.0005$,
[$\alpha$/Fe]=0.4, and masses from 3.0 $M_{\odot}$ to 5.0 $M_{\odot}$
in steps of $0.2\, M_{\odot}$. Few selected values of the mass (in $M_{\odot}$) are indicated nearby the corresponding model.
Panels of the left row: all models share the same AGB phase prescriptions 
(our reference case  $M13$), but for the rate of
$^{22}$Ne$(p,\gamma)^{23}$Na (see Table~\ref{tab:models}). 
Panels of the right row (from top to bottom): results obtained with
the LUNA rate, but varying other model assumptions, as described in
Table~\ref{tab_test} and marked by the corresponding capital letter
on top-left.
See the text for more explanation.}
\label{fig_nao}
\end{figure*}

The left panels of Fig.~\ref{fig_nao} (from top to bottom)
compare the results obtained with the $M13$ prescriptions but varying the rate of the $^{22}$Ne$(p,\gamma)^{23}$Na reaction applied to the low-metallicity set of stellar models.  
Each stellar model is represented by a point in the diagram, whose coordinates are the surface abundance ratios computed as weighted averages, that is
summing up the amounts of elements
ejected at each time time step and then normalizing them to the total
ejected mass.
The range of initial masses goes from 3.0
$M_{\odot}$ to 5.0 $M_{\odot}$ in steps of $0.2\, M_{\odot}$.

\begin{table*}
\centering
\caption{Properties of AGB models with initial metallicity $Z_{\rm i}=0.0005$
 and composition of their ejecta, obtained  with the LUNA rate.
The prescriptions used in the different sets of models are also described in 
Table~\ref{tab:models}.
From left to right the columns indicate: the initial stellar mass, the total number of thermal pulses, the final core mass,
the average helium abundance (in mass fraction), 
the average abundance ratios expressed as $\mathrm{<[n_i/n({\rm Fe})]>}$ (with abundances by number) in the ejecta of C, N, O,
the enhancement factor of the CNO content, and the average abundance ratios  of Na, Mg, and Al. As to Li, the corresponding average abundance is expressed as
$\mathrm{\log [n(Li)/n(H)]+12}$.}
\label{tab_ejecta}
\begin{tabular}{cccccccccccc}
\multicolumn{12}{c}{$Z_{\rm i} =0.0005$, $Y_{\rm i} =0.249$,  [$\alpha$/Fe]=0.4} \\
\hline
\multicolumn{12}{l}{Reference $M13$ prescriptions}\\
\multicolumn{12}{l}{Efficient third dredge-up}\\
M$_{\rm i}$ [$M_{\odot}$] & $N_{\rm tp}$ & M$_{\rm fin}$ [$M_{\odot}$]  &  <X(He)> &  <A(Li)> &  <[C/Fe]> &  <[N/Fe]> & <[O/Fe]> &
$R_{\rm cno}$ & <[Na/Fe]> & <[Mg/Fe]> & <[Al/Fe]>\\
\hline
    3.0 &  11 &   0.81 &   0.30 &   3.74 &   1.81 &   1.45 &  0.49 &  11.77 &   0.76 &   0.44 &   0.05\\
    3.2 &  11 &   0.82 &   0.31 &   3.95 &   1.34 &   2.21 &  0.46 &  10.75 &   0.85 &   0.43 &   0.08\\
    3.4 &  12 &   0.83 &   0.32 &   3.46 &   1.21 &   2.28 &  0.42 &  10.94 &   1.03 &   0.44 &   0.11\\
    3.6 &  13 &   0.85 &   0.33 &   3.30 &   1.03 &   2.12 &  0.35 &  10.94 &   1.08 &   0.44 &   0.15\\
    3.8 &  14 &   0.86 &   0.34 &   3.19 &   0.99 &   2.12 &  0.28 &  10.75 &   1.07 &   0.45 &   0.26\\
    4.0 &  16 &   0.88 &   0.35 &   3.12 &   0.90 &   2.14 &  0.19 &  10.23 &   0.99 &   0.45 &   0.32\\
    4.2 &  17 &   0.89 &   0.35 &   3.03 &   0.78 &   2.12 &  0.08 &   9.95 &   0.82 &   0.45 &   0.43\\
    4.4 &  20 &   0.91 &   0.36 &   2.91 &   0.70 &   2.11 & -0.04 &  10.26 &   0.75 &   0.45 &   0.56\\
    4.6 &  23 &   0.93 &   0.36 &   2.77 &   0.57 &   2.07 & -0.17 &   9.71 &   0.62 &   0.43 &   0.67\\
    4.8 &  26 &   0.94 &   0.37 &   2.72 &   0.54 &   2.07 & -0.22 &   9.76 &   0.58 &   0.42 &   0.71\\
    5.0 &  30 &   0.97 &   0.37 &   2.72 &   0.48 &   2.01 & -0.32 &   8.85 &   0.43 &   0.39 &   0.74\\
    \hline
\multicolumn{12}{l}{Models $B$: Efficient mass loss with  \citet{Bloecker_95} and $\eta=0.02$}\\
M$_{\rm i}$ [$M_{\odot}$] & $N_{\rm tp}$ & M$_{\rm fin}$ [$M_{\odot}$]  &  <X(He)> &  <A(Li)> &  <[C/Fe]> &  <[N/Fe]> & <[O/Fe]> &
$R_{\rm cno}$ & <[Na/Fe]> & <[Mg/Fe]> & <[Al/Fe]>\\
\hline
    3.0 & 10 &  0.80 &   0.30 &   4.00 &   1.37 &   2.23 &  0.46 &  10.36 &   0.76 &   0.42 &   0.05\\
    3.2 & 10 &  0.81 &   0.31 &   3.66 &   1.01 &   2.31 &  0.44 &   9.78 &   0.94 &   0.42 &   0.09\\
    3.4 & 10 &  0.82 &   0.32 &   3.38 &   1.31 &   2.22 &  0.38 &   8.86 &   1.03 &   0.42 &   0.12\\
    3.6 & 10 &  0.84 &   0.33 &   3.15 &   0.27 &   2.27 &  0.30 &   7.30 &   1.06 &   0.41 &   0.16\\
    3.8 & 11 &  0.85 &   0.34 &   2.97 &   1.19 &   2.14 &  0.23 &   7.18 &   1.01 &   0.42 &   0.26\\
    4.0 & 11 &  0.87 &   0.35 &   2.83 &   0.19 &   2.18 &  0.10 &   5.62 &   0.91 &   0.41 &   0.32\\
    4.2 & 12 &  0.89 &   0.35 &   2.70 &   0.98 &   2.04 & -0.01 &   5.23 &   0.67 &   0.41 &   0.43\\
    4.4 & 12 &  0.91 &   0.36 &   2.60 &   0.25 &   2.04 & -0.21 &   3.91 &   0.46 &   0.39 &   0.55\\
    4.6 & 13 &  0.92 &   0.36 &   2.65 &  -0.03 &   1.85 & -0.37 &   2.81 &   0.23 &   0.36 &   0.65\\
    4.8 & 13 &  0.94 &   0.37 &   2.55 &   0.07 &   1.92 & -0.45 &   2.95 &   0.17 &   0.35 &   0.72\\
    5.0 & 14 &  0.96 &   0.37 &   2.70 &   0.49 &   1.74 & -0.48 &   2.52 &   0.02 &   0.33 &   0.72\\
\hline
\multicolumn{12}{l}{Models $C$: Efficient HBB with $\alpha_{\rm ML}$=2.0}\\
M$_{\rm i}$ [$M_{\odot}$] & $N_{\rm tp}$ & M$_{\rm fin}$ [$M_{\odot}$]  &  <X(He)> &  <A(Li)> &  <[C/Fe]> &  <[N/Fe]> & <[O/Fe]> &
$R_{\rm cno}$ & <[Na/Fe]> & <[Mg/Fe]> & <[Al/Fe]>\\
\hline
    3.0  & 10 &  0.81 &   0.30 &   3.82 &   1.38 &   2.03 &  0.46 &   8.77 &   0.77 &   0.41 &   0.05\\
    3.2  & 10 &  0.82 &   0.31 &   3.41 &   1.27 &   2.11 &  0.40 &   8.82 &   1.00 &   0.42 &   0.10\\
    3.4  & 11 &  0.83 &   0.32 &   3.31 &   1.04 &   2.19 &  0.33 &   8.76 &   1.08 &   0.43 &   0.14\\
    3.6  & 12 &  0.84 &   0.33 &   3.18 &   0.72 &   2.06 &  0.26 &   8.41 &   1.03 &   0.42 &   0.18\\
    3.8  & 13 &  0.86 &   0.34 &   3.13 &   0.73 &   2.04 &  0.16 &   8.18 &   0.97 &   0.42 &   0.31\\
    4.0  & 14 &  0.87 &   0.35 &   3.05 &   0.70 &   2.04 &  0.04 &   7.82 &   0.83 &   0.42 &   0.42\\
    4.2  & 15 &  0.89 &   0.35 &   2.94 &   0.49 &   1.96 & -0.16 &   6.70 &   0.60 &   0.39 &   0.58\\
    4.4  & 17 &  0.91 &   0.36 &   2.82 &   0.47 &   1.97 & -0.24 &   7.22 &   0.57 &   0.38 &   0.66\\
    4.6  & 19 &  0.93 &   0.36 &   2.74 &   0.33 &   1.91 & -0.37 &   6.53 &   0.40 &   0.35 &   0.74\\
    4.8  & 21 &  0.94 &   0.37 &   2.72 &   0.31 &   1.90 & -0.42 &   6.42 &   0.36 &   0.34 &   0.75\\
    5.0  & 24 &  0.97 &   0.37 &   2.76 &   0.24 &   1.83 & -0.51 &   5.74 &   0.21 &   0.31 &   0.73\\    
\hline
\multicolumn{12}{l}{Models $D$: No third dredge-up ($\lambda$=0)}\\
M$_{\rm i}$ [$M_{\odot}$] & $N_{\rm tp}$ & M$_{\rm fin}$ [$M_{\odot}$]  &  <X(He)> &  <A(Li)> &  <[C/Fe]> &  <[N/Fe]> & <[O/Fe]> &
$R_{\rm cno}$ & <[Na/Fe]> & <[Mg/Fe]> & <[Al/Fe]>\\
\hline
    3.0 &  17 &  0.85 &   0.29 &  -1.14 &  -0.29 &   0.68 &  0.36 &   1.00 &   0.72 &   0.40 &   0.06\\
    3.2 &  20 &  0.86 &   0.31 &  -0.77 &  -0.29 &   0.72 &  0.35 &   1.00 &   0.79 &   0.40 &   0.08\\
    3.4 &  24 &  0.88 &   0.32 &   0.68 &  -0.29 &   0.76 &  0.34 &   1.00 &   0.85 &   0.40 &   0.11\\
    3.6 &  29 &  0.90 &   0.32 &   2.50 &  -0.30 &   0.79 &  0.33 &   1.00 &   0.89 &   0.40 &   0.13\\
    3.8 &  34 &  0.92 &   0.33 &   3.47 &  -0.89 &   0.91 &  0.32 &   1.00 &   0.93 &   0.39 &   0.22\\
    4.0 &  38 &  0.94 &   0.34 &   3.27 &  -0.99 &   1.10 &  0.17 &   1.00 &   0.96 &   0.39 &   0.25\\
    4.2 &  41 &  0.96 &   0.35 &   3.03 &  -0.81 &   1.29 & -0.43 &   1.00 &   0.51 &   0.39 &   0.48\\
    4.4 &  42 &  0.97 &   0.35 &   2.89 &  -0.75 &   1.34 & -1.08 &   1.00 &   0.15 &   0.32 &   0.79\\
    4.6 &  44 &  0.99 &   0.36 &   2.77 &  -0.72 &   1.36 & -1.50 &   1.00 &   0.00 &   0.23 &   0.89\\
    4.8 &  47 &  1.00 &   0.36 &   2.72 &  -0.70 &   1.36 & -1.64 &   1.00 &  -0.05 &   0.19 &   0.90\\
    5.0 &  50 &  1.02 &   0.37 &   2.72 &  -0.67 &   1.36 & -1.72 &   1.00 &  -0.15 &   0.14 &   0.80\\
\hline
\end{tabular}
\end{table*}

\begin{table*}
\centering
\contcaption{}
\label{tab_ejectacont}
\begin{tabular}{cccccccccccc}
\hline
\multicolumn{12}{l}{Models $E$: Efficient HBB with $\alpha_{\rm ML}=2$,
  $\lambda_{\rm max}=0.5$, $^{23}$Na$({\rm p},\alpha)^{20}$Na / 5}\\
M$_{\rm i}$ [$M_{\odot}$] & $N_{\rm tp}$ & M$_{\rm fin}$ [$M_{\odot}$]  &  <X(He)> &  <A(Li)> &  <[C/Fe]> &  <[N/Fe]> & <[O/Fe]> &
$R_{\rm cno}$ & <[Na/Fe]> & <[Mg/Fe]> & <[Al/Fe]>\\
\hline
    3.0 &  12 &  0.82 &   0.30 &   4.07 &   1.53 &   1.52 &  0.45 &   7.76 &   0.74 &   0.42 &   0.05\\
    3.2 &  13 &  0.83 &   0.31 &   3.45 &   1.15 &   1.94 &  0.40 &   7.73 &   0.91 &   0.42 &   0.09\\
    3.4 &  14 &  0.84 &   0.32 &   3.32 &   1.02 &   1.99 &  0.32 &   7.49 &   1.06 &   0.43 &   0.14\\
    3.6 &  15 &  0.86 &   0.33 &   3.22 &   0.67 &   1.83 &  0.23 &   7.18 &   1.08 &   0.43 &   0.19\\
    3.8 &  16 &  0.87 &   0.34 &   3.13 &   0.66 &   1.87 &  0.10 &   6.59 &   1.12 &   0.43 &   0.33\\
    4.0 &  18 &  0.89 &   0.34 &   3.04 &   0.51 &   1.91 & -0.05 &   6.58 &   1.08 &   0.43 &   0.45\\
    4.2 &  20 &  0.91 &   0.35 &   2.94 &   0.45 &   1.88 & -0.22 &   6.34 &   1.01 &   0.42 &   0.61\\
    4.4 &  22 &  0.92 &   0.36 &   2.82 &   0.28 &   1.84 & -0.41 &   5.73 &   0.89 &   0.39 &   0.73\\
    4.6 &  24 &  0.94 &   0.36 &   2.74 &   0.21 &   1.78 & -0.56 &   5.11 &   0.76 &   0.35 &   0.80\\
    4.8 &  25 &  0.95 &   0.37 &   2.72 &   0.15 &   1.78 & -0.63 &   4.75 &   0.71 &   0.33 &   0.81\\
    5.0 &  29 &  0.98 &   0.37 &   2.76 &   0.09 &   1.72 & -0.71 &   4.37 &   0.59 &   0.31 &   0.77\\
\hline
\multicolumn{12}{l}{Models $F$: Efficient mass loss with \citet{Bloecker_95} and $\eta=0.03$, $\lambda=0$, $^{23}$Na$({\rm p},\alpha)^{20}$Na / 3}\\
M$_{\rm i}$ [$M_{\odot}$] & $N_{\rm tp}$ & M$_{\rm fin}$ [$M_{\odot}$]  &  <X(He)> &  <A(Li)> &  <[C/Fe]> &  <[N/Fe]> & <[O/Fe]> &
$R_{\rm cno}$ & <[Na/Fe]> & <[Mg/Fe]> & <[Al/Fe]>\\
\hline
    3.0 & 17 &  0.84 &   0.29 &  -1.41 &  -0.29 &   0.68 &  0.36 &   1.00 &   0.72 &   0.40 &   0.06\\
    3.2 & 19 &  0.86 &   0.31 &  -0.21 &  -0.29 &   0.72 &  0.35 &   1.00 &   0.79 &   0.40 &   0.08\\
    3.4 & 23 &  0.88 &   0.32 &   1.59 &  -0.29 &   0.76 &  0.34 &   1.00 &   0.85 &   0.40 &   0.11\\
    3.6 & 23 &  0.89 &   0.32 &   3.46 &  -0.31 &   0.80 &  0.33 &   1.00 &   0.89 &   0.40 &   0.13\\
    3.8 & 28 &  0.91 &   0.33 &   3.38 &  -1.20 &   0.94 &  0.31 &   1.00 &   0.94 &   0.39 &   0.22\\
    4.0 & 29 &  0.92 &   0.34 &   3.09 &  -0.99 &   1.16 &  0.10 &   1.00 &   1.03 &   0.40 &   0.27\\
    4.2 & 27 &  0.93 &   0.35 &   2.87 &  -0.83 &   1.30 & -0.31 &   1.00 &   0.95 &   0.40 &   0.42\\
    4.4 & 24 &  0.94 &   0.35 &   2.76 &  -0.77 &   1.35 & -0.68 &   1.00 &   0.73 &   0.38 &   0.65\\
    4.6 & 20 &  0.95 &   0.36 &   2.74 &  -0.74 &   1.36 & -0.86 &   1.00 &   0.54 &   0.35 &   0.76\\
    4.8 & 20 &  0.96 &   0.36 &   2.73 &  -0.71 &   1.36 & -0.95 &   1.00 &   0.46 &   0.34 &   0.79\\
    5.0 & 17 &  0.98 &   0.37 &   2.89 &  -0.71 &   1.35 & -0.86 &   1.00 &   0.35 &   0.33 &   0.76\\
\hline
\end{tabular}
\end{table*}

\begin{figure*}
\centering
\begin{minipage}{0.48\textwidth}
\resizebox{0.8\hsize}{!}{\includegraphics{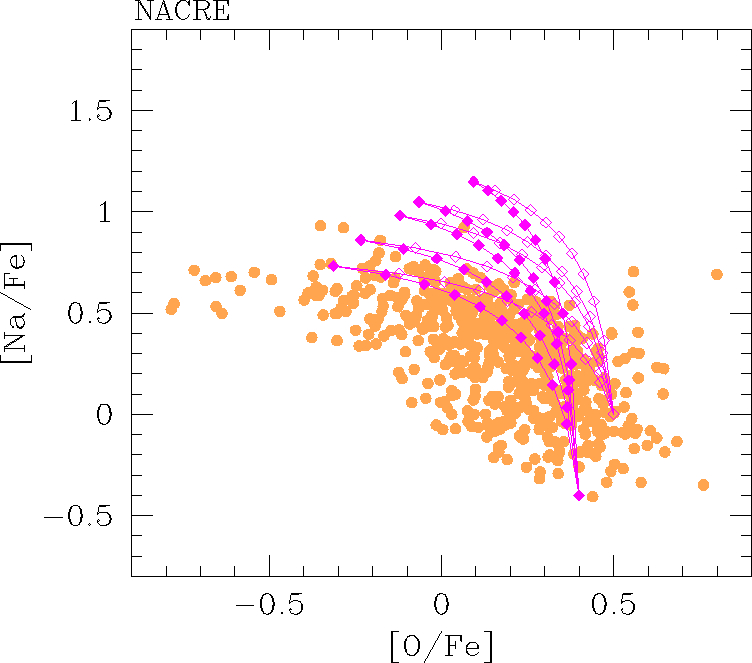}}
\end{minipage}
\hfill
\begin{minipage}{0.48\textwidth}
\resizebox{0.8\hsize}{!}{\includegraphics{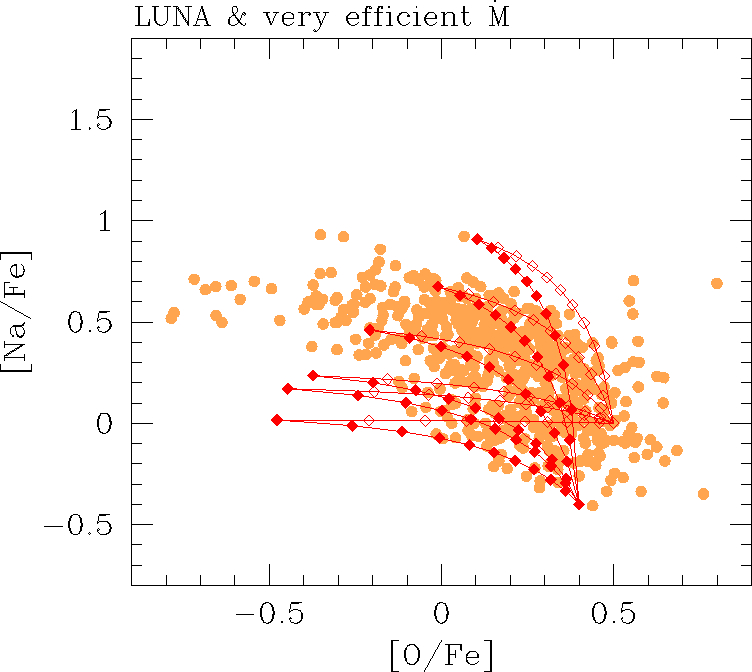}}
\put(-150,140){\Large{B}}
\end{minipage}
\hfill
\begin{minipage}{0.45\textwidth}
\resizebox{0.8\hsize}{!}{\includegraphics{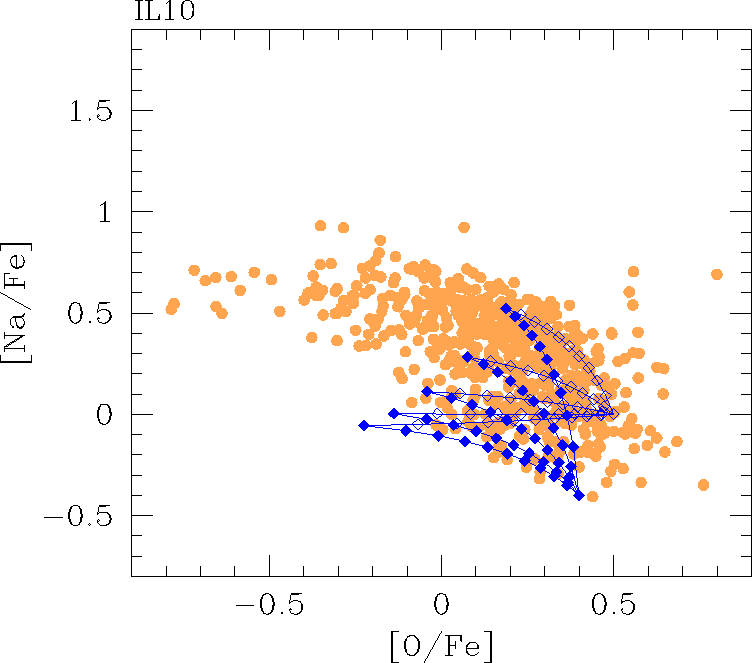}}
\end{minipage}
\hfill
\begin{minipage}{0.48\textwidth}
\resizebox{0.8\hsize}{!}{\includegraphics{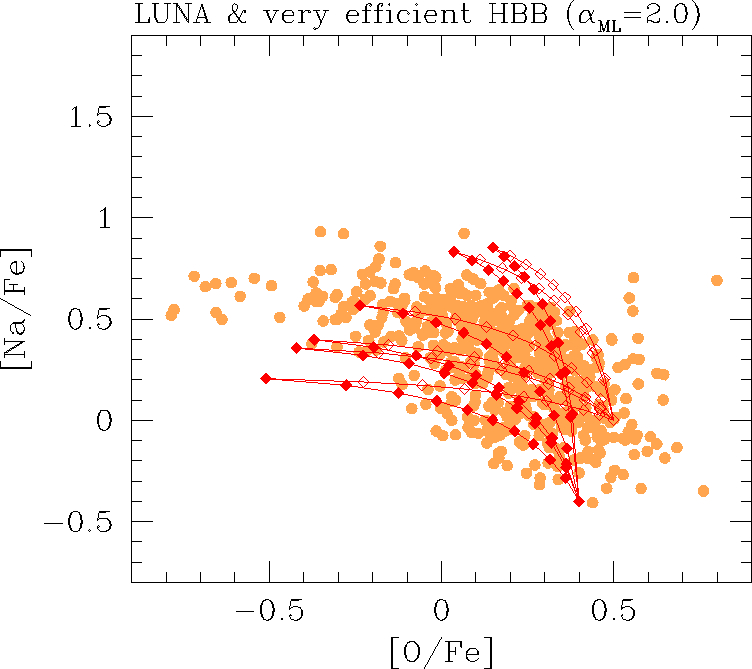}}
\put(-150,140){\Large{C}}
\end{minipage}
\hfill
\begin{minipage}{0.48\textwidth}
\resizebox{0.8\hsize}{!}{\includegraphics{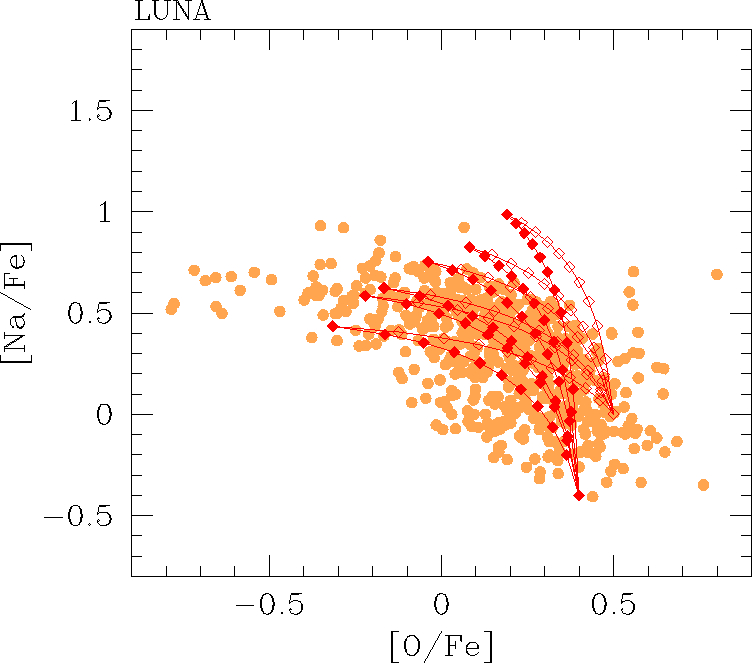}}
\put(-150,140){\Large{A}}
\end{minipage}
\hfill
\begin{minipage}{0.48\textwidth}
\resizebox{0.8\hsize}{!}{\includegraphics{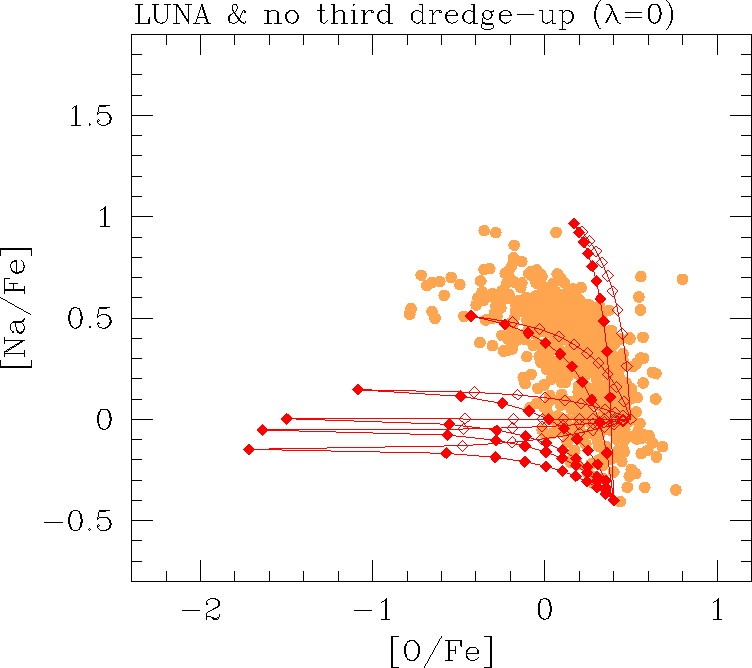}}
\put(-150,140){\Large{D}}
\end{minipage}
\hfill
\caption{O-Na anti-correlation in stars of GGCs. Observed data are the same as in figure~\ref{fig_nao}. 
The models correspond to a range of initial masses
 from 4.0 $M_{\odot}$ to 5.0 $M_{\odot}$
 in steps of $0.2\, M_{\odot}$. Lower mass models, $M_{\rm i} < 4.0\,M_{\odot}$,
 are not included because mostly too far from the observed anti-correlation.
 Following equation~(\ref{eq_dilution}) two dilution curves 
 (solid and dashed lines) have been applied to each AGB model, 
  corresponding to two choices of the
   pristine gas' composition.  Each dot along the curves refers to a
   given value of the dilution fraction $f_{p}$, which is made
   increase from 0 (pure AGB ejecta) to 1 (pristine gas) in steps of
   0.1. The models are the same as in Fig.~\ref{fig_nao}. See the text
   for more explanation.}
\label{fig_nao_dil}
\end{figure*}

\begin{figure*}
\centering
\begin{minipage}{0.48\textwidth}
\resizebox{0.8\hsize}{!}{\includegraphics{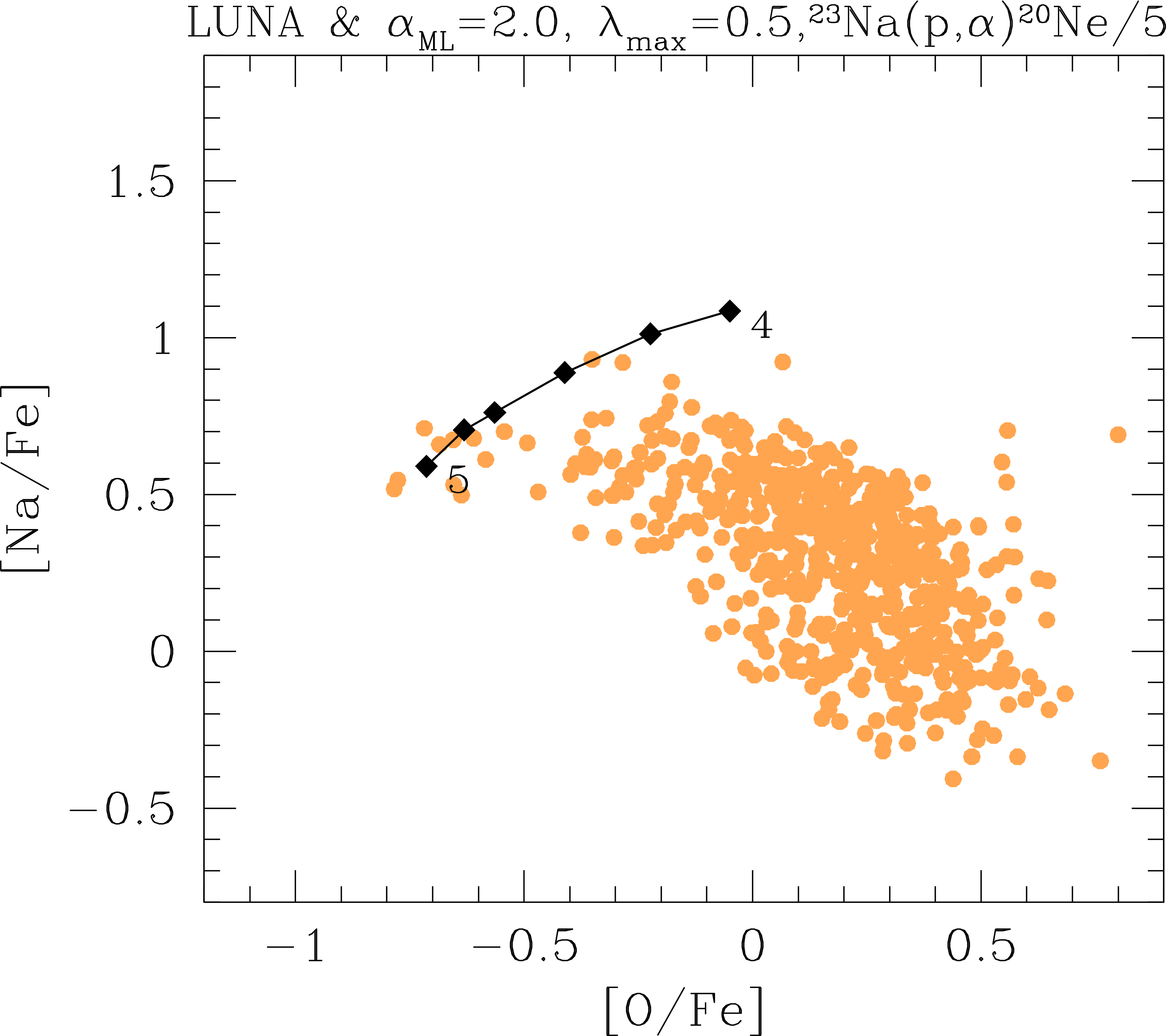}}
\put(-150,140){\Large{E}}
\end{minipage}
\hfill
\begin{minipage}{0.48\textwidth}
\resizebox{0.8\hsize}{!}{\includegraphics{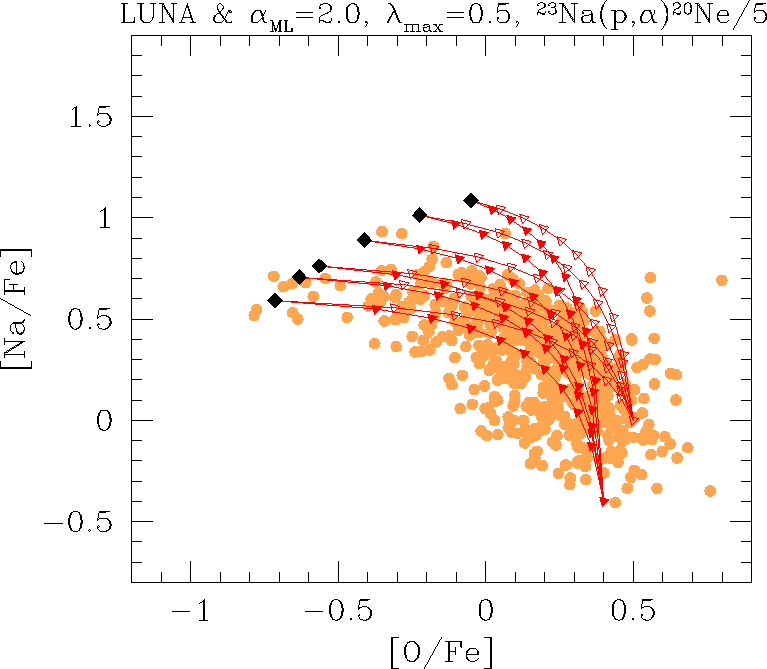}}
\put(-150,140){\Large{E}}
\end{minipage}
\hfill
\caption{The same as in Figs \ref{fig_nao} and \ref{fig_nao_dil}, but referred to the set $E$ of AGB models, characterized by a very efficient HBB, moderate third dredge-up, and a reduced rate for $^{23}$Na$(p,\alpha)^{20}$Na by a factor of 5,  so as to limit the destruction of sodium.The models correspond to a range of initial masses from 4.0 $M_{\odot}$ to 5.0 $M_{\odot}$ in steps of $0.2\, M_{\odot}$.}
\label{fig_nao_tuned}
\end{figure*}

\begin{figure*}
\centering
\begin{minipage}{0.48\textwidth}
\resizebox{0.8\hsize}{!}{\includegraphics{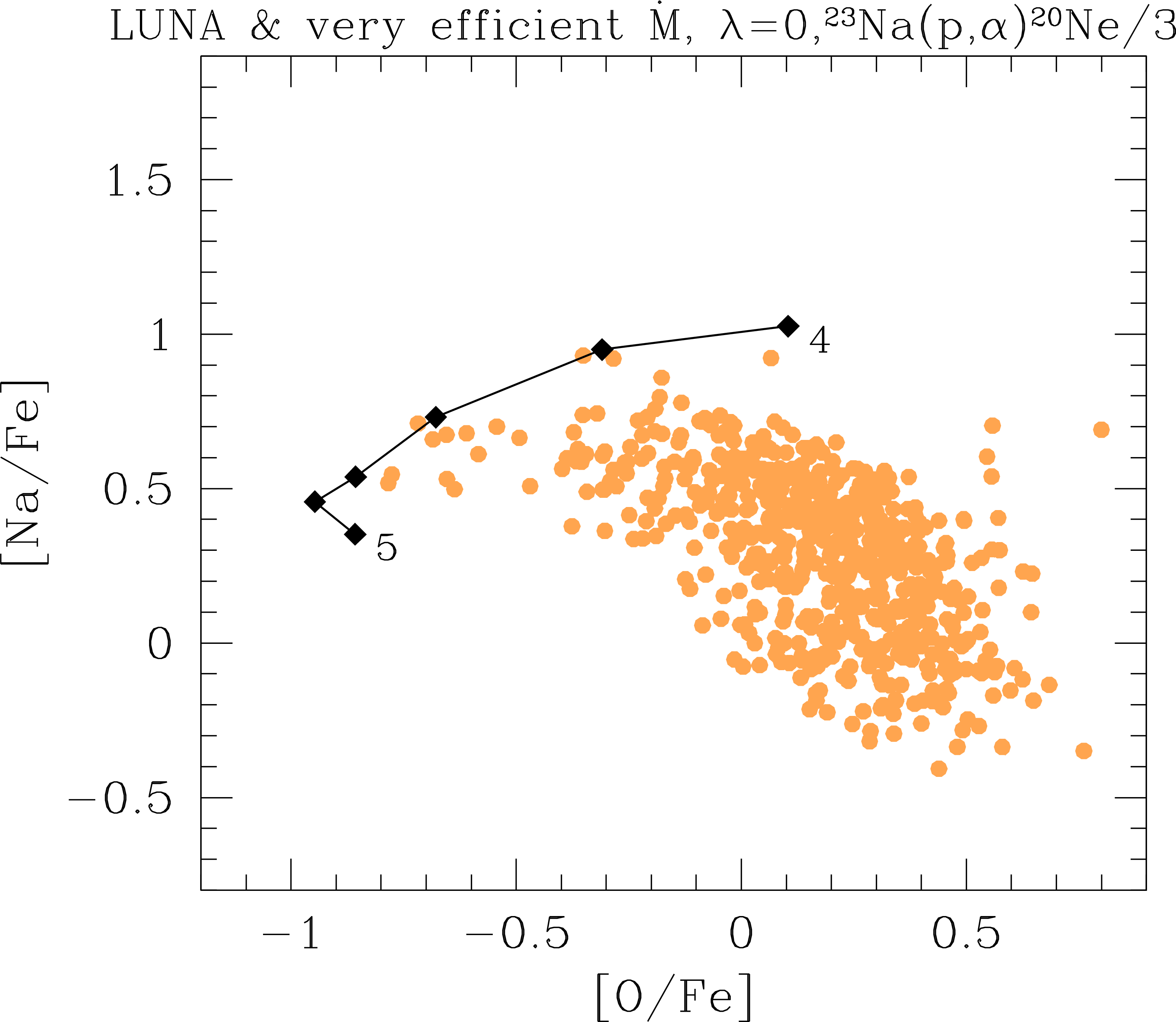}}
\put(-150,140){\Large{F}}
\end{minipage}
\hfill
\begin{minipage}{0.48\textwidth}
\resizebox{0.8\hsize}{!}{\includegraphics{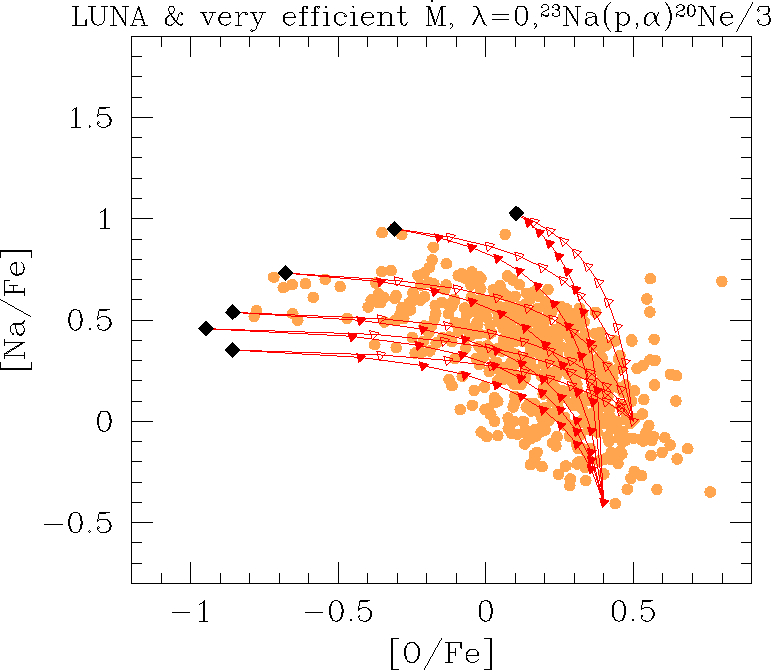}}
\put(-150,140){\Large{F}}
\end{minipage}
\hfill
\caption{The same as in Figs \ref{fig_nao} and \ref{fig_nao_dil}, but referred to the set $F$ of AGB models, characterized by efficient mass loss, no third dredge-up, and a reduced rate for $^{23}$Na$(p,\alpha)^{20}$Na by a factor of 3, so as to limit the destruction of sodium.}
\label{fig_nao_nodup_tuned}
\end{figure*}

A feature common to all panels of Fig.~\ref{fig_nao} is that,
independently of the adopted input physics, the sequence of AGB models
at increasing initial stellar mass runs crosswise the observed
anti-correlation, the higher-mass ones reaching lower [O/Fe]
values. This trend has already been  reported in the literature
\citep[see e.g.,][]{VenturaMarigo_10}.  The only way to make the
stellar models bend over the populated region is to invoke a dilution
process with gas of pristine composition, which basically shares the
same chemical pattern as the field stars of the same [Fe/H].

According to a present-day scenario the observed anti-correlation would
be the result of multiple star formation episodes within GGCs, in
which the ejecta of AGB stars from a first generation polluted the gas
involved in the subsequent secondary star formation events
\citep{VenturaDantona_08}. In this framework GGC stars that populate
the upper region of the anti-correlation (high Na, low O) would
exhibit the chemical abundances of pure AGB ejecta, while stars on the
opposite extreme (low Na, high O) would sample a pristine 
composition, typical of the first generation.  In between are all the GGC stars
born out of a mixture in which the AGB ejecta were partially diluted
into a pristine gas.

In this simplified picture low-metallicity AGB models should be found in the
upper part of the observed anti-correlation.
Looking at Fig.~\ref{fig_nao} we note that depending on the assumed rate $^{22}$Ne$(p,\gamma)^{23}$Na, the sequence of AGB models
change their location significantly. In particular,
the NACRE models are characterized by high [Na/Fe] 
and hardly intersect the data but for the highest stellar masses, 
the IL10 models cross the anti-correlation in the middle not touching the Na-richest, O-poorest points, 
the LUNA sequence attain Na abundances consistent with the upper extreme of the
anti-correlation,
but fails to reach the points with the lowest oxygen abundance,
i.e. [O/Fe]$<-0.4$.
We address this point in Section~\ref{ssec_onarep}.

It is now interesting to examine the behavior of the models when a simple dilution model is adopted.
The dilution effect can be mimicked according to the formula \citep{Conroy_12}:
\begin{equation}
[Y_{\rm i}/{\rm Fe}] = \log\left((1-f_p)10^{[Y_{\rm i}/{\rm Fe}]_o}
+f_p 10^{[Y_{\rm i}/{\rm Fe}]_p]}\right)\,,
\label{eq_dilution}
\end{equation}
where the subscripts $o$ and $p$ refer to the original pristine gas and 
the pure AGB ejecta, and  $f_{p}$ is the fraction of the AGB ejecta mixed into the gas.

For each set of models we applied Eq.~\ref{eq_dilution} to draw a
dilution curve, which starts at $[Y_{\rm i}/{\rm Fe}]_p$ given by the AGB
evolutionary calculations (with $f_p=0$) and ends at a point having coordinates
($\mathrm{[O/Fe]}_o$, $\mathrm{[Na/Fe]}_o$); with $f_p=1$).  For this
latter we assume two combinations (0.4,-0.3) and (0.5, 0.0) to mimic
some dispersion in the [Na/Fe]$_o$ and [O/Fe]$_o$ ratios, which is
present in the observed data.

By eye, the set of LUNA models at the bottom-left panel of
Fig.~\ref{fig_nao_dil} seems to reproduce better  
the trend of O-Na anti-correlation,
compared to the other cases. However, we note that data at lower [Na/Fe] are not completely 
covered by our most massive TP-AGB models (up to $M_{\rm i} = 5 M_{\odot}$). 
In this respect the impact of other AGB model prescriptions (i.e. efficiency of mass loss, HBB, and third dredge-up) 
may be important and are analyzed later in this section.

We caution that the relatively good match  of our reference
LUNA models cannot be taken as a full support to the AGB star
hypothesis. In fact, these models are characterized by an efficient third dredge-up, 
which produces a net increase in the CNO abundance in the ejecta, at
variance with the observational indication that in various GGCs 
stars  that belong to the first and second populations, have constant CNO,
within the errors, or relatively similar \citep[e.g.,][]{Ivans_etal99, Carretta_etal05}.
Recent spectroscopic observations \citep[e.g., ][]{Yong_etal09, Yong_etal15} have revealed 
a much more complex situation: 
there are stars belonging to the same clusters that 
exhibit non-negligible variations of the CNO, others
that show a constant CNO abundance. 
Given this intricate picture we analyse various degrees 
of CNO enrichment in Section~\ref{ssec_onarep}.

In order to keep the increase of the CNO abundance low in the AGB
envelopes a possibility is to invoke that almost no third dredge-up took
place at thermal pulses. In this way the ejecta would exhibit
the nucleosynthesis fingerprint of an (almost) pure NeNa cycle. In the
models this condition can be obtained assuming a very efficient
mass-loss rate and/or imposing that depth of the third dredge-up events
was small (low $\lambda$).

To explore the impact of these assumptions let us analyze the set of
TP-AGB evolutionary calculations referred to as $B$, $C$, and $D$ in Table~\ref{tab:models}. 
Relevant properties of the ejecta are presented in
Table~\ref{tab_ejecta}.

The quantity $R_{\rm cno}$ is defined as the ratio between
the average CNO abundance
in the ejecta and the initial value at the time the star formed.
We note that in our adopted definition of  $R_{\rm cno}$  
the abundances are expressed by number and not by mass fraction since during 
CNO cycle operation what is conserved is the number
of the catalysts and not their mass.

The results of our calculations 
are shown in Figs~\ref{fig_nao} and \ref{fig_nao_dil} 
(see the label at the top of each panel for identification). 
As to the sets $B$ and $C$, they are both characterized by a shorter
TP-AGB evolution, which reduces the number of TPs, hence limiting the CNO
increase at the surface.  At the same time, the shortcoming is that
the most massive AGB models considered here ($M_{\rm i} \ge 3.8\,
M_{\odot}$) tend to produce sodium ejecta that are lower than the standard case,
and do not reach the upper extreme of the anti-correlation. This would imply that the O-Na anti-correlation is caused by AGB stars within a very narrow mass range,  
which requires an extremely fine-tuned initial-mass function.  

In the case $D$ with $\lambda=0$ the CNO abundance is unchanged,
but on the O-Na diagram 
the agreement is poor as the most massive AGB stars experience a 
significant depletion of oxygen, whereas their sodium abundance becomes even lower.
In fact  
no fresh $^{22}$Ne is injected into the envelope at TPs and 
when the $^{22}$Ne$(p,\gamma)^{23}$Na reaction is reactivated during HBB 
no significant amount of $^{23}$Na is synthesized.
Moreover, as already mentioned in Sec.~\ref{ssec_evunc}, models without third dredge-up tend to have longer TP-AGB lifetimes (mass loss is less efficient because of their higher effective temperatures), so that a larger amount of oxygen is burnt into nitrogen.

\subsection{Can we recover the  Na-rich, O-poor extreme of the anti-correlation? }
\label{ssec_onarep}
All AGB models described so far are not able to extend into the O-poor extreme 
of the anti-correlation, matching the sodium abundances at the same time. 
The inability of AGB models to 
reach   [O/Fe]$< -0.5$ has been already reported 
by \citet{Dercole_etal12} who invoked the occurrence  of 
an extra-mixing process during the red giant branch phase of GGC stars.
 
More generally, 
examining the available AGB ejecta in the literature we realize that 
three main 
issues affect their suitability to represent the extreme composition
of the first stellar generation in GGCs \citep[see also][]{Dantona_etal16}.
Namely, to our knowledge, 
no existing AGB (or super-AGB) model has shown to fulfill the whole
set of conditions:
\begin{itemize}
\item $\mathrm{[O/Fe]< -0.5}$
\item $\mathrm{0.5 \la [Na/Fe] \la 0.8}$
\item$R_{\rm cno} \la 3-4$, or more stringently,  $R_{\rm cno} \simeq 1$. 
\end{itemize}  

The first two conditions, which apply to the upper extreme 
of the anti-correlation, are difficult to meet since a more efficient 
destruction of oxygen via the ON cycle is usually accompanied by an efficient 
destruction of sodium through the $\mathrm{^{23}Na(p,\alpha)^{20}Ne}$ reaction, 
and to a lesser extent through the $\mathrm{^{23}Na(p,\gamma)^{24}Mg}$.
This trend is more pronounced with increasing stellar mass, as clearly 
shown in all panels of Fig.~\ref{fig_nao}.

A way to increase the overall sodium production is to assume an efficient third
dredge-up, so that newly synthesized  $^{22}$Ne can be injected into 
the envelope and later burnt into $^{23}$Na. But this brings along the 
problem of increasing the CNO abundance, yielding $R_{\rm cno} >>1$,
as shown in Table~\ref{tab_ejecta}.

An alternative  possibility is that of lowering 
the destruction of sodium, by reducing the current rate for 
$\mathrm{^{23}Na(p,\alpha)^{20}Ne}$ reaction. This suggestion has been 
put forward by \citet{VenturaDantona_06}, and more recently by 
\citet{Dantona_etal16, Renzini_etal15, Dorazi_etal13}.

In view of the above, we single out an optimal set of
AGB model prescriptions that best reproduce  
the chemical constraints on Na, O, and CNO content,  which
characterize the upper extreme of the  anti-correlation. 

  To achieve this goal we follow a sort of ``calibration path'',
  which requires several model calculations and tests.  
  For a given level of third dredge-up efficiency, we first adjust 
  the mixing-length parameter and the mass loss to obtain the right 
  temperature evolution at the base of the convective envelope that produces 
  the right O-depletion in the average ejecta. Clearly, some mild degeneracy 
  between convection and  mass loss efficiencies is present, but the 
  uncertainty range  is small for reasonable choices of the parameters.
  Then, we reduce
  the destruction rate of $^{23}$Na$(p,\alpha)^{20}$Ne by the suitable 
  factor that allows to reach the required Na enrichment.
 
We summarize here the final results of our investigation.
Let us start from the constraint on the CNO abundance, and consider
two possible requirements expressed by $R_{\rm cno} \la 3-4$ and
$R_{\rm cno}=1$, respectively. They define two classes of
TP-AGB models.

The requirement $R_{\rm cno} \la 3-4$ implies that some dredge-up is
allowed to take place during the TP-AGB evolution.  Under these
conditions, our best set of models (named $E$ in Table~\ref{tab_test})
is calculated assuming a moderate third dredge-up, with a maximum
efficiency $\lambda_{\rm max}=0.5$, which produces $R_{\rm cno} \la
4-5$ for initial masses $M_{\rm i} \ge 4.4\, M_{\odot}$.  We are able
to reach the lowest [O/Fe] by increasing the mixing length parameter to
$\alpha_{\rm ML}=2.0$, which causes a very efficient HBB.  At the same
time, we prevent a large destruction of sodium by reducing the
IL10 rate for $\mathrm{^{23}Na(p,\alpha)^{20}Ne}$ by a factor of $5$.
All other prescriptions are the same as in our reference $M13$ set.
 
The results are presented in Fig.~\ref{fig_nao_tuned} and the relevant
characteristics of the ejecta are listed in Table~\ref{tab_ejecta}.
This set of AGB models is  able, for the first time, 
to reproduce the Na-rich, O-poor extreme of the O-Na anti-correlation,
while keeping a mild CNO increase.
The most massive AGB models, with 
$M_{\rm i}=4.6-5.0\,M_{\odot}$ reach the stars with
the lowest [O/Fe] as a consequence of a suitable combination of efficient 
HBB and mass loss,  without the need of invoking extra-mixing episodes 
as suggested by \citet{Dercole_etal12}.
At the same time, we confirm previous suggestions \citep{Dantona_etal16, Renzini_etal15} 
about the need of decreasing the destruction rate of sodium.

The requirement $R_{\rm cno}=1$ implies that no third dredge-up occurred.
Under this stringent assumption, our best performing set of models (named $F$ in Table~\ref{tab_test}),
is calculated with  $\alpha_{\rm ML}=1.74$, adopting a more efficient mass-loss prescription
\citep[][with $\eta=0.03$]{Bloecker_95}, and reducing the $\mathrm{^{23}Na(p,\alpha)^{20}Ne}$ 
rate by a factor of 3. As before, all other prescriptions are the same as in $M13$.
The results are shown in Fig.~\ref{fig_nao_nodup_tuned} and the properties of the corresponding 
ejecta are summarized in Table~\ref{tab_ejecta}. The upper extreme of the anti-correlation and its
dispersion is also well described by the average abundance of the AGB models with initial masses 
$4.0\, M_{\odot} \la M_{\rm i} \la 5.0\, M_{\odot}$.

Compared to the set $E$ with $R_{\rm cno}>1$, in models $F$ we apply a few changes in the input prescriptions which
are explained as follows.
The absence of dredge-up episodes in models $F$ makes both the atmospheres
and the convective envelopes 
somewhat hotter, as a consequence of the lower opacities\footnote{Equation of state and detailed Rosseland mean opacities are computed with 
the \texttt{\AE SOPUS} at each time step during the evolution, consistently with the chemical composition.}.
This leads to increase the strength of HBB, so that  $\alpha_{\rm ML}=1.74$ (instead of 2) already allows us to obtain the required oxygen depletion.
At the same time, the TP-AGB evolution is a little shorter which prevents
an excessive destruction 
of both oxygen and sodium. Also in this case 
we have to limit the consumption of sodium by reducing
the nuclear rate of proton captures.

In this context model predictions are heterogeneous.  
On one hand, relatively lower efficiencies of the third dredge-up are predicted at increasing core 
mass as a consequence of the weaker thermal pulses\footnote
{when the maximum He-burning luminosity attained during thermal pulses is lower.}
\citep{VenturaDantona_08, Cristallo_etal15}.
In addition, the combined action of hot dredge up
\citep{GorielySiess_04} and hot bottom burning limits the occurrence of
the third dredge up in stars with initial mass $>\, 5-6\,M_{\odot}$ 
\citep[see the discussion in][]{ Straniero_etal14}.
Interestingly, independent indications towards a modest third dredge-up in stars with $M_{\rm i} \approx 3-4 M_{\odot}$ are also derived from the
analysis of the Galactic initial-final mass relation \citep{Kalirai_etal14}.
On the other hand, other AGB models predict that the efficiency of the 
 the third dredge-up increases with the stellar mass 
\citep[e.g.,][]{KarakasLattanzio_14, Herwig_04, Karakas_etal02}. 
On observational grounds, the high Rb abundances measured 
in luminous AGB stars in the Magellanic Clouds and in the Galaxy hint that stars with HBB    do experience the third dredge-up
\citep{Zamora_etal14, Garcia-Hernandez_etal09, Garcia-Hernandez_etal06}. 
It follows that quantifying the  efficiency of the third dredge-up in massive AGB stars is still an open issue and it can be reasonably treated as a free parameter in AGB models to explore the impact of various assumptions, in a way similar to what we performed in this study.

\begin{figure}
\centering
\resizebox{\hsize}{!}{\includegraphics{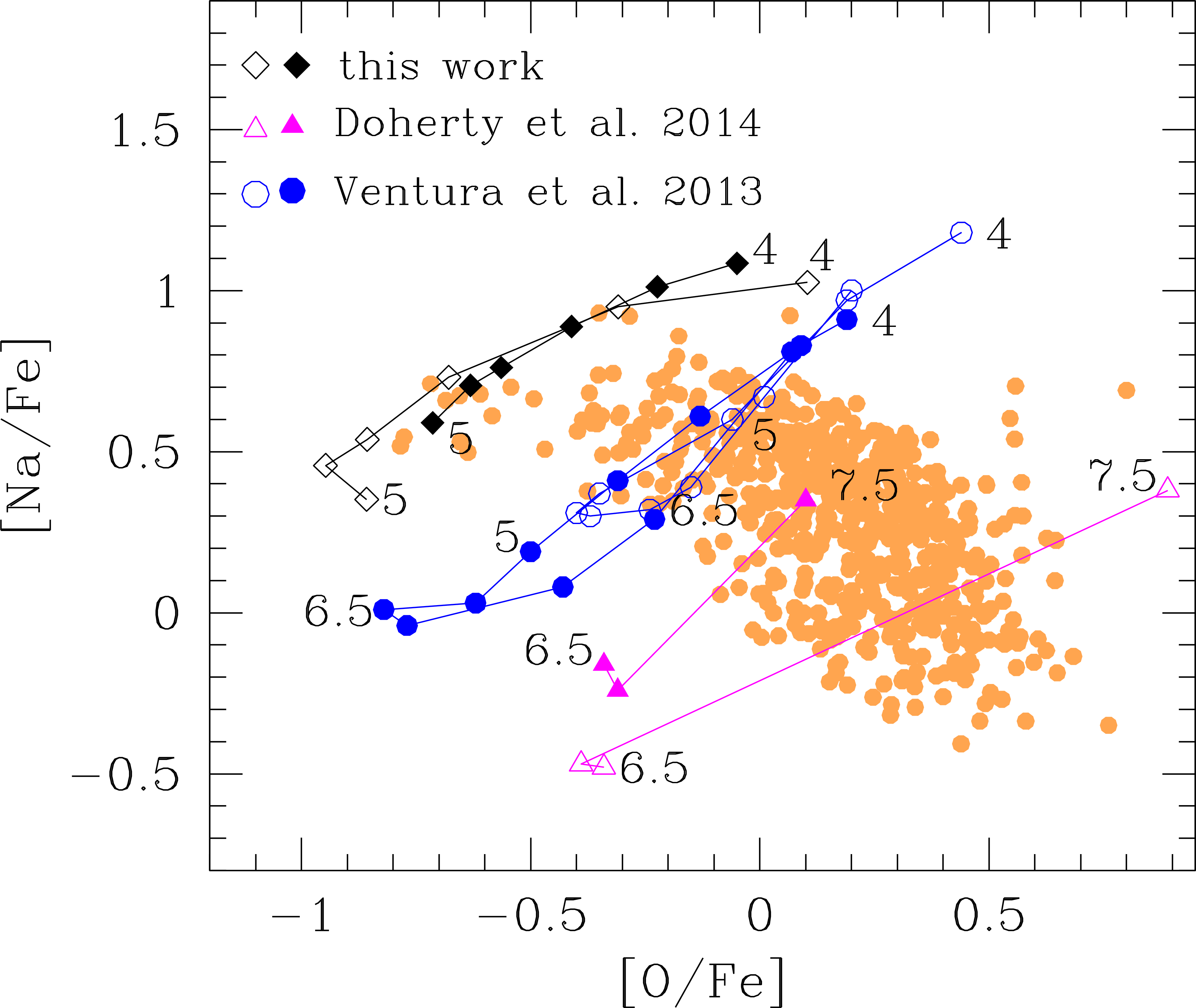}}
\caption{Comparison of mean oxygen and sodium abundances  in the
  AGB and super-AGB ejecta computed by various authors.
  Our best fitting models ($E$ and $F$) are shown together with the predictions
  of \citet{Ventura_etal13} (initial masses in the range $4.0-8.0\,M_{\odot}$; filled circles for $Z_{\rm i}=0.0003$; empty circles for $Z_{\rm i}=0.001$);
  \citet[][; initial masses in the range $6.5-7.5\,M_{\odot}$; filled triangles for $Z_{\rm i}=0.001$; empty triangles for $Z_{\rm i}=0.0001$; mass loss prescription: \citet{Bloecker_95} with $\eta=0.02$]{Doherty_etal14b}.
A few selected values of $M_{\rm i}$ are indicated (in $M_{\odot}$) nearby the corresponding models.}
\label{fig_naocomp}
\end{figure}

For comparison, in Fig.~\ref{fig_naocomp} we show our best-fitting
models ($E$ and $F$) together with the predictions of other two  theoretical
studies,
namely \citet{Ventura_etal13}, and \citet{Doherty_etal14b}, which
include AGB and super-AGB models.
We note that quite different abundances characterize the different
sets of models, even when sharing the same, or similar, initial mass
and metallicity.
In particular, as already discussed by these authors, the O-poor and
Na-rich extreme of the anti-correlation is not reached by the models,
in the framework of their adopted prescriptions.
As already mentioned, \citet{Dercole_etal12} suggested that
this difficulty may be overcome assuming deep mixing  during
the RGB phase of the second generation stars forming in a gas
with high helium abundance.

On the other hand, our analysis shows that the extreme of the
O-Na anti-correlation
may, in principle, be reproduced with pure ejecta of AGB stars, without invoking extra-mixing episodes in other phases.

In particular, our calculations demonstrate quantitatively that a sizable reduction (by a factor of 3-5) 
of the rate of the reaction $\mathrm{^{23}Na(p,\alpha)^{20}Ne}$ is necessary to prevent an excessive sodium destruction when the third dredge-up is not efficient or even absent. 
We should caution, however, that such a drastic change in the rate is not supported by recent nuclear cross section studies \citep[][]{Cesaratto_etal13, Iliadis_etal10a}. The present lower-limit estimates allow to reduce the recommended rate by a factor of $\sim 1.2 - 1.3$ at the largest. 

We did not attempt to fulfill
additional chemical constraints, such as those related to the Mg-Al anti-correlation \citep{Carretta_etal15}.
We have verified that no significant magnesium destruction is predicted in AGB  models with the adopted  set of nuclear rates. In this respect we note that our reference rate for $\mathrm{^{25}Mg(p,\gamma)^{26}Al}$ is taken from IL10, while a recent revision with LUNA has increased it by roughly a factor
of $\simeq 2$ at the temperatures relevant for HBB \citep{Straniero_etal13}. 
We plan to adopt the latter rate and to extend our chemical investigation of the Mg and Al isotopes in a follow-up study.  

Also, as shown in Table~\ref{tab_ejecta}, our massive TP-AGB models
exhibit a large helium content in their ejecta (mainly determined by
the second dredge-up on the E-AGB), which would correspond to an
increase of $\Delta Y \simeq 0.1-0.12$ with respect to the assumed
initial value, $Y_{\rm p} =0.2485$. These values are larger than the
typical range $\Delta Y_{\rm max} \simeq 0.01-0.05$ reported by
\citet{Milone_etal14} for a group of GGCs, and may represent a severe
issue to the AGB star scenario \citep{Bastian_etal15}.  We note,
however, that our analysis is focused on the Na-rich, O-poor extreme
of the anti-correlation, which is mainly populated by the stars of the
cluster NGC~2808. For this cluster the helium spread is large, $\Delta
Y_{\rm max} \simeq 0.14$ \citep{Milone_etal12}, consistent with our predictions.

A deeper scrutiny of all these additional chemical constraints requires
a dedicated study on each specific cluster, as well as to extend the analysis to other metallicities, and it is  beyond the original aim of the present paper.

\section{Summary and conclusions}
\label{sec_sum}
In this theoretical study we analyzed the ejecta of $^{22}$Ne and
$^{23}$Na contributed by intermediate-mass stars during their entire
evolution. In particular, we focused on the impact of the new LUNA
measurements of the astrophysical S-factor for the reaction
$^{22}$Ne$(p,\gamma)^{23}$Na.  The new experimental set-up and the
discovery of three new resonances have led to a significant reduction 
in the uncertainty of the rate, which drops from factors of $\simeq 100$ down
to just a few.  At the temperatures most relevant for stellar
evolutionary models the new LUNA rate is significantly lower than
the previous estimate provided by NACRE, but somewhat larger than that of
\citet{Iliadis_etal10a}.

In order to evaluate the current uncertainties that still affect the
ejecta of $^{22}$Ne and $^{23}$Na, and to disentangle those associated
to nuclear physics from those related to other evolutionary aspects,
we calculated a large grid of stellar evolutionary models with initial
masses in the interval from $3\,M_{\odot}$ to $5-6\,M_{\odot}$, for
three values of the initial composition.  For each stellar model, the
entire evolution, from the pre-main sequence to ejection of the
complete envelope, was computed varying a few key model prescriptions,
namely the rate of $^{22}$Ne$(p,\gamma)^{23}$Na, the rate of mass-loss
on the AGB, the efficiency of the third dredge-up, and the
mixing-length parameter used in our adopted theory of convection.

In the light of the results obtained with the new LUNA rate for
$^{22}$Ne$(p,\gamma)^{23}$Na, we provide below a recapitulation
of the most relevant processes that affect the ejecta of $^{22}$Ne and
$^{23}$Na from intermediate-mass stars, the main uncertainty sources,
and the implications we derived in relation to the Na-rich,
O-poor extreme of the O-Na anti-correlation in GGCs.

\begin{itemize}
\item The second dredge-up on the early-AGB causes a significant increase of the surface abundance of $^{23}$Na, up to a factor of $\simeq 10$ in stars with 
high mass and low metallicity.
Correspondingly, the surface concentration of $^{22}$Ne is diminished by
$\approx 30\%$.
These elemental changes hardly depend on the adopted rate for $^{22}$Ne$(p,\gamma)^{23}$Na, while are controlled by other physical parameters,
e.g. the efficiency of mixing and the extension of convective overshoot applied to the inner border of the convective envelope.
During the subsequent TP-AGB phase significant changes in the surface abundances of $^{22}$Ne and $^{23}$Na are caused by the occurrence of third dredge-up events and HBB. 
\item The main effect of the third dredge-up is the injection of fresh $^{22}$Ne into the envelope at thermal pulses, which will be later involved in the NeNa cycle during the next inter-pulse period. The process of HBB 
leads to an initial depletion of $^{23}$Na, followed by an increase of its abundance -- through the reaction  $^{22}$Ne$(p,\gamma)^{23}$Na -- when 
$^{23}$Na and $^{24}$Mg reach the nuclear equilibrium.
The quantitative details of these general trends critically depend on the
rate assumed for $^{22}$Ne$(p,\gamma)^{23}$Na.
\item Comparing the results for $^{22}$Ne and $^{23}$Na obtained with our reference set of input prescriptions for the AGB evolution, but varying the rate  for $^{22}$Ne$(p,\gamma)^{23}$Na, we find that the $^{23}$Na ejecta predicted with the LUNA data are quite lower than those derived with NACRE, and somewhat larger than with IL10. The opposite behavior applies to $^{22}$Ne.
\item Comparing the results for $^{22}$Ne and $^{23}$Na obtained with the recommended LUNA rate as well as the associated lower and upper limits, we estimated the current 
  uncertainties of the chemical ejecta directly ascribed to the nuclear S-factor.
At low metallicity the amplitudes of the largest error bars reach factors of $\simeq 2$ for $^{23}$Na and $\simeq 10-30\%$ for $^{22}$Ne. These uncertainties are significantly lower than those reported in past studies.

\item Other reactions involved in the NeNa cycle may contribute to the
  nuclear uncertainties of the $^{22}$Ne and $^{23}$Na ejecta, in
  particular the destruction rates for sodium,
  i.e. $\mathrm{^{23}Na(p,\alpha)^{20}Ne}$ and
  $\mathrm{^{23}Na(p,\gamma)^{24}Mg}$. Our present-day knowledge,
  based on nuclear cross section experiments \citep{Iliadis_etal10a,
    Cesaratto_etal13}, indicates that destruction of sodium is largely
  dominated by the $\mathrm{^{23}Na(p,\alpha)^{20}Ne}$ reaction at the
  temperatures relevant for HBB $\mathrm{0.07\, GK \la T \la 1.1\,
    GK}$. The estimated lower and upper limit uncertainties for this
  rate are, however, relatively low, not exceeding $20-30 \%$.

\item The remaining uncertainties of the chemical ejecta for $^{22}$Ne
  and $^{23}$Na are mainly dominated by stellar evolutionary aspects,
  in particular the efficiency of convection, mass loss, and third
  dredge-up events.  While the efficiencies of mass loss and
  convection mainly control the duration of HBB and the activation of the
  nuclear cycles, the third dredge-up has a direct effect on the total
  abundance of the isotopes that enter in the cycles.  In fact, the
  amount of material that is dredged-up to the surface determines the
  amount of new $^{22}$Ne that is added into the envelope and later
  converted into $^{23}$Na by the $^{22}$Ne$(p,\gamma)^{23}$Na reaction.
  Our tests indicate only varying the efficiency of the third
  dredge-up in low-metallicity AGB stars from high values ($\lambda
  \simeq 1$) to zero ($\lambda =0$) causes a reduction of the
  $^{22}$Ne ejecta by factors of 10-20, as well a reduction $^{23}$Na
  ejecta by factors of 4-5.

\item We examined our results in relation to the hypothesis that the
  observed O-Na anti-correlation observed in GGCs' stars is due to 
  processed material in the ejecta of low-metallicity AGB stars.  The
  ejecta obtained with the LUNA rate, together with our reference AGB
  model prescriptions, are able to recover the most Na-enriched stars of the
  anti-correlation, which are expected to exhibit the chemical
  composition of pure AGB ejecta.  By adopting a simple dilution
  model, the general morphology of the anti-correlation is also
  satisfactorily reproduced.  At the same time, however, we predict a
  sizable increase of the CNO content in the AGB ejecta (caused
  by the efficient third dredge-up assumed in the models), a feature that is
  at variance with the observations.

  On the other hand,
  assuming no or weak third dredge-up, hence no or little $^{22}$Ne
  enrichment in the envelope, models are not able to produce the
  highest [Na/Fe] values on the upper extreme of the anti-correlation.
  This difficulty holds also under the assumptions of very high mass
  loss and/or strong HBB, as in both cases the TP-AGB phase is
  shortened and no significant replenishment of $^{22}$Ne is
  predicted.  The contribution from super-AGB stars, not explicitly
  treated in this work, is likely not to improve the situation since
  sodium ejecta tend to decrease at increasing stellar mass
  \citep{Doherty_etal14b, Dercole_etal10}.

\item Starting from our reference AGB models, we changed various input
  prescriptions to verify whether the chemical constraints on sodium,
  oxygen and CNO content can be simultaneously fulfilled.  After
  several tests, we singled out two optimal sets of AGB model
  assumptions under which the Na-rich, O-poor extreme of the
  anti-correlation is, for the first time, reproduced by pure AGB ejecta 
  (without invoking external processes such as extra-mixing on the RGB).

In the first set of models we allow a moderate third dredge-up, so
that the CNO abundance increases by a factor $\la 4-5$.  Matching the
oxygen and sodium abundances requires an efficient HBB and a
significant reduction, by a factor of 5, of the rate for
$\mathrm{^{23}Na(p,\alpha)^{20}Ne}$, in combination with the LUNA rate
for $^{22}$Ne$(p,\gamma)^{23}$Na.

In the second set of models we impose the absence of any third
dredge-up event, in order to keep the total CNO abundance constant. 
In this case the extreme of the anti-correlation is
also reached by adopting moderately different prescriptions for the
mass loss, HBB, and the sodium destruction rate (with a reduction by a
factor of 3).

\item Such "calibrated" modifications (by a factor of 3-5) of the
  nuclear rate for $\mathrm{^{23}Na(p,\alpha)^{20}Ne}$ 
  confirm quantitatively earlier suggestions by independent studies 
  \citep{Dantona_etal16, Renzini_etal15, VenturaDantona_06}. 
  At the same time, they appear to be
  too large if one considers that present lower-limit estimates of the
  nuclear cross section allow a maximum reduction by a
  factor of $\simeq 1.3$.  At present, this poses a severe problem that
  undermines the suitability of the AGB star solution in
  the context of the GGCs anti-correlations.  Future
  nuclear experiments will be of key relevance to quantify more
  precisely the extent of sodium destruction in the stellar sites
  where the NeNa cycle operates.

\item Other constraints, such as the magnesium depletion and the
  helium spread of different stellar populations, are not explicitly
  considered in the chemical calibration.  
    We note that our AGB ejecta at low metallicity, likewise many
    other sets in the literature, are highly enriched in helium as a
    consequence of the second dredge-up. In the framework of a simple
    dilution model, this would likely imply a large helium spread
    between stars of the first and second generations, and  
    therefore may represent
    a serious difficulty to the AGB scenario, as discussed by
    \citet[][but see also \citet{Chantereau_etal16} for a different approach]{Bastian_etal15}.

In conclusion, the AGB star hypothesis still deserves  further
quantitative analyses, which may be performed through stellar
evolution experiments similar to those we have carried out in this
study.

\end{itemize}

\section*{Acknowledgments}
This research is mainly supported by the University of Padova and by the 
ERC Consolidator Grant funding scheme ({\em project STARKEY}, G.A. n.~615604). 
The LUNA experiment was supported by INFN, DFG (BE 4100-2/1), NAVI (HGF VH-VI-417) and OTKA (K101328).
We thank Maria Lugaro for helpful comments and discussion.
%%%%%%%%%%%%%%%%%%%% REFERENCES %%%%%%%%%%%%%%%%%%

\bibliographystyle{mn2e}
\bibliography{biblio_luna} 

\end{document}